\documentclass[11pt]{article}
\pdfoutput=1

\usepackage{euscript}
\usepackage{amssymb}
\usepackage{amsfonts}
\usepackage{amsbsy}
\usepackage{epsfig}
\usepackage{amsthm}
\usepackage{amscd}
\usepackage{amstext}
\usepackage{verbatim}
\usepackage{amsmath}
\usepackage{cancel}
\usepackage{capt-of}
\usepackage{empheq}
\usepackage{subfigure}

\usepackage[nosort]{cite}
\usepackage{bm}
\usepackage{authblk}
\usepackage{esint}
\usepackage[T1]{fontenc}
\usepackage{mathdots}
\usepackage[centertableaux]{ytableau}

\usepackage[utf8]{inputenc}
\usepackage{graphicx}
\usepackage{physics}
\usepackage{mathtools}
\usepackage{bbold}

\usepackage{setspace}
\usepackage{colortbl}
\usepackage{xcolor}

\usepackage[pdftex,linktoc=all]{hyperref}
\hypersetup{ 
colorlinks=true, 
linkcolor=black, 
citecolor=red, 
}

\def\ben{\begin{equation}}
\def\een{\end{equation}}

\def\nn{\nonumber}

\let\pa=\partial
\def\be{\begin{equation}}
\def\ee{\end{equation}}
\def\beq{\begin{equation}}
\def\eeq{\end{equation}}
\def\ba{\begin{array}}
\def\ea{\end{array}}

\def\dalemb#1#2{{\vbox{\hrule height .#2pt
       \hbox{\vrule width.#2pt height#1pt \kern#1pt
               \vrule width.#2pt}
       \hrule height.#2pt}}}

\newcommand{\bea}{\begin{eqnarray}}
\newcommand{\eea}{\end{eqnarray}}
\def\ep{{\epsilon}}
\def\vep{{\varepsilon}}

\makeatletter
\newcommand*\bigcdot{\mathpalette\bigcdot@{.5}}
\newcommand*\bigcdot@[2]{\mathbin{\vcenter{\hbox{\scalebox{#2}{$\m@th#1\bullet$}}}}}
\makeatother

\renewcommand{\eqref}[1]{(\ref{#1})}

\def\Z{{{\mathbb Z}}}

\def\N{{{\mathbb N}}}

\title{Mixmaster chaos in an AdS black hole interior}
\author{Marine~De~Clerck, Sean~A.~Hartnoll and Jorge~E.~Santos}
\affil{\it Department of Applied Mathematics and Theoretical Physics, \\
\it University of Cambridge, Cambridge CB3 0WA, UK
}

\date{}

\begin{document}

\maketitle

\begin{abstract}

We derive gravitational backgrounds that are asymptotically Anti-de Sitter, have a regular black hole horizon and which deep in the interior exhibit mixmaster chaotic dynamics. The solutions are obtained by coupling gravity with a negative cosmological constant to three massive vector fields, within an Ansatz that reduces to ordinary differential equations. At late interior times the equations are identical to those analysed in depth by Misner and by Belinskii-Khalatnikov-Lifshitz fifty years ago. We review and extend known classical and semiclassical results on the interior chaos, formulated as both a dynamical system of `Kasner eras' and as a hyperbolic billiards problem. The volume of the universe collapses doubly-exponentially over each Kasner era. A remarkable feature is the emergence of a conserved energy, and hence a `time-independent' Hamiltonian, at asymptotically late interior times. A quantisation of this Hamiltonian exhibits arithmetic chaos associated with the principal congruence subgroup $\Gamma(2)$ of the modular group. We compute a large number of eigenvalues numerically to obtain the spectral form factor. While the spectral statistics is anomalous for a chaotic system, the eigenfunctions themselves display random matrix behaviour.

\end{abstract}

\newpage

\begin{spacing}{1}
\tableofcontents
\end{spacing}

\section{Introduction}

\begin{figure}[h]
    \centering
   \includegraphics[width=0.5\textwidth]{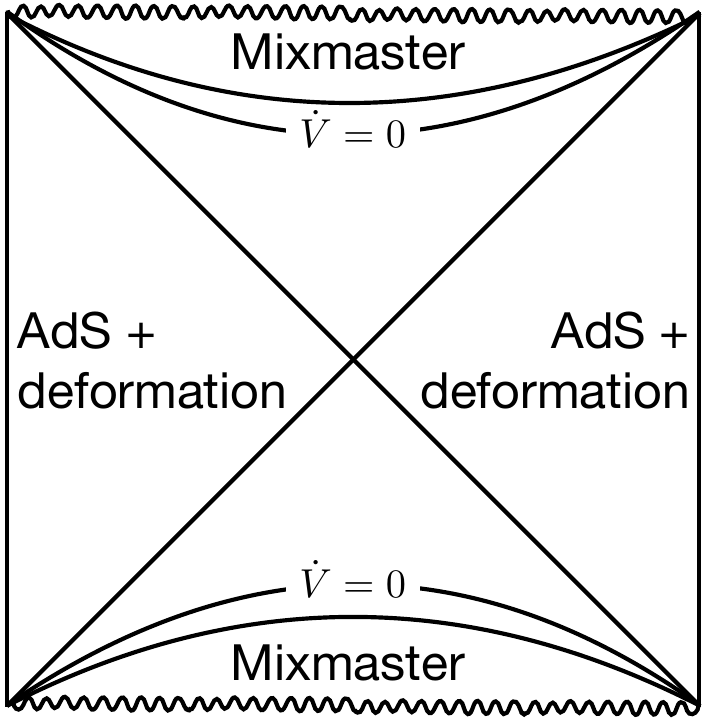}
    \caption{Schematic Penrose diagram for the solutions that we will construct. The thermofield double state of the boundary CFT is deformed by three relevant vector operators, as described in \S\ref{sec:setup} below, leading to a holographic renormalisation group flow away from AdS. The flow deforms the interior of the black hole. Beyond the maximal volume ($\dot V = 0$) slice of the interior, the deformation triggers an infinite chaotic sequence of Kasner epochs as the volume of space collapses towards the singularity.}
    \label{fig:penrose}
\end{figure}

Among the most fascinating behaviours to arise from Einstein’s equation of General Relativity is the onset of chaotic dynamics in the approach to certain cosmological singularities \cite{BKL, lifshitz1971asymptotic, PhysRevLett.22.1071, PhysRev.186.1319}. Such cosmological singularities can arise, in particular, in the interior of black holes. We are not aware, however, of an explicit realisation of so-called `mixmaster' or `BKL' chaotic dynamics in a black hole interior. One achievement of this work will be to obtain mixmaster chaos inside a four dimensional asymptotically AdS planar black hole. The setup is illustrated in Fig.~\ref{fig:penrose}.
The entire evolution from AdS boundary to singularity is captured by a small number of ordinary differential equations. These equations can be solved numerically and all the key features determined analytically. That is, we have found a simple holographic “home” for gravitational chaos.

The AdS/CFT correspondence has long held out the prospect of shedding light on black hole singularities \cite{Kraus:2002iv, Fidkowski:2003nf, Festuccia:2005pi}. However, while the dual CFT offers a well-grounded framework to consider eg.~singularity resolution, the holographic meaning of the black hole interior remains unclear. From one perspective, the entire interior is dual to the late time limit of the CFT thermofield double state \cite{Maldacena:2001kr, Maldacena:2013xja, Belin:2018bpg}. Interior time evolution is then merely a bulk gauge transformation. While the interior bulk geometry should be reconstructable in principle from the CFT --- see e.g.~\cite{deBoer:2022zps} for a recent discussion --- various natural boundary probes of the interior \cite{Hartman:2013qma, Stanford:2014jda} avoid the region close to the singularity \cite{Wall:2012uf, Engelhardt:2013tra}. This is intimately related to the fact that the singularity is a “sub-AdS” scale phenomenon. Indeed, the negative cosmological constant becomes irrelevant in the equations of motion close to the singularity.

To find the singularity in the dual CFT, it will presumably be helpful to understand what one is looking for. Until recently, the vast majority of holographic explorations of the black hole interior have considered the Schwarzschild-AdS interior. However, once the interior space starts contracting towards the singularity, interior Schwarzschild-AdS is a highly non-generic solution to the classical equations of motion. A simple holographic manifestation of this fact was given in \cite{Frenkel:2020ysx}: homogeneous, relevant scalar sources at the AdS boundary change the Kasner exponents towards the interior singularity. Recall that Kasner cosmologies have the form ${\rm d}s^2 = -{\rm d}l^2 + \sum_i l^{2 p_i} {\rm d}x_i^2$, with $l$ the proper time to the singularity and $p_i$ the Kasner exponents. The Schwarzschild singularity corresponds to the particular values $\vec p = \{-\frac{1}{3},\frac{2}{3},\frac{2}{3}\}$.
A second simple phenomenon is that such boundary deformations remove Cauchy horizons from the interior of charged black holes \cite{Hartnoll:2020rwq}, replacing them with Kasner regimes similar to those of neutral black holes.

Kasner singularities are characterised by their scaling exponents, and hence 
one might hope to identify these exponents in the dual field theory.
Complicating this picture, subsequent work rediscovered, in a holographic context, the well-known phenomenon of Kasner bounces, in which the Kasner exponents abruptly change \cite{Hartnoll:2020fhc,Cai:2020wrp,VandeMoortel:2021gsp,Li:2023tfa}. See also \cite{Dias:2021afz}. This leads to the notion of a Kasner epoch, of finite duration. Several holographic works found scenarios in which the sequence of epochs connected by bounces continues indefinitely, such that a final asymptotic Kasner regime is never reached \cite{Cai:2021obq, Henneaux:2022ijt, An:2022lvo, Hartnoll:2022rdv, Sword:2022oyg, Cai:2023igv}. These studies were able numerically to evolve the dynamics through a number of Kasner epochs. However, as was understood in the original work from the Landau Institute \cite{BKL, lifshitz1971asymptotic}, reviewed in detail in \S\ref{sec:bkl1} and \S\ref{sec:bkl2} below, the chaotic nature of the dynamics is revealed in the very long-term evolution of so-called Kasner eras. A Kasner era is a potentially large number of Kasner epochs that involves bounces between only two neighbouring `walls’. Numerical evolution through multiple Kasner eras is not straightforward (although see \cite{Berger:2002st}).

The classic papers instead obtained a closed-form recursion relation for the Kasner exponents between eras and established a beautiful connection between the sequence of Kasner eras and the Gauss map \cite{BKL, lifshitz1971asymptotic}. The Gauss map is given in (\ref{eq:b}) below and is a canonical example of a chaotic dynamical system. Our work will realise this same Gauss map in a holographic interior. The Gauss map entails specific stochastic properties about the late time interior evolution. These properties may be robust targets for a dual field theory understanding. For example, the spatial volume of the universe collapses by a doubly exponential amount over a Kasner era, with the average rate of collapse determined by the Kolmogorov-Sinai entropy of the Gauss map --- see (\ref{eq:eq}) and Fig.~\ref{fig:ks} below.

A crucial fact is that the ergodic dynamics of the far interior emerges together with a time-independent Hamiltonian. This is seen most clearly using a formulation in which the evolution of the metric components is mapped onto the motion of a particle on the hyperbolic plane, with the bounces caused by reflections from fixed walls that constrain the motion to within a hyperbolic triangle \cite{Damour:2002et, belinski_henneaux_2017}. To avoid possible confusion: this hyperbolic plane is not in space but rather in the `superspace' of metric components. The walls are exponential potentials that become infinitely steep in the late time limit. In that limit there is no time dependence left in the Hamiltonian for the particle motion. The associated conserved energy allows 
a more refined study of the ergodic dynamics, with interior solutions labelled by their emergent conserved energy. In fact, the equilateral triangular region defined by the potential walls turns out to be half of the fundamental domain of the principal congruence subgroup $\Gamma(2)$ of the modular group. See Fig.~\ref{fig:domain} below. This means that the far interior evolution is described by the mathematically rich field of arithmetic chaos \cite{Bogomolny:1992cj, sarnak1993arithmetic}.

To characterise the arithmetic chaos of the emergent late interior Hamiltonian we quantise the hyperbolic billiard dynamics in \S\ref{sec:WDW}. The eigenfunctions are given by odd automorphic forms of $\Gamma(2)$. To analyse statistical properties of the spectrum, the forms must further be classified according to their transformation under rotations and reflections of the equilateral triangle. This leads to three different symmetry sectors. Forms in each sector turn out to be associated to distinct modular groups that contain $\Gamma(2)$ as a subgroup. One of these sectors is given by the odd automorphic forms of $SL(2,\Z)$. These have previously been studied numerically in \cite{Hejhal1991, PhysRevA.44.1491, em/1048610117, PhysRevLett.69.2188, hejhal1993fourier, Steil:1994ue}. The other symmetry sectors are instead associated to odd newforms of the Hecke congruence subgroups $\Gamma_0(2)$ and $\Gamma_0(4)$, and have been less studied. In \S\ref{sec:numerical} we compute a large number of eigenvalues numerically to study the spectrum. Using these eigenvalues,
in \S\ref{sec:properties} we recover and extend known results on the spectrum: the eigenvalue density obeys the Weyl formula, the differences of neighbouring eigenvalues follow a Poisson distribution and the spectral form factor displays an exponential ramp. These last two behaviours, both reminiscent of integrable systems, are due to the existence of Hecke operators that commute with the Hamiltonian. Finally, 
we verify that the prime Fourier coefficients of the eigenfunctions are distributed according to the Wigner semi-circle law.

The emergence of a late interior chaotic Hamiltonian suggests both the possibility of a dual description of the interior and also the possibility of accounting for the black hole microstates in a systematic way. Pursuing either of these directions will require going beyond our cohomogeneity-one backgrounds and allowing for full spatial inhomogeneity in the interior. This may not be a hopeless task --- it was argued by BKL \cite{BKL} that upon approach to the singularity different points in space decouple and evolve independently, which is a huge simplification of the dynamics. For a recent discussion see \cite{Damour:2002et}, and see \cite{Garfinkle:2020lhb} for numerical evidence. We emphasise that the spatial decoupling of points towards the singularity is logically distinct from the chaotic dynamics. In the present work we have embedded the chaotic dynamics into a holographic context, but we are not considering inhomogeneities. We make some comments about inhomogeneities in \S\ref{sec:inhomo}. A more microscopic description may also require understanding of stringy or matrix degrees of freedom in the near-singularity regime. One important question is whether the modular structure of the dynamics extends to the full microscopic theory as an organising principle (cf.~\cite{Kleinschmidt:2009hv, Kleinschmidt:2009cv,Kleinschmidt:2022qwl}).

The same mathematical structure, harmonic analysis on a modular domain, has recently been used to study two-dimensional CFT partition functions \cite{Benjamin:2021ygh}. Arithmetic quantum chaos may provide a useful perspective on CFT quantum chaos \cite{Haehl:2023tkr,Haehl:2023wmr}.
Furthermore, the modular subgroup $\Gamma(2)$ has previously appeared in the context of the CFT bootstrap \cite{Maloney:2016kee}. It would be very interesting, of course, to establish any kind of connection between the near-singularity gravitational dynamics and scrambling in the dual CFT. We should emphasise, however, to avoid potential confusion, that the chaos that emerges towards gravitational singularities is distinct from the chaos associated to the infinite blueshift that can be experienced by an observer close to a horizon, which has been studied in depth in recent years following \cite{Shenker:2013pqa}. However, similar chaotic can also arise towards a timelike singularity, and this may potentially be more accessible to the boundary theory \cite{Shaghoulian:2016umj}.

\section{Setup with massive vector fields}
\label{sec:setup}

The first step is to write down the simplest possible action that admits {\it (i)} AdS asymptotics, {\it (ii)} a black hole horizon and {\it (iii)} chaotic mixmaster dynamics towards the interior singularity. In particular, we wish to realise all of these things with a {\it (iv)} planar, homogeneous slicing of the spacetime. The following theory with three massive vector fields coupled to gravity and a cosmological constant seems to be well-suited
\be\label{eq:action}
S = \int {\rm d}^4x\,\sqrt{-g}\left[R + 6 - \sum_{i=1}^3 \left(\frac{1}{4} F_i^2 + \frac{\mu_i^2}{2} A_i^2\right) \right] \,.
\ee
The cosmological constant ensures the AdS asymptotics, with unit AdS radius. Near the singularity the mass terms and the cosmological constant will become subleading. In this far interior regime, without the mass terms, the field strengths are fixed by a Gauss law while producing `electric walls' that bound the evolution of the metric components \cite{Belinsky:1981vdw, Damour:2000th}. We will need three Maxwell fields to fully bound the metric components and ensure chaotic dynamics towards the singularity.
Although will not pursue a string-theoretic realisation of (\ref{eq:action}), we may note that multiple Maxwell fields arise, for example, in gauged supergravity theories \cite{Cvetic:1999xp}.
While the mass terms drop out towards the singularity, they are necessary for this theory to have a planar black hole exterior solution with all three vector fields nonzero, within the Ansatz that we now introduce. The masses will also ensure the absence of Cauchy horizons in the interior. We will choose the masses of the vector fields to be such that the dual vector operators are relevant deformations of the CFT. Sourcing this operator therefore preserves the AdS asymptotics.

The metric is
\be
{\rm d}s^2 = \frac{1}{z^2} \left(- F e^{-2 H} {\rm d}t^2 + \frac{{\rm d}z^2}{F}  +  e^{- 2 G} {\rm d}x^2 +  e^{2 G} {\rm d}y^2 \right) \,, \label{eq:met}
\ee
with $F,H,G$ functions of $z$, while the vector fields will be
\be\label{eq:vec}
A_1 = \phi_t\,{\rm d}t \,, \quad A_2 = \phi_x\,{\rm d}x \,, \quad A_3 = \phi_y\,{\rm d}y \,.
\ee
Here again the $\phi_i$ are functions of $z$. It is important that (\ref{eq:vec}) does not include a longitudinal component, {\it i.e.}~$\vec \nabla \cdot \vec A_i = 0$ on the constant $z$ slices, as a longitudinal component is not bounded by electric walls \cite{Henneaux:2022ijt}. Substituting the Ansatz into the equations of motion,
\begin{align}\label{eq:full}
R_{\mu\nu} - \frac{R}{2} g_{\mu\nu} - 3 g_{\mu\nu} & = \frac{1}{2}\sum_{i=1}^3 \left(F_{i \,\rho \mu} F^{\rho}_{i \, \nu} - \frac{1}{4} g_{\mu\nu} F_i^2  + \mu_i^2 A_{i\, \mu} A_{i \, \nu} - \frac{\mu_i^2}{2} g_{\mu\nu} A_i^2\right) \,, \\
\nabla_\mu F_i^{\mu \nu} - \mu_i^2 A_i^\nu & = 0 \,, \label{eq:massV}
\end{align}
the equations are solved so long as the $\phi_i$ and $G$ obey second order ordinary differential equations, while $F$ and $H$ obey first order equations. See Appendix \ref{app:eqs} for the equations.

For the purposes of giving numerical illustrations of our setup we will set the three masses equal with the choice $\mu_i^2 = - \frac{3}{16}$. With this choice of negative mass squared, which is above the Breitenlohner-Freedman bound for vectors in AdS spacetime, the asymptotic behavior of the fields as $z \to 0$ at the AdS boundary is
\be\label{eq:bdyexp}
F \to 1 \,, \qquad G,H \to 0 \,, \qquad \phi_i \to \phi_i^{(0)} z^{1/4} + \phi_i^{(1)} z^{3/4} \,.
\ee
The fact that the leading source term $\phi_i^{(0)}$ goes to zero as $z \to 0$ ensures the survival of the AdS asymptotics.\footnote{As noted, the negative mass squared that we have chosen is above the Breitenlohner-Freedman bound for a massive vector field. This ensures that the scaling dimension $\Delta$ of the dual operator is real and the AdS vacuum is stable against classical perturbations. However, the scaling dimension corresponding to (\ref{eq:bdyexp}) is $\Delta = \frac{7}{4}$ which is below the CFT unitarity bound of $\Delta = 2$ for a spin 1 operator in three dimensions. This means that our solutions do not describe deformations of a unitary CFT. It is unclear to us whether the boundary conditions (\ref{eq:bdyexp}), which include a source, are necessarily pathological from the bulk point of view. Certainly the equations of motion that we will be solving, following from the Ansatz (\ref{eq:met}) and (\ref{eq:vec}), are well-behaved. The skeptical reader may note that we could also have chosen a relevant deformation with a positive mass squared and hence consistent with the unitarity bound. The qualitative behaviour of the solutions is not expected to be different. We have found the choice in the main text more convenient for illustrative numerical purposes --- for instance, as we shall see shortly, Cauchy horizons are excluded a priori with a negative mass squared --- even while perhaps the immediate holographic interpretation is less clear.}

The black hole horizon occurs at $F=0$. The black hole interior has $F < 0$, so that $t$ becomes spacelike and $z$ timelike. An argument given in \cite{Hartnoll:2020rwq} implies that when $\mu_x^2 < 0$ or $\mu_y^2 < 0$ there cannot be a further inner (Cauchy) horizon where $F$ would again change sign. Specifically, suppose that both inner and outer horizons were present, at $z_I$ and $z_O$ respectively. At these points $F(z_I) = F(z_O) = 0$. The massive vector equations (\ref{eq:massV}) with $i=x,y$ then imply that 
\be
0 = \int_{z_O}^{z_I} \frac{{\rm d}}{{\rm d}z} \left(F e^{\pm 2G-H} \frac{{\rm d} \phi_i}{{\rm d}z} \phi_i \right) {\rm d}z = \int_{z_O}^{z_I} e^{\pm 2G-H} \left[\frac{\mu_i^2}{z^2} \phi_i^2  + F \left(\frac{{\rm d} \phi_i}{{\rm d}z}\right)^2 \right] {\rm d}z \,. \label{eq:bad}
\ee
The $+$ sign is for $\phi_x$ and $-$ for $\phi_y$. 
So long as $\phi_i$ does not vanish identically --- which it doesn't, because it has been sourced at the boundary ---  the final term in (\ref{eq:bad}) is strictly negative because $F < 0$ in between the horizons and by assumption $\mu_i^2 < 0$. This contradicts the vanishing of the integral in (\ref{eq:bad}) and hence the inner horizon cannot exist.

We evolve the equations of motion in two steps. Firstly, we solve for the exterior geometry using variables $F$, $G$, $H$, and $\hat{\phi}_i\equiv \phi_i/z^{1/4}$. These variables admit a power series expansion in $\sqrt{z}$ and $(1-z)$ near the conformal boundary and black hole event horizon, respectively. We may then introduce a coordinate $\hat{z}\equiv \sqrt{z}$ with respect to which all functions admit analytic expansions near the conformal boundary. To solve the equation of motion, we employ a collocation method and discretise the system on a Gauss-Lobatto grid. Subsequently, the resulting algebraic equations are solved via a relaxation method using a standard Newton-Raphson routine. These methods are extensively reviewed in \cite{Dias:2015nua}.

Solving for the interior geometry presents a significantly greater challenge. We first express the equations of motion in first-order form by introducing new variables $\tilde{G}\equiv G^\prime$ and $\tilde{\hat{\phi}}\equiv \hat{\phi}^\prime$. These variables are then evolved in the interior using the horizon values of all variables previously determined upon solving for the exterior geometry. To achieve long term stability, we use an adaptive implicit Runge-Kutta method of the Radau type \cite{HAIRER199993}. Without the use of an adaptive method, these simulations would be nearly impossible to perform. Specifically, the required time step to accurately evolve all equations varies widely. The final challenge arises from the need to evolve the system over an extended period, involving extremely small numerical values. To address this issue, we employ octuple precision. This challenge proves to be the primary bottleneck in our computations, primarily due to its demanding memory requirements. A typical simulation, at its peak usage, can demand up to 150GB of RAM and over 10GB of storage.

\begin{figure}[h]
    \centering
   \includegraphics[width=0.95\textwidth]{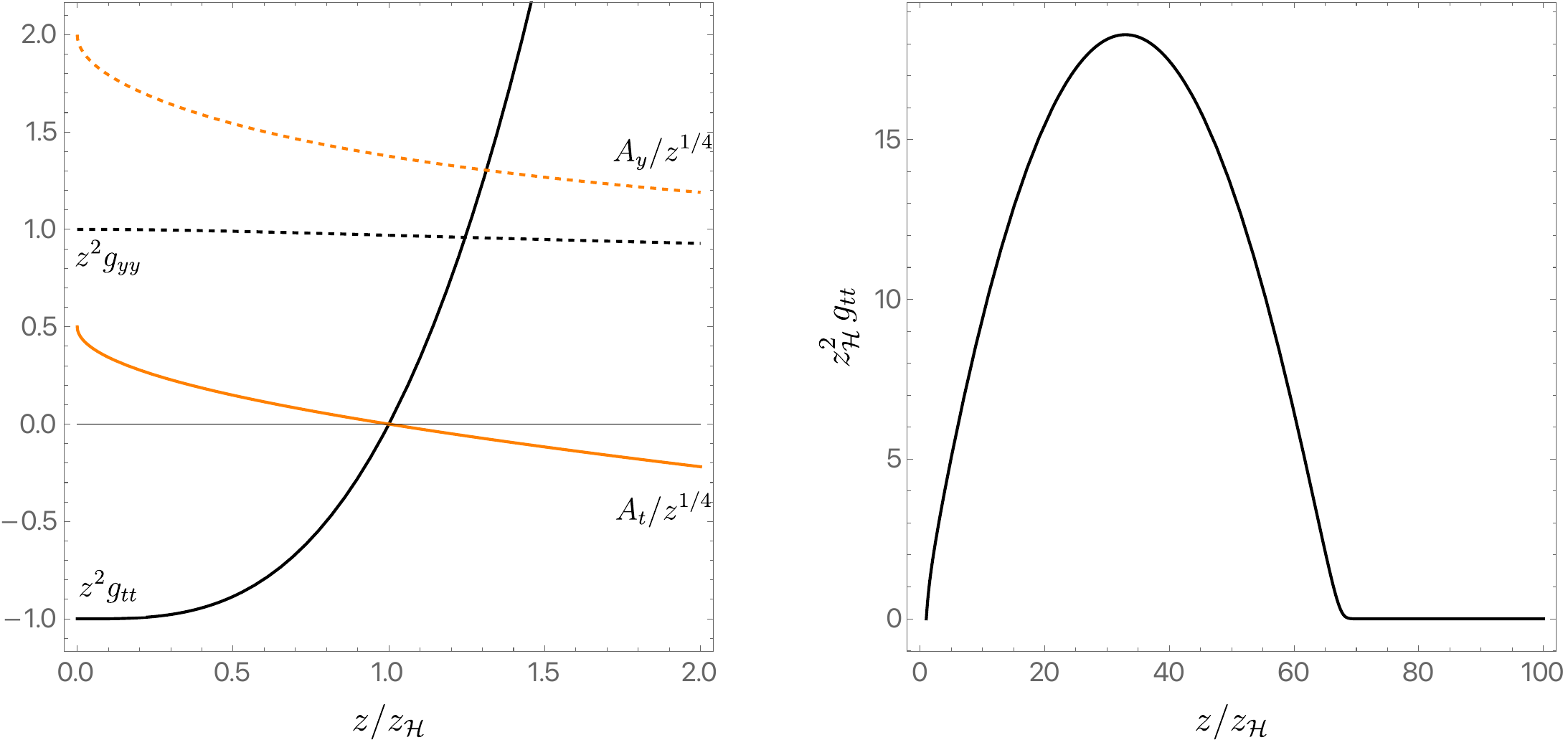}
    \caption{Left: evolution of metric and vector fields from the boundary through the horizon. Right: first interior maximum of $g_{tt}$, collapse of the Einstein-Rosen bridge, and absence of an inner horizon. Boundary conditions used: $\phi_t^{(0)} \approx 2.88867 \, T^{5/4}$, $\phi_x^{(0)}=2\phi_t^{(0)}$ and $\phi_y^{(0)}=4\phi_t^{(0)}$.}
    \label{fig:continuity}
\end{figure}

Fig.~\ref{fig:continuity} illustrates the evolution of the metric and massive vector fields from the AdS boundary (at $z=0$) and through the horizon (at $z=z_\mathcal{H}$).
The left hand plot of Fig.~\ref{fig:continuity} shows the smooth evolution of the fields across the horizon. We note that $g_{tt}$ vanishes at the horizon, by definition, and that furthermore $A_t$ must vanish at the horizon for regularity. The square root behavior of $A_i/z^{1/4}$ towards the boundary is due to the subleading near-boundary term in (\ref{eq:bdyexp}). The right hand plot in the figure shows that while $g_{tt}$ reaches a maximum in the interior and starts to decrease, the occurrence of an inner horizon is thwarted by the argument around (\ref{eq:bad}) above. Instead of vanishing again, $g_{tt}$ becomes exponentially small. This phenomenon was called the `collapse of the Einstein-Rosen bridge' in \cite{Hartnoll:2020rwq}.
Our interest here is the subsequent evolution in the far interior. We recall in Appendix \ref{sec:geodesic}, however, that the maximum of $g_{tt}$ seen in Fig.~\ref{fig:continuity}, before the collapse, makes it difficult to probe the far interior evolution from the AdS boundary.

In the absence of an inner horizon, there will be a cosmological interior evolution towards a singularity at $z \to \infty$. As noted above and discussed extensively below, we expect a chaotic evolution towards the singularity. The following Fig.~\ref{fig:bounces} is an illustrative example of the far interior evolution.
\begin{figure}[h]
    \centering
   \includegraphics[width=\textwidth]{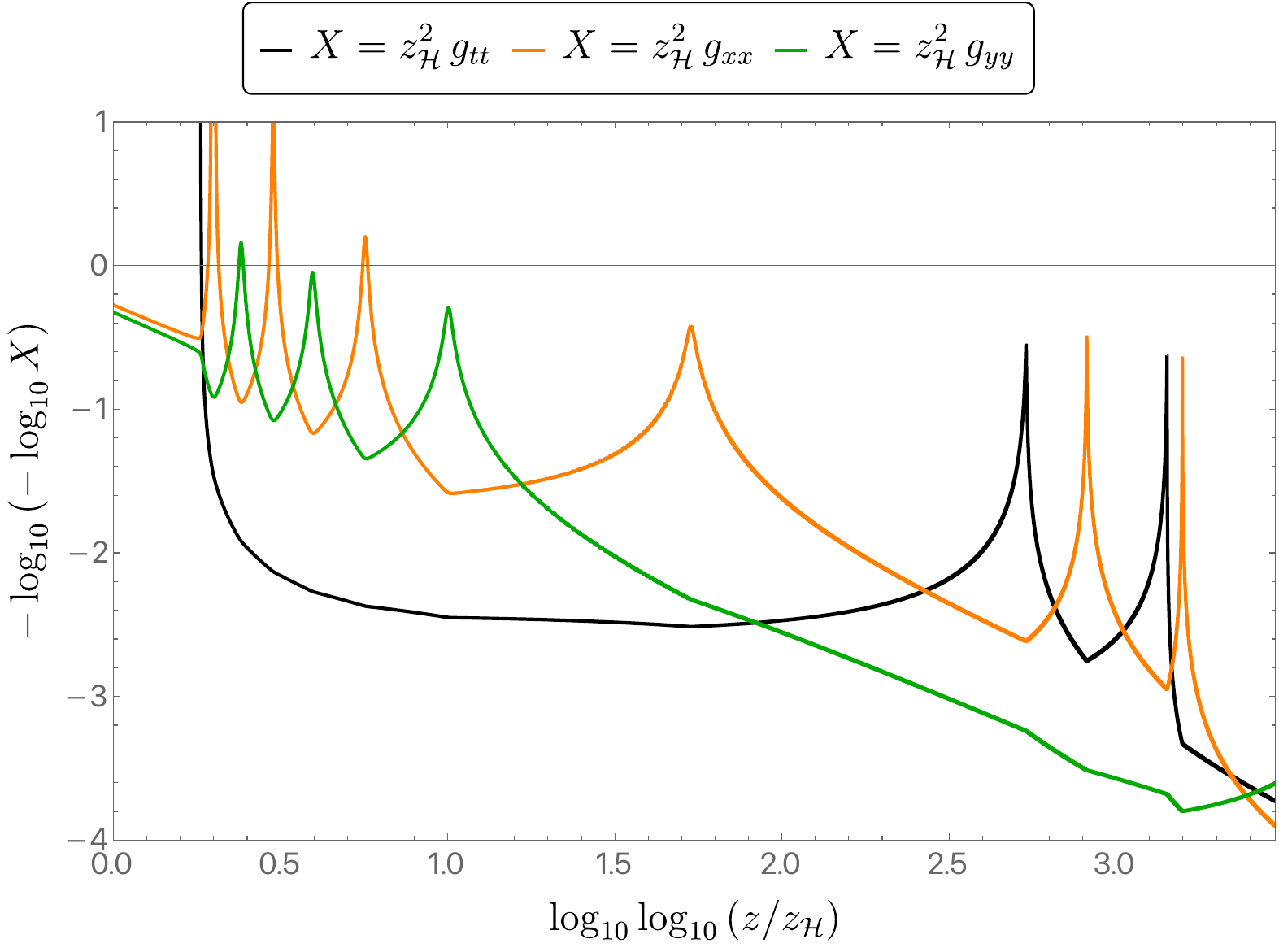}
    \caption{The far interior evolution shows a sequence of Kasner epochs connected by bounces. A succession of epochs in which two fixed metric components alternate dominance is called an era. Two full eras are visible in this plot. These numerics have the same AdS boundary conditions as in Fig.~\ref{fig:continuity}.}
    \label{fig:bounces}
\end{figure}
To capture all of the key dynamical features in a single plot we have found it necessary to resort to doubly-logarithmic axes. This fact gives a sense of the accuracy needed for the numerics. We will understand the appearance of doubly logarithmic scales analytically in \S\ref{sec:bkl2} below. 
As we elaborate in the following, Fig.~\ref{fig:bounces} shows a sequence of Kasner epochs connected by bounces. The first Kasner epoch emerges out of the collapse of the Einstein-Rosen bridge shown in Fig.~\ref{fig:continuity}. In each Kasner epoch one metric component grows while the other two shrink, even while the overall volume collapses. A Kasner era is a sequence of epochs
in which two given metric components alternate dominance, while the third continues to collapse. We will recall below that the chaotic nature of the interior evolution is contained in the transitions between different eras. In our setting, all of these phenomena are emerging in the interior of an asymptotically AdS black hole. The late interior time dynamics of bounces, epochs and eras can be captured analytically \cite{BKL, lifshitz1971asymptotic}. In the following section we will rederive these classic results in our context.

\section{Approach to the singularity: Kasner epochs and bounces}
\label{sec:bkl1}

\subsection{Equations of motion}

In the far interior, the cosmological constant term and the mass terms for the vector fields become negligible in the equations of motion. Intuitively this is because these terms are more relevant, in the sense of the renormalisation group, than the Einstein and Maxwell terms in the action, respectively. They therefore become small as the volume of space goes to zero close to the singularity. We verify this fact explicitly in Appendix \ref{app:eqs}. It follows that the approach to the singularity is described by the simplified equations of motion:
\be\label{eq:simp}
R_{\mu\nu} - \frac{R}{2} g_{\mu\nu} = \frac{1}{2}\sum_{i=1}^3 \left(F_{i \,\rho \mu} F^{\rho}_{i \, \nu} - \frac{1}{4} g_{\mu\nu} F_i^2 \right) \,, \qquad \nabla_\mu F_i^{\mu \nu} = 0 \,.
\ee
Here we have dropped the mass terms and cosmological constant term in (\ref{eq:full}). Almost all of our remaining analyses will use the simplified equations (\ref{eq:simp}). The role of the full equations (\ref{eq:full}) is to connect from the AdS boundary to the interior, up to and including the collapse of the Einstein-Rosen bridge in Fig.~\ref{fig:continuity}. The far interior dynamics in Fig.~\ref{fig:bounces} is, we shall see, perfectly described by (\ref{eq:simp}).

The volume of spatial slices decreases monotonically close to the singularity (see again Appendix \ref{app:eqs}). It is therefore convenient to use the spatial volume as a coordinate in the metric, rewriting the Ansatz (\ref{eq:met}) as
\be \label{eq:metric}
{\rm d}s^2 = e^{-\rho} e^{-2 h} {\rm d}t^2 - n^2 {\rm d}\rho^2 + e^{-\rho} e^{h} \left[e^{-\sqrt{3} g} {\rm d}x^2 +  e^{\sqrt{3} g} {\rm d}y^2 \right] \,.
\ee
Here $g,h,n$ are functions of $\rho$. We see that $\rho$ controls the volume of constant $\rho$ spatial slices while $h$ and $g$ determine their shape. Specifically, the spatial volume is proportional to $e^{-\frac{3}{2}\rho}$. The factors of $2$ and $\sqrt{3}$ in (\ref{eq:metric}) are helpful to `diagonalise' the equations of motion for $h$ and $g$ \cite{PhysRevLett.22.1071, PhysRev.186.1319}, as we will see shortly. Towards the singularity the volume collapses and $\rho \to \infty$. The spatial volume also vanishes at the horizon and therefore the coordinates (\ref{eq:metric}) are useful (\emph{i.e.}~single valued) beyond the maximal volume slice in the black hole interior, shown in Fig.~\ref{fig:penrose}. All of the interesting cosmological dynamics occurs beyond that slice, once the interior universe has started contracting.

The vector fields are again given by (\ref{eq:vec}), where
the $\phi_i$ are now functions of $\rho$. In the absence of mass terms, the equations of motion (\ref{eq:simp}) for the $\phi_i$ may immediately be integrated up in terms of the constant fluxes $f_i$:
\be\label{eq:flux}
\dot \phi_t = f_t\,n\,e^{\rho/2-2 h}\,, \quad \dot \phi_x = f_x\,n\,e^{\rho/2-\sqrt{3} g+h} \,, \quad \dot \phi_y = f_y\,n\,e^{\rho/2 + \sqrt{3} g+h} \,.
\ee
Here dot denotes derivative with respect to $\rho$. Using (\ref{eq:flux}) to eliminate the vector fields from the Einstein equation in (\ref{eq:simp}), one may solve for the lapse $n$ and obtain second order differential equations for $g$ and $h$. If we introduce the potential
\be\label{eq:pot}
V(g,h) = \frac{1}{2} \log \left[f_t^2 e^{-2 h} + e^h \left(f_x^2 e^{-\sqrt{3} g} + f_y^2 e^{\sqrt{3} g} \right) \right] \,,
\ee
(we should note that this is the logarithm of what is commonly referred to as an effective potential in this context, cf.~\cite{PhysRevLett.22.1071, PhysRev.186.1319})
then the lapse $n$ obeys
\be\label{eq:lapse}
\frac{1}{3} n^2 e^{2\rho+2V} + \dot g^2 + \dot h^2  =  1 \,,
\ee
and the differential equations are
\be\label{eq:deqs}
\frac{\ddot g}{1 - \dot g^2 - \dot h^2} = \frac{1}{2} \dot g - \frac{\partial V}{\partial g} \,, \qquad
\frac{\ddot h}{1 - \dot g^2 - \dot h^2} = \frac{1}{2} \dot h - \frac{\partial V}{\partial h} \,.
\ee
The remainder of this section will be an analysis of the equations (\ref{eq:deqs}). Our interest is in the behaviour of the solutions towards the singularity as $\rho \to \infty$. For completeness, in Appendix \ref{sec:large} we also describe the behaviour of the large volume $\rho \to -\infty$ limit. The large volume limit is not reached in the black hole interiors we have considered.

The equations (\ref{eq:deqs}) are similar to those of a relativistic particle moving in two dimensions $\{g,h\}$ in the presence of a potential $V$ and subject to an accelerating `anti-friction' term. This is only an analogy, however, as the kinetic term is a little different to the relativistic particle case. Indeed, the equations imply that
\be\label{eq:non}
\dot g^2 + \dot h^2 = \frac{{\rm d}}{{\rm d}\rho} \left[2 V - \log(1 - \dot g^2 - \dot h^2) \right] \,.
\ee
This equation expresses the nonconservation of the particle's `energy' due to the anti-friction terms.
The anti-friction pushes the particle up the potential leading, as we will see, to a sequence of bounces off the potential.

In the numerical evolution of Fig.~\ref{fig:bounces} we saw a sequence of `Kasner epochs' punctuated by abrupt bounces. We will now derive this behaviour analytically from the equations (\ref{eq:deqs}). It is first convenient to introduce notation to characterise this sequence. The bounces occur at interior times $\rho_n$. Each bounce will be off one of three `walls' in the effective potential, and hence can be labelled by $s_n \in \{t,x,y\}$. The Kasner epochs in between bounces will be characterised by a single number $u_n$ that determines the Kasner exponents. Thus we will have a sequence
\be
\cdots \quad \xrightarrow[\phantom{\text{ Kasner }}]{\phantom{u_n}} \quad \rho_{n-1}, s_{n-1} \quad \xrightarrow[\text{ Kasner }]{u_{n-1}} \quad \rho_n, s_n \quad \xrightarrow[\text{ Kasner }]{u_n} \quad \rho_{n+1}, s_{n+1}  \quad \xrightarrow[\phantom{\text{ Kasner }}]{\phantom{u_n}} \quad \cdots  \,. \nonumber
\ee
The important results in this section will be two known formulae. The first, (\ref{eq:unmap}) below, relates $u_n$ and $u_{n+1}$. This expression is at the heart of the chaotic behavior. The second, \eqref{eq:rr} below,
relates the locations of the bounces $\rho_n$ and $\rho_{n+1}$. This will characterise the rate at which chaos onsets. We have found it useful to rederive these results, first given in \cite{BKL, lifshitz1971asymptotic}.

\subsection{Kasner epochs}

During the Kasner epochs, the first term in (\ref{eq:lapse}) is subleading. This leads to null motion with $\dot g^2 + \dot h^2 = 1$. The equations of motion (\ref{eq:deqs}) therefore require $\ddot g = \ddot h = 0$. We write the $n$th Kasner epoch as
\be\label{eq:kasner}
g = g_{n} + v_n (\rho - \rho_{n}) \,, \qquad h = h_{n}+ w_n (\rho - \rho_{n}) \,, \qquad v_n^2 + w_n^2 = 1 \,.
\ee
Here $g_n,h_n,v_n,w_n$ and $\rho_{n}$ are constants, with $g_{n} = g(\rho_{n})$ and $h_{n} = h(\rho_{n})$. It will be convenient to parameterise the epochs in terms of three `Kasner exponents' $\{p^t_n,p^x_n,p^y_n\}$:
\be\label{eq:vel}
 3 p^t_n \equiv 1 + 2 w_n \,, \qquad  3 p^x_n \equiv 1 + \sqrt{3} v_n - w_n  \,, \qquad 3 p^y_n \equiv  1 - \sqrt{3} v_n - w_n \,.
\ee
This parameterisation requires 
\be\label{eq:pi1}
p_n^t + p_n^x+ p_n^y  = 1 \,.
\ee
Furthermore, the null motion in the Kasner epochs is seen to require that
\be\label{eq:pi2}
(p_n^t)^2 + (p_n^x)^2+ (p_n^y)^2 = 1  \,.
\ee
These are the familiar two constraints on Kasner exponents \cite{kasner}, and hence there is only one independent exponent. For example, we may consider $p^t_n$ as the independent exponent.

It will be helpful later to explicitly write down the metric in the Kasner epoch. The equations of motion as written above are singular in the strict Kasner limit (we will include perturbations about Kasner shortly). That is because an exact Kasner solution requires turning off the $f_i$ fluxes, which sends $V \to - \infty$ in (\ref{eq:pot}). To obtain the Kasner metric it is simplest to rewrite (\ref{eq:lapse}), using the full equations of motion, as
\be\label{eq:ndotn}
\frac{\dot n}{n} = - \frac{1}{2} \left(2 + \dot g^2 + \dot h^2\right) \,.
\ee
It follows that in a Kasner epoch, with $\dot g^2 + \dot h^2 = 1$,
the lapse obeys $\dot n = - \frac{3}{2} n$. One should not confuse the lapse $n$ with label $n$ of the Kasner epochs. Therefore on the $n$th Kasner epoch the metric is
\be\label{eq:kmetric}
{\rm d}s^2 = e^{-3 p^t_n \rho} \frac{{\rm d}t^2}{t_n^2}  - e^{- 3 \rho} \frac{{\rm d}\rho^2}{z_n^2} + e^{-3 p^x_n \rho} \frac{{\rm d}x^2}{x_n^2} + e^{-3 p^y_n \rho} \frac{{\rm d}y^2}{y_n^2} \,.
\ee
Here $t_n, x_n, y_n, z_n$ are constants.

The constraints (\ref{eq:pi1}) and (\ref{eq:pi2}) imply that one, and only one, of the Kasner exponents is necessarily negative. In the metric (\ref{eq:kmetric}) we see that the spatial direction with negative exponent is expanding while the other two directions are collapsing. Let us denote the negative exponent by $p^-_n$, the larger of the positive exponents by $p^+_n$ and the middle exponent by $p^0_n$. The following parametrisation will turn out to be very convenient \cite{LK}
\be\label{eq:un}
p^-_n = - \frac{u_n}{1+u_n+u_n^2}\,, \qquad p^0_n = \frac{1 + u_n}{1+u_n+u_n^2}\,, \qquad p^+_n = \frac{u_n(1 + u_n)}{1+u_n+u_n^2}\,.
\ee
Here we take $u_n \geq 1$. Depending on the Kasner epoch, $p^t_n$ could be any of these.

\subsection{The end of a Kasner epoch}
\label{sec:kasner}

The Kasner epochs are visible as the curves of growth and collapse in Fig.~\ref{fig:bounces}. These epochs come to an end when the first term in (\ref{eq:lapse}) is no longer negligible. Using the fact, from above, that $n^2 \propto e^{-3\rho}$ on Kasner epochs we see that, to leading exponential accuracy at large $\rho$, the potential becomes important at $\rho_n$ such that
\be\label{eq:rV}
\rho_n = 2 V(\rho_n) \equiv 2 V_n \,.
\ee
The potential is also large and, coming out of a given Kasner epoch, will be dominated by one of the three terms in (\ref{eq:pot}). That is, to leading exponential accuracy we may write
\be\label{eq:pot2}
V_n = \text{max} \left(- h_n, \frac{h_n - \sqrt{3} g_n}{2}, \frac{h_n + \sqrt{3} g_n}{2}\right) \equiv \text{max} \left(V_n^t, V_n^x, V_n^y \right)\,.
\ee
The potential therefore defines three `walls' and a Kasner epoch ends when the universe hits a wall. Towards the asymptotic interior, the walls confine the motion of the metric components to the triangular region defined by
\be\label{eq:walls}
\frac{h}{\rho} \geq - \frac{1}{2} \,, \qquad \frac{h}{\rho} \pm \sqrt{3} \, \frac{g}{\rho} \leq 1 \,.
\ee
These constraints follow from (\ref{eq:lapse}), (\ref{eq:rV}) and (\ref{eq:pot2}).
The following Fig.~\ref{fig:triangle}
\begin{figure}[h]
    \centering
   \includegraphics[width=\textwidth]{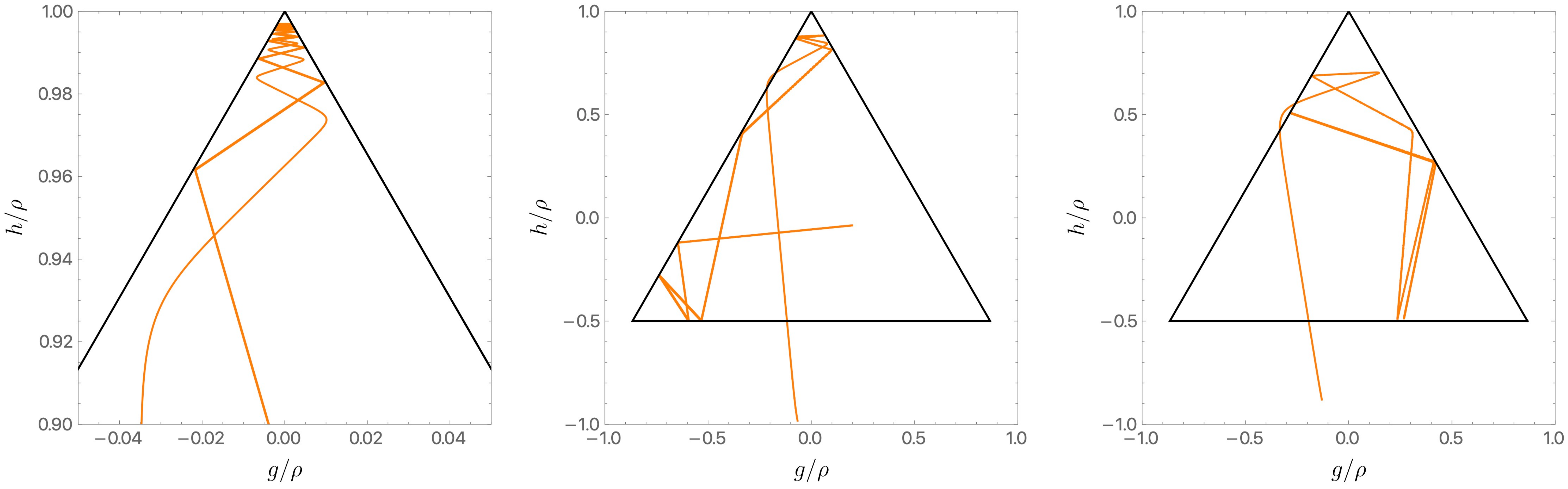}
    \caption{The late time interior dynamics of $g(\rho)$ and $h(\rho)$ is confined to the triangle (\ref{eq:walls}). The plots show three different illustrative trajectories, generated numerically using the full equations of motion (\ref{eq:full}) and (\ref{eq:massV}). The straight lines correspond to Kasner epochs. An era is given by a sequence of epochs bouncing between two given walls. The earliest bounces extend slightly outside the triangle as the exponential walls have not yet become infinitely steep. The leftmost plot is zoomed in to a tip of the triangle, showing an era of many epochs. Left: $\phi_t^{(0)}\approx1.49011 \,T^{5/4}$, $\phi_x^{(0)}=3\phi_t^{(0)}$ and $\phi_y^{(0)}=4\phi_t^{(0)}$. Middle: $\phi_t^{(0)}\approx2.88867\, T^{5/4}$, $\phi_x^{(0)}=2\phi^{(0)}_t$ and $\phi_y^{(0)}=4\phi^{(0)}_t$. Right: $\phi_t^{(0)}\approx5.50918 \, T^{5/4}$, $\phi_x^{(0)}=2\phi_t^{(0)}$ and $\phi_y^{(0)}=3\phi_t^{(0)}$.}
    \label{fig:triangle}
\end{figure}
shows three examples of the bounded evolution of the universe.
It is clear from the figure that each Kasner epoch ends in a `bounce' to a subsequent Kasner epoch.
We will characterise such bounces in detail in the following sections.

While the potential (\ref{eq:pot}) is not identical to the original Bianchi IX mixmaster potential from \cite{BKL, lifshitz1971asymptotic, PhysRevLett.22.1071, PhysRev.186.1319}, the walls in (\ref{eq:walls}) are the same as the dominant walls in that case. In the Bianchi IX model the walls are curvature walls, whereas in our model --- which has flat spatial slices and no curvature walls --- they are electric walls. The sub-dominant curvature walls in the Bianchi IX dynamics, absent in our potential, are outside of the triangle (\ref{eq:walls}) and are therefore irrelevant for the asymptotic interior dynamics. This is why we will recover the known recursion relations between Kasner epochs.

The breakdown of a Kasner epoch, due to the fluxes,
can be seen explicitly by considering perturbations of $g$ and $h$ about the Kasner behaviour (\ref{eq:kasner}). In Appendix \ref{sec:kaspert} we show that the perturbations grow exponentially due to one of the Kasner exponents being negative.

\subsection{Reflection rule}
\label{sec:reflect}

As can be seen in Fig.~\ref{fig:bounces} or Fig.~\ref{fig:triangle}, the bounces are characterised by a sudden change of $\dot g$ and $\dot h$ while $g$ and $h$, and hence $\pa_g V$ and $\pa_h V$ in the equations of motion (\ref{eq:deqs}), remain constant. The values of these derivatives of the potential at the bounce follow directly from the asymptotic behaviour of the potential in (\ref{eq:pot}) and depend on which of the three walls is involved in the bounce:
\be\label{eq:Wi}
\left\{
\begin{array}{lll}
t\;  : & \pa_g V \to 0 \,, & \pa_h V \to -1 \\
x \;  : & \pa_g V \to -\frac{\sqrt{3}}{2} \,,  & \pa_h V \to \frac{1}{2} \\
y \; : &\pa_g V \to \frac{\sqrt{3}}{2} \,, & \pa_h V \to \frac{1}{2} 
\end{array}
\right. \,.
\ee

Using the constant values of the potential gradients in (\ref{eq:Wi}), the functions $g$ and $h$ no longer appear undifferentiated in the differential equations (\ref{eq:deqs}). These may now be considered as first order equations for $\dot g$ and $\dot h$ and solved analytically. In Appendix \ref{sec:bouncesol} we obtain an explicit solution for the bounce by solving both of the equations in (\ref{eq:deqs}) with $\pa_g V$ and $\pa_h V$ constant. The solution is seen to extend to the Kasner regimes on either side of the bounce. The most important information to extract is the reflection rule. This can be done directly without using the full solution, as we now show.

The equations of motion (\ref{eq:deqs}) imply that
\be\label{eq:bounce}
\frac{{\rm d} \dot g}{{\rm d} \dot h} = \frac{\dot g - 2 \pa_g V}{\dot h - 2 \pa_h V} \,.
\ee
Here $\pa_g V$ and $\pa_h V$ are understood to be constant, with values given in (\ref{eq:Wi}). This equation may be solved to obtain a linear relation between $\dot g$ and $\dot h$ across the bounce:
\be\label{eq:linear}
\frac{\dot g - 2 \pa_g V}{v_n - 2 \pa_g V} = \frac{\dot h - 2 \pa_h V}{w_n - 2 \pa_h V} \,.
\ee
Here we fixed the constant of integration in terms of the Kasner exponents $\{v_n,w_n\}$ on one side of the bounce. We may now furthermore apply (\ref{eq:linear}) to the Kasner regime on the other side of the bounce, setting $\dot g = v_{n+1}$ and $\dot h = w_{n+1}$. The equation then relates the Kasner exponents before and after the bounce. Using the further fact that the Kasner epochs obey $\dot g^2 + \dot h^2 = 1$ we thereby obtain the following `reflection rules':\be\label{eq:mess}
\left\{
\begin{array}{ll}
t: v_{n+1} = \frac{3 v_n}{5 + 4 w_n} \,, & w_{n+1} = - \frac{4 + 5 w_n}{5 + 4 w_n} \\[10pt]
x: v_{n+1} = \frac{2 \sqrt{3} (2 - 5 w_n)(w_n-1) - 3 v_n (1+2 w_n)}{13 + 4 w_n(-5+4 w_n)} \,, & w_{n+1} = \frac{10 w_n^2 + w_n - 6 \sqrt{3} v_n (w_n-1)-2}{13 + 4 w_n(-5+4 w_n)}  \\[10pt]
y: v_{n+1} = \frac{-2 \sqrt{3} (2 - 5 w_n)(w_n-1) - 3 v_n(1+2 w_n)}{13 + 4 w_n(-5+4 w_n)} \,, & w_{n+1} = \frac{10 w_n^2 + w_n + 6 \sqrt{3} v_n(w_n-1)-2}{13 + 4 w_n(-5+4 w_n)}  
\end{array} \right.
\,.
\ee
In the formulae above it is understood that $w_n^2+v_n^2 = 1$. We have checked (\ref{eq:mess}) against the numerics. We may also verify that the transformations square to the identity, so that they are indeed reflections.

The messy formulae in (\ref{eq:mess}) are more transparent in terms of the $p_n^i$ exponents from (\ref{eq:vel}):
\be\label{eq:nice}
t: \frac{p^y_{n+1}}{p^x_{n+1}} = - \frac{p^y_{n}}{p^x_{n}} \,, \qquad x:\frac{p^t_{n+1}}{p^y_{n+1}} = - \frac{p^t_{n}}{p^y_{n}} \,, \qquad y: \frac{p^t_{n+1}}{p^x_{n+1}} = - \frac{p^t_{n}}{p^x_{n}} \,.
\ee
The meaning of (\ref{eq:nice}) can be seen as follows. Suppose that in approaching the bounce $p^t_n < 0$ while $p^x_n, p^y_n > 0$. This will lead to a `$t$' bounce at $\rho_{n+1}$. According to (\ref{eq:nice}), after the bounce either $p^x_{n+1}$ or $p^y_{n+1}$ (but not both) will be negative and hence $p^t_{n+1} > 0$. The motion after the bounce is therefore away from the $t$ wall and towards either the $x$ or $y$ wall.

The correct geometrical framework to understand the reflections in (\ref{eq:nice}) is in terms of hyperbolical billiard motion \cite{Damour:2002et, belinski_henneaux_2017}. In general, hyperbolic reflections form a Coxeter group. In the present case there are no relations among the different reflections, because the corresponding underlying triangle in the hyperbolic plane has edges that are asymptotically parallel. We shall make essential use of the hyperbolic description in our discussion of the Wheeler-DeWitt equation and arithmetic quantum chaos in \S\ref{sec:WDW} below.

A final simplification is possible in terms of the $u_n$ variables introduced in (\ref{eq:un}). Recall that $u_n \geq 1$. The map (\ref{eq:nice}) now becomes the same for all three types of bounce:
\be\label{eq:unmap}
u_{n+1} = \max\left(u_n - 1, \frac{1}{u_n-1} \right) \,.
\ee
These are now seen to be --- as anticipated -- precisely the same reflections as are described in the original papers \cite{BKL,lifshitz1971asymptotic} for the Bianchi IX mixmaster universe. As we will elaborate further below, the two different possibilities in (\ref{eq:unmap}) correspond to transitions between Kasner epochs and Kasner eras, respectively. In deriving (\ref{eq:unmap}) from (\ref{eq:nice}) one finds that the Kasner exponents before and after the transition at $\rho_{n+1}$ exchange roles as follows:
\be\label{eq:ppp}
\left\{ \begin{array}{c}
p_n^- \to p_{n+1}^0 \\
p_n^0 \to p_{n+1}^- \\
p_n^+ \to p_{n+1}^+
\end{array} \right.
\qquad \text{or} \qquad
\left\{ \begin{array}{c}
p_n^- \to p_{n+1}^+ \\
p_n^0 \to p_{n+1}^- \\
p_n^+ \to p_{n+1}^0
\end{array} \right. \,.
\ee
Within a given Kasner era, the first case in (\ref{eq:ppp}), the universe bounces between the same two walls, with $p^-$ and $p^0$ alternating between each epoch while the $p^+$ direction continually collapses. Bouncing into a new era involves a transition to a different pair of walls.

\subsection{Evolution of the volume within an era}
\label{sec:evo}

The metric (\ref{eq:metric}) together with the potential (\ref{eq:pot2}) shows that $e^{-\rho_n + 2 V_{n}^i}$ is the value of the $i$th metric component at the $n$th bounce. The bounces at $\rho_n$ and $\rho_{n+1}$ are connected by the Kasner epoch with metric (\ref{eq:kmetric}). We may therefore equate the change in the metric components over the Kasner era with the difference in the metric components at the bounce on each side. To leading exponential accuracy:
\be\label{eq:tp}
- 3 p_n^i (\rho_{n+1} - \rho_n) = (2 V_{n+1}^i - \rho_{n+1}) - (2 V_n^i - \rho_n) = 2 \left(V_{n+1}^{i} - V_{n}^i + V_{n}^{s_n}- V_{n+1}^{s_{n+1}} \right)\,.
\ee
For the second equality we used the bounce condition (\ref{eq:rV}), with $V_n = V_n^{s_n}$ to leading exponential order. Recall that $s_n$ denotes the wall that is involved in the bounce.

The relation (\ref{eq:tp}) holds for any pair of bounces. Specialising now to within a given Kasner era, the bounces alternate between two walls. This means that $s_{n-1} = s_{n+1}$. We may now put $i = s_{n+1}$ and also $i = s_{n}$ in (\ref{eq:tp}). This leads to the ratio
\be\label{eq:rr}
\frac{\rho_{n+1} - \rho_n}{\rho_n - \rho_{n-1}} = - \frac{p_{n-1}^{s_{n-1}}}{p_n^{s_{n+1}}} = \frac{1 + u_{n-1}}{u_{n}} \frac{1 + u_{n} + u_{n}^2}{1 + u_{n-1} + u_{n-1}^2} \,. \qquad  \quad \text{(intra era)}
\ee
In the final equality we have used the parametrisation (\ref{eq:un}) together with the fact that $p_n^{s_{n+1}} = p_n^-$, the growing exponent, while $p_{n-1}^{s_{n-1}} = p_{n-1}^0$ from 
the left hand (intra era) transformation in (\ref{eq:ppp}).
The expression (\ref{eq:rr}) can be found in \cite{lifshitz1971asymptotic} and implicitly in \cite{BKL}.

Below (\ref{eq:metric}) we noted that $\rho$ is proportional to the logarithm of the volume. Iterating the recursion relation (\ref{eq:rr}) therefore determines the change in volume over an era. Let $\rho_0$ be the first bounce of a new Kasner era, after the `big bounce' where $u \to 1/(u-1)$. The first epoch of the new era therefore has exponent $u_{-1}$, while the second epoch of the new era has exponent $u_0$. This second epoch is where a new Kasner exponent becomes negative (see Fig.~\ref{fig:era} in Appendix \ref{sec:fiddly}). We can write
\be\label{eq:uo}
u_0 = k - 1 + x \,,
\ee
where $x \equiv u_0 - \left \lfloor{u_0}\right \rfloor$, the fractional part of $u_0$, and consequently, from (\ref{eq:unmap}), $k \in \N$ is the number of epochs in the era, before another big bounce occurs. Solving the recursion relation (\ref{eq:rr}) one obtains, as first found in \cite{BKL, lifshitz1971asymptotic},
\be\label{eq:length}
\rho_k - \rho_0 = \Delta_{0} \, k \left(k+x+\frac{1}{x} \right)\,.
\ee
With $\Delta_0$ defined below. We should note here that the next big bounce occurs at $\rho_{k-1}$. However, as we have already noted, the `third' Kasner exponent doesn't become negative until the second epoch of the subsequent era. This allows us to use the recursion relation (\ref{eq:rr}) up to $\rho_k$, which is the first bounce in the subsequent era. The formula (\ref{eq:length}) therefore gives the change in $\rho$ over the era, $\rho_k - \rho_0$,
in terms of the change
\be\label{eq:Delta0}
\Delta_{0} \equiv \frac{u_0 (\rho_1 - \rho_0)}{1+u_0+u_0^2} = - p^-_0 (\rho_1 - \rho_0)\,,
\ee
of the logarithm of the growing metric component over the second epoch of the era, as the universe shrinks from $\rho_0$ to $\rho_1$. More precisely, from the Kasner metric (\ref{eq:kmetric}), we see that (\ref{eq:Delta0}) is one third of the change of the logarithm of the metric over the Kasner epoch.

The expression (\ref{eq:length}) explicitly solves the dynamics of a given era. That expression will be the starting point for an analysis of inter era dynamics in the following \S\ref{sec:erad}. It is, however, necessary to establish one more intra-era result beforehand. This is a more detailed evolution of the spatial metric components through the era. We have relegated the details to Appendix \ref{sec:fiddly}. It is again convenient to work with the logarithm of the metric functions as these evolve linearly over Kasner epochs (\ref{eq:kmetric}). The key quantity will be a generalisation of (\ref{eq:Delta0}) to (one third of) the change in the logarithm of the growing metric component over the $n$th epoch
\be
\Delta_n \equiv - p_n^- (\rho_{n+1} - \rho_n) \,.
\ee
In Appendix \ref{sec:fiddly} we show that evolution through the $k$ epochs of an era leads to
\be\label{eq:ddk}
\Delta_k = \rho_0 + \Delta_0 (k^2 +kx - 1) \,.
\ee
We will use this formula, first given in \cite{BKL,lifshitz1971asymptotic}, in the following section.

\section{Classical dynamical system of Kasner Eras}
\label{sec:bkl2}

\subsection{Era dynamics}
\label{sec:erad}

We have seen that a useful quantity to keep track of is the $u$ exponent of the second Kasner epoch of an era. For the $N$th Kasner era, generalising (\ref{eq:uo}), we may write this quantity as
\be\label{eq:ukx}
u_N = k_N - 1 + x_N \,,
\ee
with, as previously, $x_N \equiv u_N - \left \lfloor{u_N}\right \rfloor$ and $k_N \in \N$. The exponent transition rule (\ref{eq:unmap}), applied $k_N$ times to traverse the era, then implies that
\begin{align}
k_{N+1} & = \left\lfloor{\frac{1}{x_N}}\right \rfloor \,, \label{eq:a} \\
x_{N+1} & = \frac{1}{x_N} - \left\lfloor{\frac{1}{x_N}}\right \rfloor \,. \label{eq:b}
\end{align}
The second of these is called the Gauss map, a well-studied chaotic map.\footnote{The Gauss map generates the continued fraction representation of $x_0$ \cite{BARROW19821}. If $x_0$ is rational this representation terminates after a finite number of steps. These are a measure zero set of universes. Furthermore, if $x_0$ is rational eventually one reaches an epoch with $u_n=1$, which is mapped to $\infty$ by (\ref{eq:unmap}). The Kasner metric in this limit is Rindler spacetime, with the solution heading towards a Cauchy horizon (and towards a corner of the triangle in Fig.~\ref{fig:triangle}). The argument in (\ref{eq:bad}), excluding Cauchy horizons, thus implies that the subleading mass terms in the equations of motion will nudge the solution away from this fine-tuned behaviour.} It maps $x_N \in [0,1]$ to $x_{N+1} \in [0,1]$. The location $\rho_N$ of the first bounce in the $N$th era is given by (\ref{eq:length}) as
\be
\rho_{N+1} = \rho_N + \Delta_N k_N\left(k_N + x_N + \frac{1}{x_N}\right) \,, \label{eq:c}
\ee
while the change $\Delta_N$ in the logarithm of the metric component is given by (\ref{eq:ddk}) as
\be
\Delta_{N+1} = \rho_N + \Delta_N (k_N^2 + k_N x_N  - 1) \,. \label{eq:d}
\ee
The recursion relations (\ref{eq:a}) -- (\ref{eq:d}) govern the evolution of Kasner eras in the far interior \cite{BKL,lifshitz1971asymptotic}.

\begin{figure}[h]
    \centering
   \includegraphics[width=\textwidth]{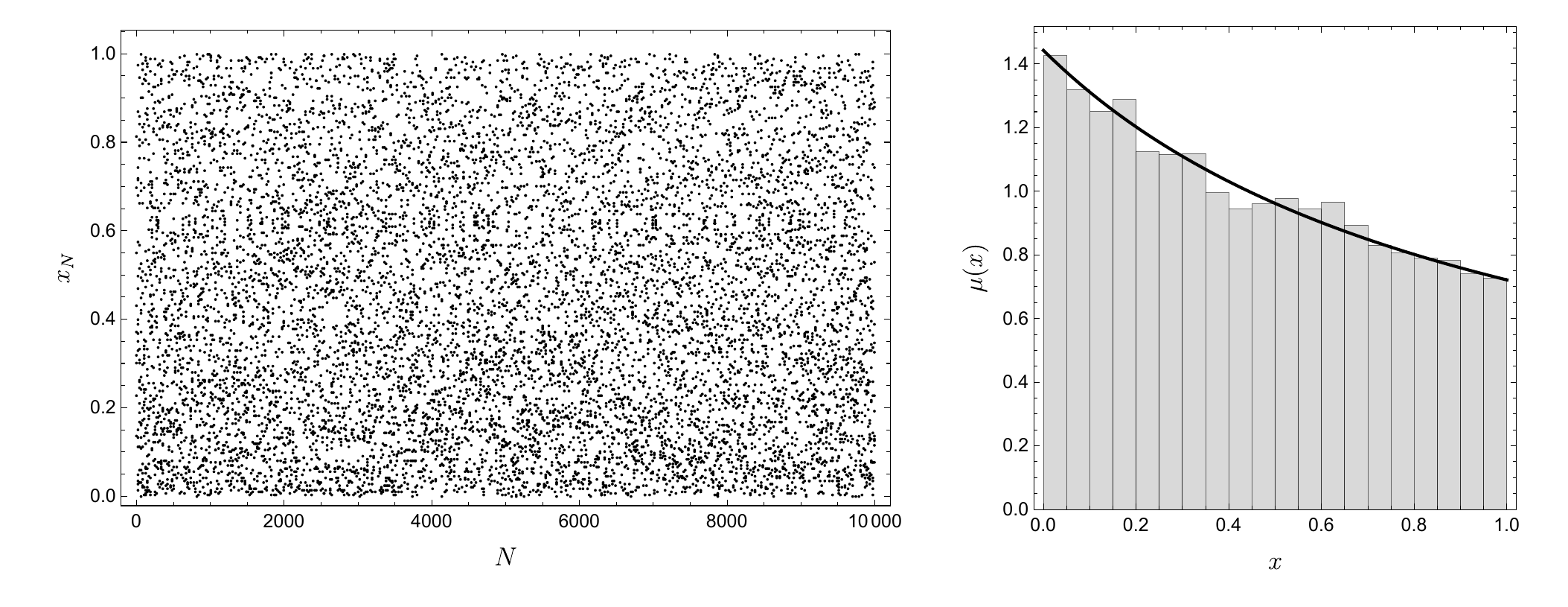}
    \caption{Left plot shows 10000 points generated using the Gauss map (\ref{eq:b}), starting from $x_0 = \sqrt{3}/\pi$ (it is important to start with an irrational number). To run the numerics to large $N$ it is important also to work with $x$ to a large number of decimal places. Right plot shows a histogram of the numerical values of $x_N$ generated together with the predicted equilibrium distribution (\ref{eq:mu}).}
    \label{fig:gm}
\end{figure}

It is well-known that the Gauss map (\ref{eq:b}) for $x_N$ leads to a late-time steady state distribution of values of $x$ given by
\be\label{eq:mu}
\mu(x) = \frac{1}{\log 2} \frac{1}{1+x} \,.
\ee
See e.g.~\cite{BKL,lifshitz1971asymptotic,BARROW19821}. We have illustrated this distribution in Fig.~\ref{fig:gm}, using a numerical sequence of points generated using the Gauss map.

There are two important `timescales' associated to the Gauss map that relate to the equilibrium distribution (\ref{eq:mu}). Both are reviewed in \cite{BARROW19821}. The first is the rate of mixing in equilibrium. This can be quantified using the Kolmogorov-Sinai entropy expressed as an average over Lyapunov exponents. If we write the Gauss map as $x_{N+1} = T(x_N)$ then the derivative $|T'(x)| = 1/x^2$ is a measure of how rapidly nearby points diverge under the map. The chaotic nature of the map leads to exponential divergences and so the Lyapunov exponent is introduced as $\log |T'(x)|$. The averaged exponent is then the entropy
\be\label{eq:h}
h = \int \log |T'(x)|\,\mu(x)\,{\rm d}x = - 2  \int_0^1 \frac{\log x }{(1+x) \log 2 } \,{\rm d}x = \frac{\pi^2}{6 \log 2} \approx 2.37 \,.
\ee
The chaotic dynamics therefore erases initial conditions in an order one number of eras.

A second timescale is defined by the rate at which the equilibrium distribution (\ref{eq:mu}) is reached. This also turns out to be very fast. Suppose that $\mu_N(x)$ is a distribution of values of $x$ that has been fed iteratively through the Gauss map $N$ times. This can be shown to converge to the equilibrium distribution at the rate \cite{wirsing1974theorem}
\be
|\mu_N(x) - \mu(x)| \sim e^{-1.19 N} \,.
\ee

The two rates of mixing considered above are both in terms of the number of eras $N$. A more physical `clock' is given by the spatial volume. The volume can be obtained as a function of $N$
as follows \cite{Khalatnikov1985}. We outline the steps but do not reproduce the details. Firstly, change variables in the recurrence relations 
(\ref{eq:a}) and (\ref{eq:b}) to
\be\label{eq:deltaN}
\xi_{N} \equiv \log \rho_N \qquad \text{and} \qquad \delta_N \equiv \frac{\Delta_N}{\rho_N}\,.
\ee
This leads to the relation, from (\ref{eq:a}),
\be\label{eq:xi}
\xi_{N+1} - \xi_N = \log \left[1 + \delta_N k_N \left( k_N + x_N + \frac{1}{x_N} \right) \right] \,.
\ee
The equilibrium distribution (\ref{eq:mu}) for $x$ may be seen to imply equilibrium distributions for $k$ and $\delta$. Using these distributions on the right hand side of (\ref{eq:xi}) shows that, in equilibrium,
\be\label{eq:eq}
\left\langle \log \frac{\rho_N}{\rho_0} \right\rangle = \langle \xi_N - \xi_0 \rangle = N \langle \xi \rangle = h N \,,
\ee
where $h$ is given in (\ref{eq:h}). We verify this relation numerically in the following Fig.~\ref{fig:ks}.
\begin{figure}[h]
    \centering
   \includegraphics[width=\textwidth]{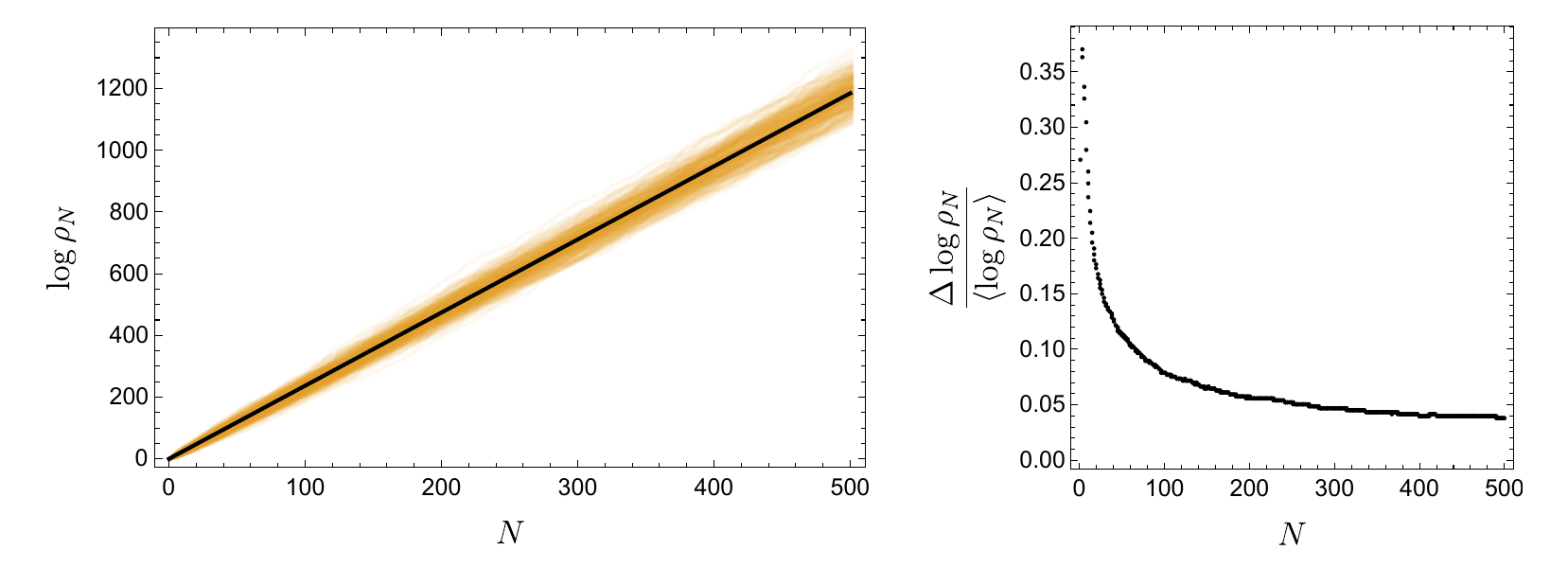}
    \caption{Left: Collapse of the interior as a function of the number of eras $N$. Black line is the equilibrium prediction (\ref{eq:eq}). Orange curves are 300 solutions to the recursion relations (\ref{eq:a}) -- (\ref{eq:d}) with the initial $x_0$ randomised between 0 and 1. In performing the numerics it is important to take $x_0$ randomised to a high number of decimal places.
    The remaining initial values were taken as $\rho_0 = 1$, $k_0 = 3$ and $\Delta_0 = 0.4$. Right: The standard deviation of $\log \rho_N$ relative to the mean, computed using the 300 trajectories shown on the left, decays like $N^{-1/2}$. This is in agreement with a computation in \cite{Khalatnikov1985}.}
    \label{fig:ks}
\end{figure}
Remarkably, then, the rate of collapse of the spatial volume towards the singularity is directly given in terms of the Kolmogorov-Sinai entropy of the underlying dynamical system. More precisely, from (\ref{eq:eq}) and below (\ref{eq:metric}), the double logarithm of the volume
\be
\left\langle \log\left(- \frac{2}{3} \frac{\log V_N}{\rho_0} \right) \right\rangle = h N \,,
\ee
at late interior times. Equivalently, we may say that there is a doubly exponential decrease in volume over an era. This is quite consistent with the full numerical evolution in Fig.~\ref{fig:bounces}.

The results above show that after a very short number of eras, the initial conditions of the interior are erased and the interior cosmological evolution is well-described by considering an ensemble of interior universes drawn from the distribution (\ref{eq:mu}). For astrophysical black holes, the doubly exponential decrease of the volume implies that the interior will reach the Planck scale within an order one number of eras. At the Planck scale the classical description will break down and it is not known whether the chaotic dynamics extends as an underlying principle to the quantum gravitational regime --- a point we will return to below. Even if there are only a small number of eras, however, we have noted that this is enough for chaos to onset. In a holographic context, the volume at the initial bounce can likely be made arbitrarily large in Planck units by taking the AdS radius to be large in Planck units. This should allow an arbitrarily large number of eras within the classical regime.

\subsection{Lapse dynamics}

It remains to characterise the late interior evolution of the lapse $n$. This will allow us to 
demonstrate the emergence of a conserved interior energy in the following \S\ref{sec:energy}.

On the $n$th Kasner epoch we have already established in (\ref{eq:kmetric}) that, up to exponentially small corrections, the lapse is
\be\label{eq:nkasner}
n = \frac{e^{-\frac{3}{2}\rho}}{z_n} \,.
\ee
Here $z_n$ is a constant.
We may obtain the change in $z_n$ between two sequential epochs by integrating (\ref{eq:ndotn}) across the $n$th bounce:
\be\label{eq:logz}
\log \frac{z_{n}}{z_{n-1}} = -\frac{1}{2} \int_n (1 - \dot g^2 - \dot h^2)\,{\rm d}\rho = -\frac{1}{2} \int_n \frac{\ddot g\,{\rm d}\rho}{\frac{1}{2} \dot g - \pa_g V} \,.
\ee
In the second step we used the equation of motion (\ref{eq:deqs}). We have explained in \S\ref{sec:reflect} that across a bounce $\pa_g V$ is a constant, given in (\ref{eq:Wi}). We may write this constant as $\pa_g V^{s_n}$ where $s_n = \{t,x,y\}$ labels the wall involved in the bounce. With $\pa_g V$ constant the last integral in (\ref{eq:logz}) may be performed to yield
\be\label{eq:z2}
\frac{z_n}{z_{n-1}} = \frac{v_{n-1} - 2 \pa_g V^{s_n}}{v_{n} - 2 \pa_g V^{s_n}} = \frac{1 - u_{n-1} + u_{n-1}^2}{1+ u_{n-1} + u_{n-1}^2} \,. \qquad (\text{intra era})
\ee
Recall that $\dot g = v_n$ on the $n$th Kasner epoch. The second step above assumes an intra era transition with $u_{n} = u_{n-1}-1$, and uses (\ref{eq:Wi}), (\ref{eq:vel}), (\ref{eq:un}) and (\ref{eq:unmap}). The final expression holds for bounces off any of the three walls.

The recursion relation (\ref{eq:z2}) can be solved within an era to give
\be\label{eq:z3}
z_n = z (1 + u_n + u_n^2) \,.
\ee
Here $z$ is a constant over the era. The `big bounce' from one era to the next, wherein $u_{k-1} = 1/(u_{k-2}-1)$, also follows (\ref{eq:z2}). Repeatedly using (\ref{eq:z2}), therefore, we can obtain $z_k$ in terms of $z_0$. This gives the inter era transition rule
\be\label{eq:z4}
z_{N+1} = z_N \frac{1 - x_N + x_N^2}{1 - k_N - x_N + (k_N + x_N)^2} \,.
\ee
This equation may be added to the recursions relations (\ref{eq:a}) -- (\ref{eq:d}). In this way we achieve a full characterisation of the interior metric. To our knowledge (\ref{eq:z3}) and (\ref{eq:z4}) did not appear in the classic papers from the Landau Institute. The relation (\ref{eq:z4}) is necessary to establish the emergence of a conserved energy, as we describe in the following section.

The proper time to the singularity, which may be accessible to boundary probes \cite{Grinberg:2020fdj}, can be obtained by integrating the lapse (\ref{eq:lapse}) over all of the Kasner epochs. However, because of the doubly exponential collapse of the volume in the interior, the contribution of large $N$ eras to this proper time is tiny.

\subsection{Conserved energy}
\label{sec:energy}

In this section we will establish that the quantity
\be\label{eq:cen}
\vep \equiv \frac{3 e^{-\frac{3}{2}\rho}}{n} \left(g \dot g + h \dot h - \rho  \right) \,,
\ee
is conserved under the dynamical system we have obtained above for epochs and eras. In \S\ref{sec:en} below we will
see that (\ref{eq:cen}) is a conserved energy that emerges in the limit
where the bounces become abrupt and the cosmological dynamics can be described as geodesic motion on a quotient of the hyperbolic plane. We will also demonstrate in \S\ref{sec:revisit} that the
`time' conjugate to this conserved energy is simply the number of eras.

Firstly, we must express (\ref{eq:cen}) in terms of the variables that appear in the epoch and era recursion relations. On the $n$th Kasner epoch, using (\ref{eq:kasner}) and (\ref{eq:nkasner}) in (\ref{eq:cen}):
\begin{align}
\vep_n & = 3 z_n (g_n v_n + h_n w_n - \rho_n) \label{eq:l1} \\
& = - 3 z_n \left(\sum_i V_n^i p_n^i + \rho_n\right) \\
& = - \frac{9 z_n}{2} \frac{u_n [2 + u_n] \Delta_n + \rho_n}{1 + u_n + u_n^2} \label{eq:use} \\
& = - \frac{9 z}{2} \left(u_n [2 + u_n] \Delta_n + \rho_n\right) \qquad (\text{within an era}) \,. \label{eq:l3}
\end{align}
To obtain the second line we used (\ref{eq:vel}) and (\ref{eq:pot2}). To obtain the third line we expressed $p_n^\pm$ and $p_n^0$ in terms of $u_n$ using (\ref{eq:un}), and $V^0_n = \rho_n/2$, $V^-_n = (\rho_n - 3 \Delta_n)/2$ and $V^+_n = - (2\rho_n - 3 \Delta_n)/2$. These last relations follow from the arguments given in Appendix \ref{sec:fiddly} and shown in Fig.~\ref{fig:era} of that Appendix. The final line then follows from (\ref{eq:z3}).

The first line (\ref{eq:l1}) already shows that $\vep$ is indeed constant on a given Kasner epoch. Using the intra era transitions rules
\be\label{eq:rules}
u_{n+1} = u_{n} - 1 \,, \quad \Delta_{n+1} = \frac{1 + u_n}{u_n} \Delta_n \,, \quad \rho_{n+1} = \rho_n + \frac{(1 + u_n + u_n^2)}{u_n} \Delta_n \,,
\ee
in the final line (\ref{eq:l3}) one verifies that within an era
\be
\vep_{n+1} = \vep_n \,.
\ee
This shows that the energy is also conserved between epochs within a given era. The transition rules (\ref{eq:rules}) follow directly from arguments given in \S\ref{sec:evo} and Appendix \ref{sec:fiddly}.

Finally, the energy on the second epoch of the $N$th era is given by
\be
\vep_N = - \frac{9 z_N}{2} \frac{([k_N + x_N]^2 - 1)\Delta_N + \rho_N}{(k_N+x_N)(k_N + x_N-1)+1} \,.
\ee
This expression follows directly from (\ref{eq:use}) and the definition of $k_N$ and $x_N$ in (\ref{eq:ukx}). Using the inter era transformations (\ref{eq:a}) -- (\ref{eq:d}) along with the lapse transformation (\ref{eq:z4}) one verifies the conservation of energy over eras
\be
\vep_{N+1} = \vep_N \,.
\ee

The emergence of this conserved energy at late interior times is nontrivial.  It raises the possibility of a dual Hamiltonian description of the late interior that might extend beyond the classical gravity regime. Such a description could potentially provide an interior setting in which to account for the black hole microstates. Finally, we should note that the distribution $\mu(x)$ in (\ref{eq:mu}) is independent of the energy of the trajectory and is therefore an extremely universal statement about the late interior dynamics.

\section{The late interior Hamiltonian}
\label{sec:WDW}

\subsection{The Wheeler-DeWitt equation}
\label{sec:en}

In the remainder of the paper we will discuss in detail the arithmetic chaotic properties of the emergent late interior Hamiltonian. This can be done most elegantly starting from the quantum mechanical spectrum of the Hamiltonian. There is an intimate connection between this spectrum and the spectrum of classical closed geodesics \cite{Bogomolny:1992cj}. To leading order in the semiclassical limit, the macroscopic metric degrees of freedom can be quantised using the Wheeler-DeWitt (WDW) equation \cite{DeWitt:1967yk}. The WDW equation is not a complete microscopic theory and is in general somewhat intractable. However, when only a few metric degrees of freedom are macroscropic, as is the case for our spatially homogeneous interiors, the reduced `minisuperspace' theory captures the leading quantum mechanical effects \cite{Halliwell:1989myn}.\footnote{The first corrections to this leading order description are the quantum mechanical zero point energies. These are captured by functional determinants. All of the inhomogeneous metric modes contribute to the zero point energy and hence these corrections are beyond the minisuperspace description.\label{foot:det}}

The first objective is to obtain the minisuperspace WDW equation for our interior. There is a large literature on the WDW equation for mixmaster cosmologies. Some references may be found in e.g.~\cite{GRAHAM19951103}. The first step is to substitute the Ansatz
\be \label{eq:metric2}
{\rm d}s^2 = e^{-\Omega} e^{-2 h} {\rm d}t^2 - n^2 {\rm d}\rho^2 + e^{-\Omega} e^{h} \left(e^{-\sqrt{3} g} {\rm d}x^2 +  e^{\sqrt{3} g} {\rm d}y^2 \right) \,, \qquad A_i = \phi_i\,{\rm d}x^i \,.
\ee
into the action (\ref{eq:action}). Here $\{g, h, \Omega, n, \phi_i\}$ are functions of $\rho$. The usual Gibbons-Hawking boundary term should be added to remove second derivative terms from the action. We see that the Ansatz (\ref{eq:metric2}) is the same as (\ref{eq:metric}) previously, except that we have relaxed the gauge choice in (\ref{eq:metric}) that set $\Omega = \rho$. Define the Lagrangian via $S = \int d^3x L$. The momenta and Hamiltonian are defined in the usual way as
\be
\pi_A \equiv \frac{\pa L}{\pa \dot \Phi_A} \,, \qquad  H \equiv \sum_A \pi_A \dot \Phi_A - L  \,.
\ee
In these expressions $\Phi_A$ runs over $\{g,h,\Omega,\phi_i\}$. One obtains
\be
H = \frac{1}{6} \int {\rm d}\rho\,e^{\frac{3}{2}\Omega}\,n\,{\mathcal H} \,,
\ee
so that the lapse $n$ imposes the Hamiltonian constraint
\be\label{eq:cons}
{\mathcal H} \equiv - \pi_\Omega^2 + \pi_g^2 + \pi_h^2 +
3 e^{- \Omega} \left(e^{-2h} \pi_t^2 + e^{h - \sqrt{3} g} \pi_x^2 + e^{h + \sqrt{3} g} \pi_y^2  \right)
= 0 \,.
\ee
Here we have again considered late interior times so that the mass and cosmological constant terms in the action can be dropped. The momentum constraint is trivial in spatially homogeneous minisuperspace models.

The WDW equation is the canonical quantisation of the constraint (\ref{eq:cons}).
Consider wavefunctions of the form
\be
\Psi = \Psi(g,h,\Omega) e^{\frac{i}{\sqrt{3}}\sum_i f_i \phi_i} \,,
\ee
\emph{i.e.}~a Fourier basis in the space of the Maxwell potentials. Then the WDW equation becomes
\be\label{eq:wdw}
\left[\frac{\pa^2}{\pa \Omega^2} - \frac{\pa^2}{\pa g^2} - \frac{\pa^2}{\pa h^2} + e^{- \Omega + 2 V(g,h)} \right] \Psi(g,h,\Omega) = 0 \,.
\ee
Here $V$ is the potential defined previously in (\ref{eq:pot}). In this context, of course, it is more proper to think of $e^{- \Omega + 2V}$ as the potential.

In (\ref{eq:wdw}) we see that, as usual, the volume coordinate $\Omega$ has the wrong-sign kinetic term. It is therefore natural to consider $\Omega$ as the clock variable and use (\ref{eq:wdw}) to evolve the wavefunction in $\Omega$, given some initial wavefunction of $g$ and $h$ at e.g. $\Omega = 0$. From (\ref{eq:metric2}) the volume is $e^{-3 \Omega}$ so that the late-time singularity is at $\Omega \to + \infty$.

At late times it is useful to reformulate the above dynamics in terms of hyperbolic quantum billiards. This will enable the appearance of sharp walls to be used explicitly in the solution of the equation at late times. Make the change of variables \cite{thorne2000gravitation, belinski_henneaux_2017}
\be\label{eq:misner}
\Omega = e^\tau \cosh R \,,\qquad  g = e^\tau \sinh R \cos \phi \,, \qquad h = e^\tau \sinh R \sin \phi \,.
\ee
These are known as the Chitre-Misner coordinates.
The WDW equation then becomes
\be\label{eq:wdw2}
\left[e^{-2\tau} \left(\frac{\pa^2}{\pa \tau^2} + \frac{\pa}{\pa \tau} -  \nabla^2_{H^2}\right) + e^{- e^\tau \cosh R + 2 V(e^\tau \sinh R \cos\phi,e^\tau \sinh R \sin\phi)} \right] \Psi(\tau,R,\phi) = 0 \,,
\ee
with the Laplacian on the hyperbolic plane $H^2$:
\be
\nabla^2_{H^2} = \frac{\pa^2}{\pa R^2} + \coth R \frac{\pa}{\pa R} + \frac{1}{\sinh^2R} \frac{\pa^2}{\pa \phi^2} \,.
\ee
The above formulae come from changing variables in the WDW equation (\ref{eq:wdw}). One could also quantise the Hamiltonian constraint directly in terms of $\tau,R,\phi$. Both in (\ref{eq:wdw}) and (\ref{eq:wdw2}) there is an ordering ambiguity. This ambiguity is not important to leading semiclassical order. The ordering choice above gives the canonical Laplacian operator on hyperbolic space and more generally corresponds to writing the WDW differential operator as the Laplacian of the inverse DeWitt metric.

The advantage of writing the WDW equation as in (\ref{eq:wdw2}) is that as $\tau \to \infty$ the potential simply constrains the motion to lie in a domain of the hyperbolic plane. That is, as $\tau \to \infty$ the WDW equation becomes the hyperbolic motion
\be\label{eq:WDW2}
\left[\frac{\pa^2}{\pa \tau^2} + \frac{\pa}{\pa \tau} - \nabla^2_{H^2}\right] \Psi(\tau,R,\phi) = 0 \,,
\ee
with the constraint that $\Psi = 0$ outside of the hyperbolic triangle
\be\label{eq:bdy}
- \frac{1}{2} < \sin \phi \tanh R \,, \qquad \sin (\phi \pm \frac{\pi}{3}) \tanh R < \frac{1}{2} \,.
\ee
In particular, the infinite well potential described by (\ref{eq:bdy}) is independent of $\tau$. Equations (\ref{eq:WDW2}) and (\ref{eq:bdy}) define the quantum hyperbolic billiard problem.

Equations (\ref{eq:WDW2}) and (\ref{eq:bdy}) can be solved by separation of variables \cite{PhysRevD.42.2483}. First consider the modes, obeying the boundary conditions in (\ref{eq:bdy}),
\be
- \nabla^2_{H^2}\Psi_n(R,\phi)  = \left(\frac{1}{4} + \vep_n^2 \right) \Psi_n(R,\phi) \,.
\label{eq: wave eq H2}
\ee
Here we may take $\vep_n > 0$. The general solution is then
\be\label{eq:sch}
\Psi(\tau,R,\phi) = \sum_{n\pm} c_{n\pm} \Psi_n(R,\phi) e^{[- \frac{1}{2} \pm i \vep_n] \tau}  \,.
\ee
Here the $c_{n\pm}$ are constants.

As commented in footnote \ref{foot:det} above, the minisuperspace WDW equation is only trustworthy to leading order in the semiclassical limit. Beyond that limit fluctuations of inhomogeneous modes contribute to the wavefunction and ordering ambiguities become relevant. Restricting, then, to highly excited states one has from (\ref{eq:sch}) that 
\be
\Psi \approx \sum_{n\pm} c_{n\pm} \Psi_n(R,\phi) e^{\pm i \vep_n \tau} \,.
\ee
This expression is, of course, just the usual solution for the time-independent Schr\"odinger equation with energy levels $\vep_n$ and time $\tau$. The similarity to the Schr\"odinger equation arises at large $\tau$ because of an emergent `time-translation' invariance in $\tau$ at late times. In this late time limit and to leading order in the semiclassical expansion one could obtain conserved probabilities using the conventional norm of quantum mechanics. However, this norm will not be conserved away from the late time limit. Furthermore, away from the semiclassical limit the factor of $e^{-\tau/2}$ in (\ref{eq:sch}) will cause the Schr\"odinger norm to decay towards the singularity \cite{Kleinschmidt:2009cv, Kleinschmidt:2009hv, Perry:2021mch}. The more natural norm in the present context is instead the WDW norm \cite{DeWitt:1967yk}, as we now describe. The WDW norm leads to conserved probabilities even away from the late time and semiclassical limits.

Treating $\tau$ as the relational clock, the WDW norm for equation (\ref{eq:wdw2}) is
\be\label{eq:wdwnorm}
||\Psi||^2 = -\frac{i}{2} \int  \left(\Psi^\star e^{\tau} \frac{\pa \Psi}{\pa \tau} - e^{\tau} \frac{\pa \Psi^\star}{\pa \tau} \Psi \right) \sinh R \, {\rm d}R\,{\rm d}\phi \,.
\ee
This norm is conserved, $\pa_\tau ||\Psi||^2 = 0$, for states obeying (\ref{eq:wdw2}) and is positive on states built out of only the $+$ modes in (\ref{eq:sch}). Physically, this condition for positivity is usually thought of as restricting to modes on the collapsing as opposed to expanding branch of the wavefunction of the universe. Specifically, letting $c_{n+} = c_n$ and $c_{n-} = 0$ in (\ref{eq:sch}), the norm is
\be\label{eq:norm}
||\Psi||^2 = \sum_n \vep_n |c_n|^2 \int |\Psi_n(R,\phi)|^2 \sinh R \, {\rm d}R\,{\rm d}\phi \,.
\ee
States are therefore normalised by
\be
\sum_n |c_n|^2 = 1 \,, \qquad \frac{1}{\sqrt{\vep_n}} \int |\Psi_n(R,\phi)|^2 \sinh R \, {\rm d}R\,{\rm d}\phi = 1\,.
\ee
The expression (\ref{eq:norm}) is very similar to the usual Schr\"odinger norm. As we have noted, at late times the difference between the Schr\"odinger and WDW norms in the present context is subleading in the semi-classical expansion.

In the remainder we characterise the quantum chaotic nature of the hyperbolic billiard energy level spectrum $\vep_n$ and the corresponding eigenstates $\Psi_n$. For an overview of earlier work on this theme in the context of cosmological singularities, see eg.~\cite{GRAHAM19951103}. We may again emphasise that it is the emergent time-translation symmetry at late interior time that allows the solution of the WDW equation to be written in terms of modes in (\ref{eq:sch}), with conserved WDW norm (\ref{eq:norm}).
As we have noted above, it is tempting to think that this emergent Hamiltonian data could be the starting point for a dual description of the interior.

\subsection{Era dynamics revisited}
\label{sec:revisit}

Before studying the spectrum of the emergent Hamiltonian, we will briefly connect our discussion to the conserved energy of the era dynamics found in \S\ref{sec:energy}. The conserved energy (\ref{eq:cen}) is nothing but the momentum conjugate to $\tau$. From the change of variables (\ref{eq:misner}) 
\be\label{eq:taumom}
p_{\tau} = \Omega\,p_\Omega + h\,p_h + g\,p_g \,.
\ee
The momenta $p_\Omega$, $p_h$ and $p_g$ are obtained in the standard way after substituting the Ansatz (\ref{eq:metric2}) into the action (\ref{eq:action}). Finally, setting $\Omega = \rho$ leads to the expression (\ref{eq:cen}) for the energy.

The momentum (\ref{eq:taumom}) generates translations in the time variable $\tau$. The time at which an era starts can be obtained as follows. The definition of $\tau$ in (\ref{eq:misner}) directly implies that (with $\Omega = \rho$ in the coordinates being used in \S\ref{sec:energy})
\be
e^{2\tau} = \rho^2 - g^2 - h^2 \,.
\ee
At the start of an era this gives, using the same identities as used between (\ref{eq:l1}) and (\ref{eq:l3}),
\be
e^{2\tau_N} = \rho_N^2 - g_N^2 - h_N^2 = 3 \Delta_N (\rho_N - \Delta_N) = 3 \rho_N^2 \delta_N (1 - \delta_N) \,.
\ee
Recall that $\delta_N$ was introduced in (\ref{eq:deltaN}). At large $N$, $\rho_N$ continues to grow while $\delta_N$ remains bounded --- from Appendix \ref{sec:fiddly} one sees that $0 \leq \delta_N \leq 1$. In equilibrium $\delta_N$ has a fairly flat distribution between $0$ and $1$ \cite{Khalatnikov1985}. Therefore to leading order at large $N$, and using (\ref{eq:eq}),
\be
\tau_N \approx \log \rho_N \approx h N \,.
\ee
We see that the time conjugate to the conserved energy is simply the number of eras.

\subsection{\label{sec:cong}Congruence subgroups and automorphic forms}

\begin{figure}[h]
    \centering
   \includegraphics[width=0.73\textwidth]{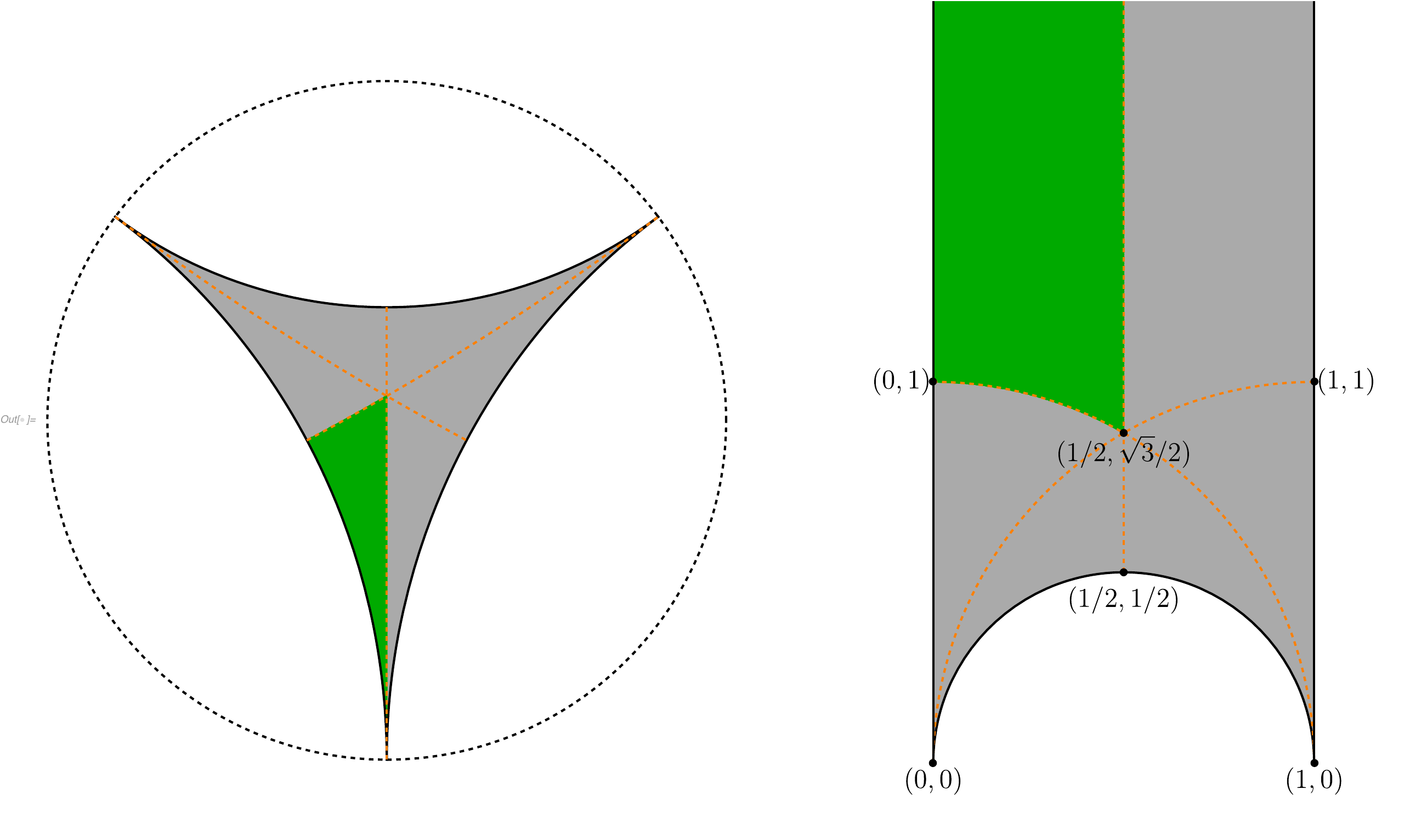}
    \caption{Left: The original domain on the Poincaré disk of the billiard motion is shown in grey. The smaller green triangle is used in the spectral statistics analysis. Right: The same two triangles mapped to the upper half plane.}
    \label{fig:domain}
\end{figure}

In this section we will show how the solutions to the hyperbolic Laplacian \eqref{eq: wave eq H2} with boundary conditions \eqref{eq:bdy} can be obtained from harmonic analysis on the fundamental domain of $\Gamma(2)$, a Fuchsian group with important arithmetic properties \cite{GRAHAM19951103}. In general, Fuchsian groups are discrete subgroups of the $PSL(2,\mathbb{R})$ isometry of the hyperbolic plane. The connection to the fundamental domain of $\Gamma(2)$ is seen most easily by mapping the domain of the billiard from the Poincaré disk $(R,\phi)$ to the upper half plane $(x,y)$ by the conformal transformation 
\begin{align}\label{eq:changecoordinates}
    z = x+ iy =  \frac{\sqrt{3}i}{2}\frac{e^{i\phi}+i\tanh(R/2)}{e^{i\phi}-i\tanh (R/2)}+\frac{1}{2} \,,
\end{align}
with $y>0$.
The metric on the upper half plane becomes
\begin{equation}
    {\rm d}s^2 = \frac{{\rm d}x^2+{\rm d}y^2}{y^2}.
\end{equation}
In these new coordinates, the walls \eqref{eq:bdy} of the triangle bounding the cosmological billiard motion are given by three geodesics, depicted in Fig.~\ref{fig:domain}, 
\begin{equation}
    x=0, \qquad x=1 \qquad \text{and} \qquad \left(x-\frac{1}{2}\right)^2+y^2=\frac{1}{4} \,.
    \label{eq: walls UHP}
\end{equation}
It will be useful to note that reflections across the three walls \eqref{eq: walls UHP} are given by
\begin{align}\label{eq:refl}
    z \rightarrow -z^* \,, \qquad z \rightarrow 2-z^* \,, \qquad z \rightarrow \frac{z^*}{2z^*-1} \,,
\end{align}
respectively.
These generate a group called $G \Gamma(2)$ \cite{belinski_henneaux_2017}. We will see shortly that this group is closely related to $\Gamma(2)$, but contains an additional reflection. 

The billiard domain (\ref{eq: walls UHP}) on the upper half plane corresponds to half of the fundamental domain of the Fuchsian group 
\begin{equation}\label{eq: gamma(2)}
    \Gamma(2) \equiv \left\{ \begin{pmatrix}
        a & b\\ c &d
    \end{pmatrix} \in SL(2,\mathbb{Z} ) \mid (a,d  = 1 \land  b,c  = 0) \mod 2 \right\} ,
\end{equation}
which is generated by the $SL(2,\mathbb{Z})$ transformations
\be
    z \rightarrow z+2 \,,\qquad
    z \rightarrow \frac{z}{2z+1}.
    \label{eq: generators gamma(2)}
\ee
The group $\Gamma(2)$ is known as the principal congruence subgroup of $SL(2,\mathbb{Z})$ of level 2. The fundamental domain of $ \Gamma(2)$, which is the grey region in Fig.~\ref{fig:domain} together with its reflection about the imaginary axis $x=0$, defines a minimal region on $H^2$ which tessellates the whole upper half plane under the action \eqref{eq: generators gamma(2)}. Harmonic analysis on this domain is concerned with automorphic eigenfunctions $\psi(z)$ of the hyperbolic Laplacian. These are waveforms that are invariant under the action of the group,
\begin{equation}\label{eq:auto}
    \psi( \gamma z) = \psi(z), \qquad \forall \gamma \in \Gamma(2) .
\end{equation}
Automorphic waveforms are well-known to have rich properties, that we will recover below. Firstly, we explain how automorphic forms (\ref{eq:auto}) provide the solutions to our spectral problem, in which we must impose vanishing on the walls \eqref{eq: walls UHP}.

The spectrum of the Laplacian on the fundamental domain of a Fuchsian group has both a continuous and a discrete part. These are known as the Eisenstein series and the Maa{\ss} cusp forms, respectively \cite{sarnak1993arithmetic}. For the purposes of Hamiltonian eigenvalue statistics, we shall focus on the discrete spectrum in this work. However, not all $\Gamma(2)$ Maa{\ss} cusp forms will vanish on \eqref{eq: walls UHP}, which is a stronger condition than (\ref{eq:auto}). To resolve this, we start by noting an obvious parity symmetry of the fundamental domain, such that waveforms may be classified into odd $\psi(-z^*) = -\psi(z)$ and even $\psi(-z^*) = \psi(z)$ under a reflection across the imaginary axis. Clearly, our Dirichlet (vanishing) boundary condition on the imaginary axis requires that we restrict to odd parity waveforms. We may now show that this restriction is sufficient to imply Dirichlet boundary conditions on the remaining two walls \eqref{eq: walls UHP} of the triangle.
Using oddness followed by invariance under the $\Gamma(2)$ generators \eqref{eq: generators gamma(2)} we obtain
\begin{align}
    \psi(2-z^*) &= - \psi(z-2 )= - \psi(z ), \label{eq:r2} \\
    \psi\left(\frac{z^*}{2z^*-1}\right) &= - \psi\left(\frac{z}{-2z+1}\right)= - \psi(z ),
    \label{eq:dirichlet semicircle}
\end{align}
We may recognise these equations as imposing oddness under the reflections (\ref{eq:refl}) about the remaining two walls. The waveform therefore vanishes on all three walls, as required. It follows that the waveforms we are interested in are precisely the odd $\Gamma(2)$ Maa{\ss} cusp forms.

There is an additional complication due to the fact that our domain is an equilateral triangle, see the left plot in Fig.~\ref{fig:domain}. This triangle has an $S_3$ symmetry group, which permutes the three corners of the triangle. In order to correctly analyze chaotic properties of the spectrum, for example in our computation of the spectral form factor below, it is essential to first separate out the waveforms that transform under different representations of this symmetry. The permutation group $S_3$ is generated by one reflection, which we can take to be $z \rightarrow 1- z^*$, this is a reflection across the middle of the right plot in Fig.~\ref{fig:domain}, and one rotation, $R$, by $2\pi/3$ in the hyperbolic plane $\phi$-coordinate. On the upper half plane
\begin{equation}
    R(z) = \frac{z-1}{z} \,,
    \label{eq: Z3 rotation}
\end{equation}
using  the change of coordinates \eqref{eq:changecoordinates}. These transformations commute with the Laplacian, as they are symmetries of the hyperbolic plane, and preserve the $\Gamma(2)$ automorphic and oddness properties of the waveforms. Oddness may be verified directly --- $S_3$ permutes the walls and is therefore compatible with imposing Dirchlet boundary conditions on all three walls --- while the automorphic property follows from the fact that the group $\Gamma(2)$ is invariant under conjugation by either generator of $S_3$.\footnote{Explicitly, let $\phi(z) \equiv \psi(R(z)) = \psi((z-1)/z)$ be the rotated waveform. Oddness follows from $\phi(-z^*) = \psi(R(-z^*)) = \psi((z^*+1)/z^*) = - \psi((z-1)/z) = - \phi(z)$. The second to last step uses the reflection (\ref{eq:r2}). The automorphic property follows from $\phi(\gamma z) = \psi(R(\gamma z)) = \psi(\gamma' R(z)) = \psi(R(z)) = \phi(z)$. Here we have used $R\gamma R^{-1} = \gamma' \in \Gamma(2)$ whenever $\gamma \in \Gamma(2)$. This last fact may be verified using the generators (\ref{eq: generators gamma(2)}).} It follows that odd $\Gamma(2)$ waveforms can be classified according to the irreducible representations of $S_3$.

The group $S_3$ has three irreducible representations: two 1-dimensional representations and one 2-dimensional representation. Waveforms transforming in the 1-dimensional representations are invariant under the $\mathbb{Z}_3$ rotations and are either even (trivial representation) or odd (sign representation) under the reflections. These two types of waveforms are determined by their form on a smaller triangle with walls at
\begin{equation}\label{eq:walls small triangle}
    x=0, \qquad x=1/2 \qquad \text{and} \qquad x^2+y^2=1,
\end{equation}
which is the smaller green triangle in Fig.~\ref{fig:domain}. The trivial and sign representations correspond to imposing 
either Neumann or Dirichlet boundary conditions on the orange lines in Fig.~\ref{fig:domain}, respectively. On the other hand, two basis waveforms in the 2-dimensional (standard) representation transform into linear combinations of each other under the discrete rotations. These waveforms can be taken to be even and odd, respectively, under a chosen reflection. 

The above classification into irreducible representations is all we will need to set up numerics in the following section. However, in order to connect with previous results on the spectra of Maa{\ss} forms (such as in the database \cite{lmfdb}) we will now show that the different representations have different arithmetic properties. Specifically, we will show that
odd $\Gamma(2)$ waveforms transforming under the different irreducible representations of $S_3$ can be realised as automorphic forms of an
additional class of congruence subgroups of $SL(2,\mathbb{Z})$, known as Hecke congruence subgroups:
\begin{equation}\label{eq: gamma0(N)}
    \Gamma_0(N) \equiv \left\{ \begin{pmatrix}
        a & b\\ c &d
    \end{pmatrix} \in SL(2,\mathbb{Z} ) \mid c  = 0 \mod N \right\} .
\end{equation}
We will be concerned with $N =1,2$, and also $N=4$. In these cases the generators are
\begin{align}
    z \rightarrow z+1 \,, \qquad z \rightarrow \frac{z}{Nz+1} \,.
    \label{eq: generators gamma0(N)}
    \end{align}
It is easy to see that the following inclusions hold:
\be\label{eq:include}
\Gamma(2) \subset \Gamma_0(2) \subset \Gamma_0(1) = SL(2,\mathbb{Z})  \,. 
\ee
The first inclusion arises because acting twice with the first generator of $\Gamma_0(2)$ in (\ref{eq: generators gamma0(N)}) gives the first generator of $\Gamma(2)$ in (\ref{eq: generators gamma(2)}). The second inclusion is seen by acting twice with the second generator of $\Gamma_0(1)$ in \eqref{eq: generators gamma0(N)}.
Bigger groups have smaller fundamental domains, so that $\Gamma_0(1) \backslash H^2  \subset \Gamma_0(2) \backslash H^2  \subset \Gamma(2) \backslash H^2$.
We now proceed to show that the set of odd newforms associated to the sequence of inclusions (\ref{eq:include}) --- these are waveforms that are not inherited from bigger groups under the inclusions ---  transform according to a particular irreducible representation of $S_3$. $\Gamma_0(1)$ will be associated to the sign representation, newforms of $\Gamma_0(2)$ to the standard representation and newforms of $\Gamma(2)$ to the trivial representation.

We may first show that an odd Maa{\ss} waveform for $\Gamma_0(1) = SL(2,\mathbb{Z})$ transforms according to the sign representation, which corresponds to setting Dirichlet boundary conditions on the small triangle \eqref{eq:walls small triangle}. Since the rotation \eqref{eq: Z3 rotation} is an element of $SL(2,\mathbb{Z})$, it immediately follows that the Maa{\ss} waveform is invariant under that rotation. Moreover, using that the waveform is odd under reflections across the $y$-axis and invariant under the transformation $z \rightarrow z + 1$, one finds it is also odd under reflections across the vertical line $x = 1/2$ (which, we recall, is one of the reflections generating $S_3$):
\begin{equation}
    \psi(1-z^*) =  -\psi(z -  1) = - \psi(z).
\end{equation}
By rotational symmetry, the odd waveforms on $SL(2,\mathbb{Z})$ are therefore also odd under the other two reflections in $S_3$.

Next, we turn to the odd Maa{\ss} newforms for $\Gamma_0(2)$. Our claim is that these comprise the waveforms of the original billiard problem that transform in the two-dimensional irreducible representation and which are odd under the $S_3$ reflection about the vertical line $x=1/2$.\footnote{The even partner waveforms can be constructed from the odd waveforms, by acting with $S_3$ generators. However, these partners are not automorphic on $\Gamma_0(2)$, as they are not invariant under $z \rightarrow z +1$.} We need to check that a newform on $\Gamma_0(2)$ vanishes on the three walls of the large triangle, and also transforms nontrivially under rotations. The generator $z \rightarrow z +1$ is common to both $\Gamma_0(2)$ and $\Gamma_0(1)$. Using this generator as in the previous paragraph, we may conclude that the waveform vanishes on the vertical walls $x=0$ and $x=1$, and is odd about $x=1/2$. The other generator, $z \to z/(2z+1)$, is common to both $\Gamma_0(2)$ and $\Gamma(2)$. We may therefore use the argument in \eqref{eq:dirichlet semicircle} to conclude that the waveform vanishes on the third, semi-circle, wall of the large triangle \eqref{eq: walls UHP}. Importantly, however, the waveform is not zero on the semicircle $x^2+y^2=1$, otherwise it would solve the Dirichlet problem on the small triangle and be a $\Gamma_0(1)$ waveform and hence not a newform of $\Gamma_0(2)$. This nonvanishing breaks rotational symmetry because the geodesic $x^2+y^2=1$ can be mapped to $x=1/2$ under a $\mathbb{Z}_3$ rotation. It follows that the waveform transforms under rotations, as required.

Finally, this leaves the waveforms transforming in the trivial representation which must therefore be newforms on $\Gamma(2)$. We may note that, by comparing the generators \eqref{eq: generators gamma(2)} and \eqref{eq: generators gamma0(N)}, $\Gamma(2)$ is seen to be conjugate to $\Gamma_0(4)$ under the action $z \to 2 z$. The trivial representation may therefore also be associated to newforms of $\Gamma_0(4)$. We have checked that our numerical eigenvalues, obtained in the following section, match eigenvalues of the corresponding newforms of $\Gamma_0(1), \Gamma_0(2)$ and $\Gamma_0(4)$ found in the database \cite{lmfdb}.

\section{Numerical methods for hyperbolic triangles}
\label{sec:numerical}

We have shown in \S\ref{sec:cong} that eigenvalues for the sign and trivial representations of $S_3$ can be obtained from the spectrum of $\nabla^2_{H^2}$ on a smaller triangular domain.\footnote{Waveforms in the `standard' representation of $S_3$ must instead be computed on the full triangular domain. Using methods analogous to those described below we have computed a handful of these eigenvalues and checked that they match eigenvalues of odd $\Gamma_0(2)$ newforms found in the database \cite{lmfdb}.} This domain is given by the interior of (\ref{eq:walls small triangle}) and is the green region in Fig.~\ref{fig:domain}. To solve the Laplace equation on this domain we must specify boundary conditions.
For all waveforms we demand
\begin{subequations}
\begin{equation}
\lim_{y\to+\infty}\psi(x,y)=0\,.
\end{equation}
The sign representation further requires Dirichlet boundary conditions on all walls,
\begin{equation}
\psi(x,y)=0\qquad\text{on}\qquad x^2+y^2=1\,,\; x=0\;\;\text{and}\;\; x=\frac{1}{2}\,,
\end{equation}
while the trivial representation requires Neumann boundary conditions on two of the walls,
\begin{equation}
\psi(0,y)=0\,,\qquad \left.\frac{\partial \psi(x,y)}{\partial x}\right|_{x=1/2} = 0\,,\qquad \left.x\frac{\partial \psi(x,y)}{\partial x}+y\frac{\partial \psi(x,y)}{\partial y}\right|_{x^2+y^2=1}=0\,.\label{eq:bcircle}
\end{equation}
\end{subequations}
We will refer to these as `Dirichlet' and `Neumann' waveforms, respectively.

We have implemented two numerical methods to obtain eigenpairs $(\varepsilon_n,\Psi_n)$. The first method allowed the computation of large numbers of eigenvalues $\varepsilon_n$, while the second allowed a very accurate determination of the waveforms $\Psi_n$. In addition to the self-consistency checks described in \S\ref{sec:checks} below, our eigenvalues agree with available results in \cite{lmfdb} and \cite{personal}. The methods we will use do not assume the Hecke relations --- these are described in \S\ref{sec:heck} below --- but rather obtain the Hecke relations as an output. This gives a further check of the results and is complementary to other methods such as those based on \cite{Hejhal1991}.

\subsection{Numerical method to determine large collections of eigenvalues}
Here we use the methods of \cite{PhysRevA.44.1491}, extended with minor adaptations to Neumann waveforms.

The general solution to the Laplace equation (\ref{eq: wave eq H2}) can be written as a sum of separable modes. For the Dirichlet case:
\begin{equation}
\Psi_n(x,y)=\sum_{m=1}^{+\infty}c_m^n \sqrt{y}\,K_{i \varepsilon_n}(2\pi m y)\sin(2\pi m x)\,,
\label{eq:oddcase}
\end{equation}
where $K_{i \varepsilon_n}$ is a modified Bessel function of the second kind. By construction, $\Psi_n(x,y)$ in (\ref{eq:oddcase}) automatically satisfies three of the boundary conditions: it vanishes on the vertical walls and towards $y \to +\infty$.
It remains to impose vanishing on the curved boundary $x^2+y^2=1$, which requires that
\begin{equation}
0=\sum_{m=1}^{+\infty}c_m^n\,(1-x^2)^{1/4}K_{i \varepsilon_n}\left(2\pi m \sqrt{1-x^2}\right)\sin(2\pi m x)\qquad \forall x\in[0,1/2]\,.
\label{eq:cond}
\end{equation}
The above condition will determine the allowed $c_m^n$ and $\varepsilon_n$, as we now explain. The set of functions $\{\sin(2\pi l x)\}$, with $l\geq1$, forms a complete basis for functions on the half unit interval that vanish at $x=0,1/2$. We may therefore
project (\ref{eq:cond}) onto this basis to obtain a linear system of equations for the $c_m^n$:
\begin{subequations}
\begin{equation}
\sum_{m=1}^{+\infty} c_m^n A_{lm}(\varepsilon_n)=0\,,
\label{eq:sum}
\end{equation}
with
\begin{equation}
A_{lm}(\varepsilon_n)\equiv4\int_0^{1/2}\mathrm{d}x\,(1-x^2)^{1/4} K_{i \varepsilon_n}\left(2\pi m \sqrt{1-x^2}\right)\sin(2\pi l x)\sin(2\pi m x) \,.
\label{eq:entries}
\end{equation}
\end{subequations}
The numerical task at hand consists in doing the following:
\begin{enumerate}
\item Truncate the sum (\ref{eq:sum}) up to some number of terms $N$ so that
\begin{equation}
\sum_{m=1}^{N} c_m^n A_{lm}(\varepsilon_n)\approx 0 \; \Leftrightarrow \; {\bf A}(\varepsilon_n)\cdot {\bf c}^n\approx 0\,.
\label{eq:sumN}
\end{equation}
\item The truncated sum is now recognised as computing the null space of the matrix ${\bf A}(\varepsilon_n)$, with entries $A_{lm}(\varepsilon_n)$ given in (\ref{eq:entries}).
\item We can determine non-trivial solutions to the null space of ${\bf A}(\varepsilon_n)$ by scanning the values of $\varepsilon_n$ until $\mathrm{det}{\bf A}(\varepsilon_n)=0$.
\item Once we know the value of $\varepsilon_n$ for which $\mathrm{det}{\bf A}(\varepsilon_n)=0$, we can determine ${\bf c}^n$ by finding the null space of  ${\bf A}(\varepsilon_n)$.
\end{enumerate}

While the procedure above could in principle accurately determine all the allowed values of $\varepsilon_n$ in a given range, in practice it leads to an unstable algorithm. This is because the matrix elements $A_{lm}(\varepsilon_n)$ quickly become too small and in general lead to a very ill-conditioned matrix ${\bf A}(\varepsilon_n)$. We must use \emph{pre-conditioners} to accurately determine when $\mathrm{det}{\bf A}(\varepsilon_n)$ changes sign.\footnote{We aim to solve $\mathbf{A}(\vep_n)\cdot{\bf c}^n=0 $, but instead we will solve $\mathbf{L}\cdot\mathbf{A}(\vep_n)\cdot \mathbf{R}\cdot \hat{{\bf c}}^n=0$ and choose $\mathbf{L}$ and $\mathbf{R}$ so that $\mathbf{A}^\prime(\vep_n)\equiv\mathbf{L}\cdot\mathbf{A}(\vep_n)\cdot \mathbf{R}$ is well-conditioned. Once we find $\hat{{\bf c}}^n$ we can determine ${\bf c}^n=\mathbf{R}\cdot\,\hat{{\bf c}}^n$.
} The pre-conditioners are chosen so that the diagonal and near-diagonal elements of the pre-conditioned matrix are large relative to the remaining entries. For the Dirichlet case, following \cite{PhysRevA.44.1491}, we take the left and right pre-conditioners to define
\begin{subequations}
\begin{equation}
A^\prime_{lm}(\varepsilon_n)=l^{3/2}\,\,A_{lm}(\varepsilon_n)\,\frac{e^{\frac{\pi \varepsilon_n}{2}+\phi_m(\varepsilon_n)}}{m^{1/2}} \,,
\end{equation}
with
\begin{equation}
\phi_m(\varepsilon_n)=\left\{
\begin{array}{cc}
0 & \text{if}\quad \sqrt{3}\pi m\leq \varepsilon_n
\\
\sqrt{3\pi^2m^2-\varepsilon_n^2}-\varepsilon_n \arccos\left(\frac{\varepsilon_n}{\sqrt{3}\pi m}\right) & \text{if}\quad \sqrt{3}\pi m> \varepsilon_n
\end{array}\right..
\label{eq:phipre}
\end{equation}
\label{eqs:precond}
\end{subequations}
Both pre-conditioners are diagonal. The ill-conditioning of $\mathbf{A}(\vep_n)$ is closely tied to the rapid decay of the modified Bessel function $K_{i\vep_n}(y)$ at large $y$. The somewhat intricate form of (\ref{eq:phipre}) is determined in  \cite{PhysRevA.44.1491} from the asymptotic behaviour of the Bessel function.

A final complication is that $\mathrm{det}{\bf A^\prime}(\varepsilon_n)$, computed for large but fixed $N$, is a highly oscillatory function of $\varepsilon_n$. To make sure that we capture all the zeros of $\mathrm{det}{\bf A^\prime}(\varepsilon_n)$ in a given range of $\varepsilon_n$ we use bisection methods. We will detail below several tests we performed on our data that strongly suggest we did not miss any of the first 2250 eigenvalues. 

We now turn to the Neumann sector, adapting the methodology of \cite{PhysRevA.44.1491}. As far as we are aware, a large number of numerical eigenvalues have not previously been obtained for this sector. In the Neumann case the mode expansion becomes
\begin{equation}
\Psi_n(x,y)=\sum_{m=1}^{+\infty}c_m^n \sqrt{y}\,K_{i \varepsilon_n}\left[\pi (2m-1) y\right]\sin[\pi (2m-1) x]\,.
\label{eq:evencase}
\end{equation}
The boundary condition on the circle $x^2+y^2=1$ in (\ref{eq:bcircle}) is then
\begin{align}
& \frac{1}{2}(1-x^2)^{1/4}\sum_{m=1}^{+\infty}\,c^n_m\Bigg\{\sin\left[\pi(2m-1) x\right]\,\Big[K_{i \varepsilon_n}\left[\pi (2m-1) \sqrt{1-x^2}\right] \label{eq:nuts}
\\
& -(1-x^2)^{1/2}  (2m-1)\pi\left(K_{i \varepsilon_n+1}\left[\pi (2m-1) \sqrt{1-x^2}\right]+K_{i \varepsilon_n-1}\left[\pi (2m-1) \sqrt{1-x^2}\right]\right)\Big] \nn 
 \\
& +2(2m-1)\pi\,x\,K_{i \varepsilon_n}\left[\pi (2m-1) \sqrt{1-x^2}\right]\,\cos\left[\pi(2m-1) x\right]\Bigg\}=0\qquad \forall x\in\left[0,1/2\right]\,. \nn
\end{align}

A useful basis of functions which vanish at $x=0$ and are even across $x=1/2$ is $\left\{\sin[(2l-1)\pi x]\right\}$ with $l=1,2,\dots$. The procedure follows the same outline as in the Dirichlet case above, and, after making appropriate truncations, we find that we need to solve
\begin{subequations}
\begin{equation}
\sum_{m=1}^{N}B_{lm}(\varepsilon_n) c^n_m\approx 0 \,,
\end{equation}
with
\begin{multline}
B_{lm}(\varepsilon_n)\equiv 2\int_0^{1/2}\mathrm{d}x\,\Bigg\{\sin\left[\pi(2m-1) x\right]\,\Big[K_{i \varepsilon_n}\left[\pi (2m-1) \sqrt{1-x^2}\right]
\\
-(1-x^2)^{1/2}  (2m-1) \pi\left(K_{i \varepsilon_n+1}\left[\pi (2m-1) \sqrt{1-x^2}\right]+K_{i \varepsilon_n-1}\left[\pi (2m-1) \sqrt{1-x^2}\right]\right)\Big]
 \\
+2(2m-1)\pi\,x\,K_{i \varepsilon_n}\left[\pi (2m-1) \sqrt{1-x^2}\right]\,\cos\left[\pi(2m-1) x\right]\Bigg\}\sin\left[\pi(2l-1) x\right]\,.
\end{multline}
\end{subequations}
As previously, the matrix $\mathbf{B}(\vep_n)$, with components $B_{lm}(\vep_n)$, is ill-conditioned.
Finding suitable pre-conditioners might appear challenging, as 
the boundary condition (\ref{eq:nuts}) involves multiple Bessel functions that need to balance against each other. However, after
some trial and error we have found that it is sufficient to define a matrix $\mathbf{B}^\prime(\vep_n)$ just as in (\ref{eqs:precond}), but with $m$ in (\ref{eq:phipre}) replaced by $(m-1/2)$. For the Neumann case, computing the matrix $\mathbf{B}^\prime(\vep_n)$ is more daunting, and we have only managed to accurately determine the first 1200 eigenvalues. Again, below, we will provide tests of our data suggesting that we have captured all the Neumann eigenvalues in the range $0 < \varepsilon_n \lesssim 170.31(7)$.

In both Dirichlet and Neumann cases, we assess the numerical accuracy by varying $N$ to determine up to which decimal place we can trust our numerical method. All the numerical data reported here is determined with precision up to at least the $10^{\rm th}$ decimal place.

To accurately determine the precise value of a given $\varepsilon_n$ to a desired number of decimal places and the corresponding coefficients $c^n_m$ in the waveform, we employed a different numerical scheme, which we will outline in the following section.

\subsection{Numerical scheme for finding $\varepsilon_n$ and few thousand $c^n_m$\label{sec:fewthousandcns}}
With a given $\varepsilon_n$ known to some precision, we can change gears and use a different numerical method to determine $\varepsilon_n$ and the corresponding $c^n_m$ more accurately. In this method one solves for the eigenpair $\{\Psi_n,\varepsilon_n\}$ directly on a discretised grid. First, we change to a new set of coordinates wherein $(x,y)$ are mapped to coordinates $(X,Y)$,
\begin{equation}
X= x \quad\text{and}\quad Y = \frac{y}{\sqrt{1+y^2}} \,,
\label{eq:coordXY}
\end{equation}
that take finite values on the integration domain. The interior of the walls (\ref{eq:walls small triangle}) becomes
\begin{equation}
X^2+\frac{Y^2}{1-Y^2}\geq 1\,,\quad0\leq X\leq1/2\,,\quad\text{and}\quad \sqrt{\frac{3}{7}}\leq Y\leq 1\,.
\label{eq:nontrivial}
\end{equation}

The integration domain is still non-trivial, see Fig.~\ref{fig:int} below, but has four boundaries. To proceed, we use the methods of \cite{Dias:2015nua}, and in particular the so-called Coons patches.
\begin{figure}[h]
    \centering
   \includegraphics[width=0.43\textwidth]{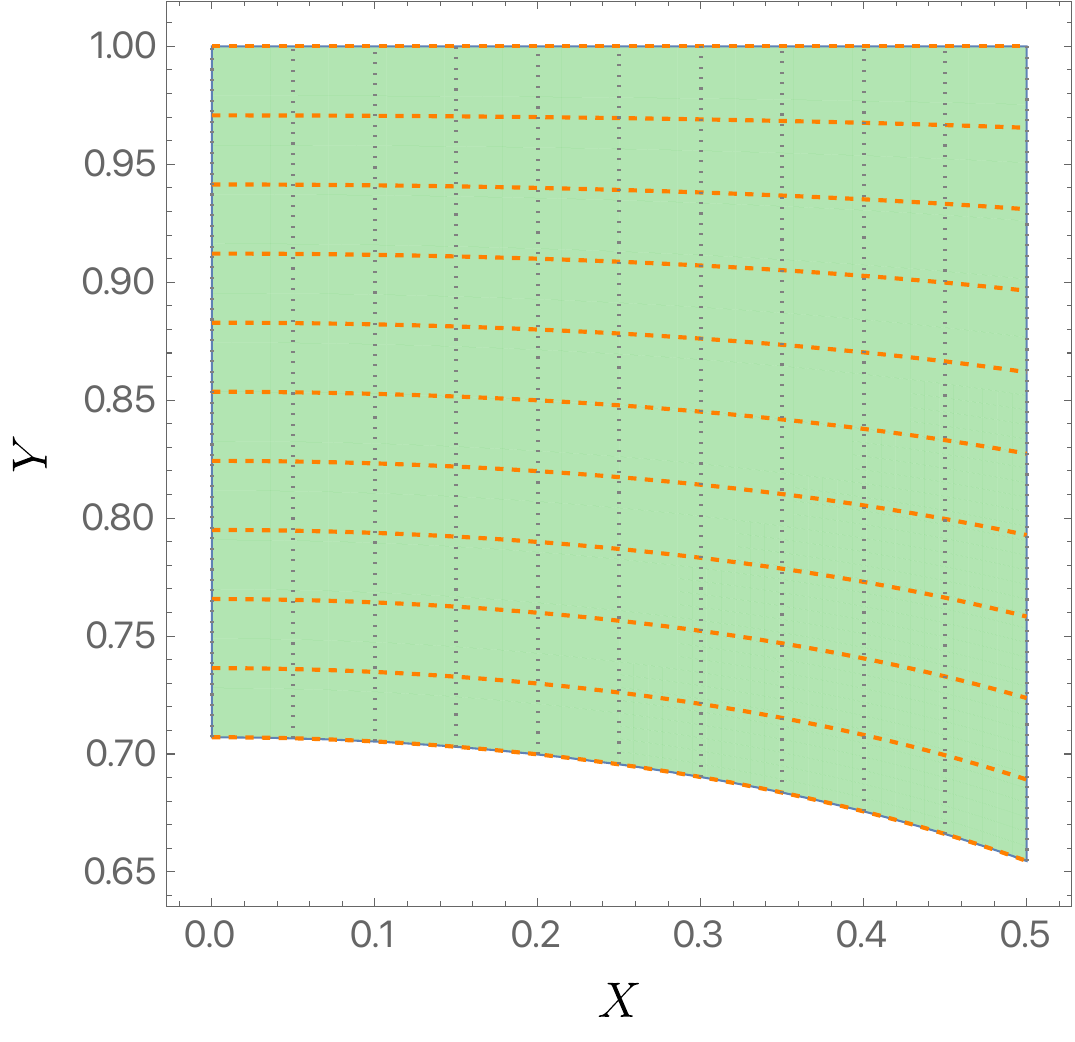}
    \caption{The domain of integration as a function of the $(X,Y)$ coordinates defined in (\ref{eq:nontrivial}). The logical space introduced in \cite{Dias:2015nua} is spanned by Cartesian coordinates $(u,v)$ represented above as dashed orange lines (constant $u$) and dotted grey lines (constant $v$).}
    \label{fig:int}
\end{figure}
The Coons patch technique revolves around introducing a logical space $(u,v)\in(0,1)^2$ and expressing all numerical derivatives with respect to $X$ and $Y$ in terms of derivatives with respect to $u$ and $v$, which exist within a standard Cartesian grid. In Fig.~\ref{fig:int}, we depict lines of constant $u$ as dashed orange lines and lines of constant $v$ as dotted grey lines. Along each Cartesian direction in logical space, we implement a Gauss-Lobatto grid with $\mathcal{N}$ nodes and solve the resulting discretised equations using a standard Newton-Raphson algorithm (see \cite{Dias:2015nua} for various examples of solving eigenvalue problems using Newton-Raphson).

Once we determine $\{\varepsilon_n, \Psi_n(X, Y)\}$, we interpolate to find $\Psi_n$ on a constant $y=y_c$ slice via the map (\ref{eq:coordXY}). Now, we have a function $F(x) = \Psi_n(x, y_c)$ whose Fourier decomposition we aim to obtain. To achieve this, we utilise FFTs for quarter Fourier grids and extract the corresponding $c^n_m$. In this process, we need to choose a value for $y_c$, and our numerical results indicate that selecting $y_c$ at the location of the maximum of $|\Psi(X, Y)|$ yields the best outcomes. Additionally, it is crucial to evaluate $K_{i \varepsilon_n}(\chi)$ to derive the final form of the $c^n_m$. For sufficiently large $\chi$ and/or $\varepsilon_n$, computing these Bessel functions poses numerical challenges. Fortunately, we can overcome this hurdle using the methods outlined in \cite{booker_strömbergsson_then_2013}.

\subsection{Validating the numerical data}
\label{sec:checks}

The so-called Weyl formula allows a stringent test of our numerical data, checking that we have not omitted any eigenvalues. In the following section the Weyl formula will also allow the spectrum to be `unfolded', ensuring a mean eigenvalue spacing of unity in order to align with standard level spacing statistics. In essence, the Weyl formula counts the smoothed number of eigenvalues $\bar{N}(\mu)$ below or equal to a given eigenvalue $\mu\equiv\varepsilon^2+\frac{1}{4}$.

For the Dirichlet sector, the Weyl formula is \cite{PhysRevA.44.1491}:
\begin{equation}
\bar{N}_D(\mu)=\frac{1}{4 \pi }\left[\frac{\pi \mu }{6}-\sqrt{\mu }\,\log \mu+(2- \log 8) \sqrt{\mu }+\frac{23 \pi }{36}+\mathcal{O}(\mu^{-1/2})\right]\,,
\label{eq:weylF}
\end{equation}
where the subscript ${}_D$ is for `Dirichlet'. The leading term in the expansion is determined by the area of the billiard and is independent of the boundary conditions or shape of the billiard.\footnote{Locally one may consider the Laplace equation to be in flat space with eigenstates $e^{i k\cdot x}$ and eigenvalues $\mu = k^2$. Given a region of area $A$ one can, by the usual counting of plane wave states, fit $N = A \int d^2k/(2\pi)^2$ states into the region. Setting $d^2k = 2 \pi k dk$ and integrating up to $k =\sqrt{\mu}$ gives $N(\mu) = \frac{A}{4 \pi} \mu$.\label{foot:count}} The area of the small triangle defined by the interior of (\ref{eq:walls small triangle}) is simply $\pi/6$.

\begin{figure}[h]
    \centering
   \includegraphics[width=0.75\textwidth]{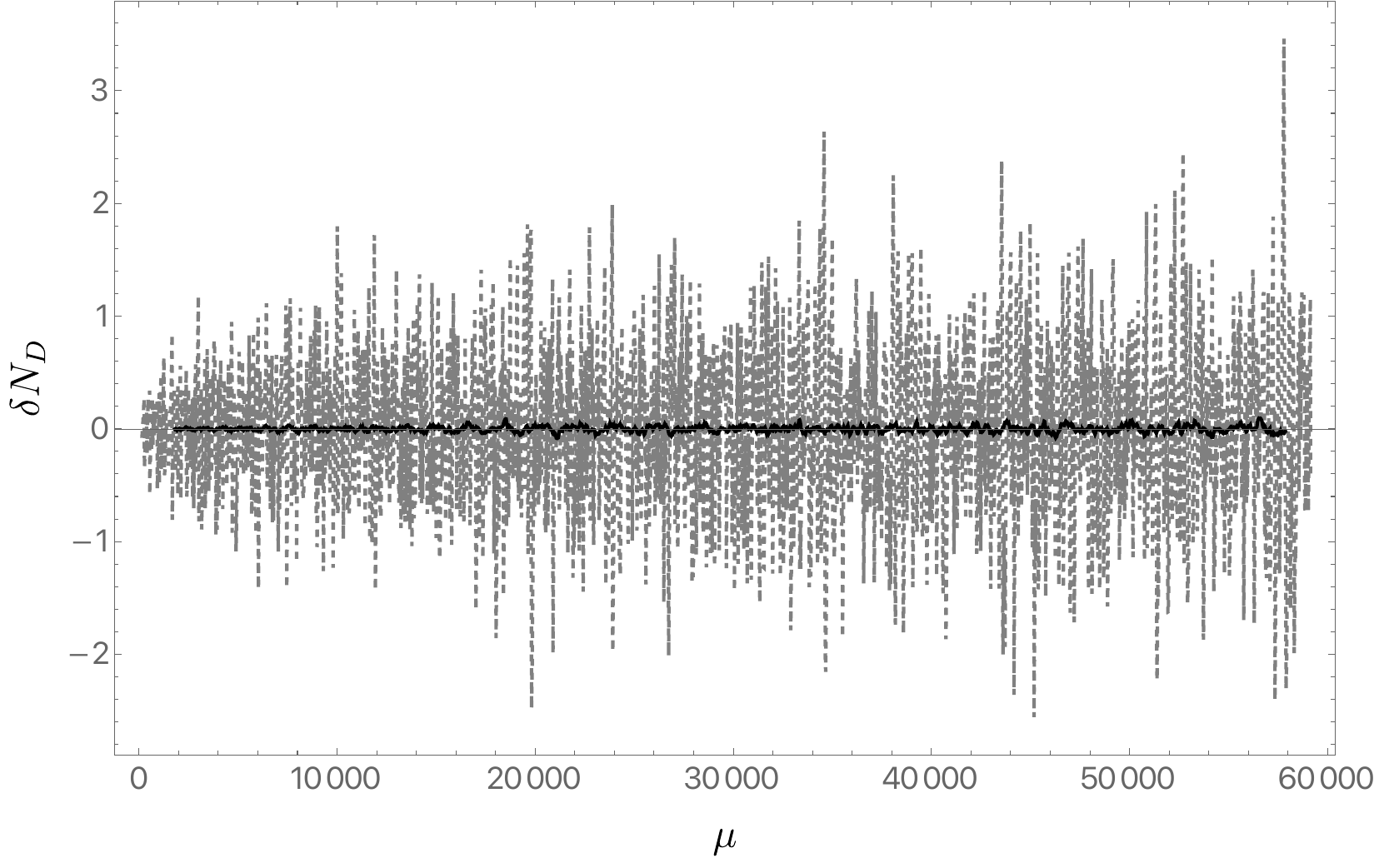}
    \caption{$\delta N_D(\mu)$ as a function of $\mu$. The dashed grey line is the raw numerical data. Solid black line is the moving average of $\delta N_D(\mu)$ using the 100 nearest elements.}
    \label{fig:deltaNo}
\end{figure}

From our data we can compute the number $N_D(\mu)$ of eigenvalues below $\mu$. Let
\begin{equation}\label{eq:dND}
\delta N_D(\mu) =N_D(\mu)-\left[\bar{N}_D(\mu)-\frac{1}{2}\right]\,.
\end{equation}
If our data is accurate, we expect $\delta N_D(\mu)$ to oscillate around $0$ and not have large deviations. In Fig.~\ref{fig:deltaNo} we plot $\delta N_D(\mu)$ as a function of $\mu$ as a dashed grey line.
$\delta N_D(\mu)$ oscillates around zero for the first 2250 eigenvalues, indicating that we have not missed any eigenvalues. As an even stronger test we also plot a moving average $\langle\delta N_D(\mu)\rangle$, using the nearest 100 elements, of $\delta N_D(\mu)$. This is shown as a black solid line. If we have found all eigenvalues, the moving average should be almost zero and indeed it is down relative to $\delta N_D(\mu)$ by a factor of fifty.

\begin{figure}[h]
    \centering
   \includegraphics[width=0.8\textwidth]{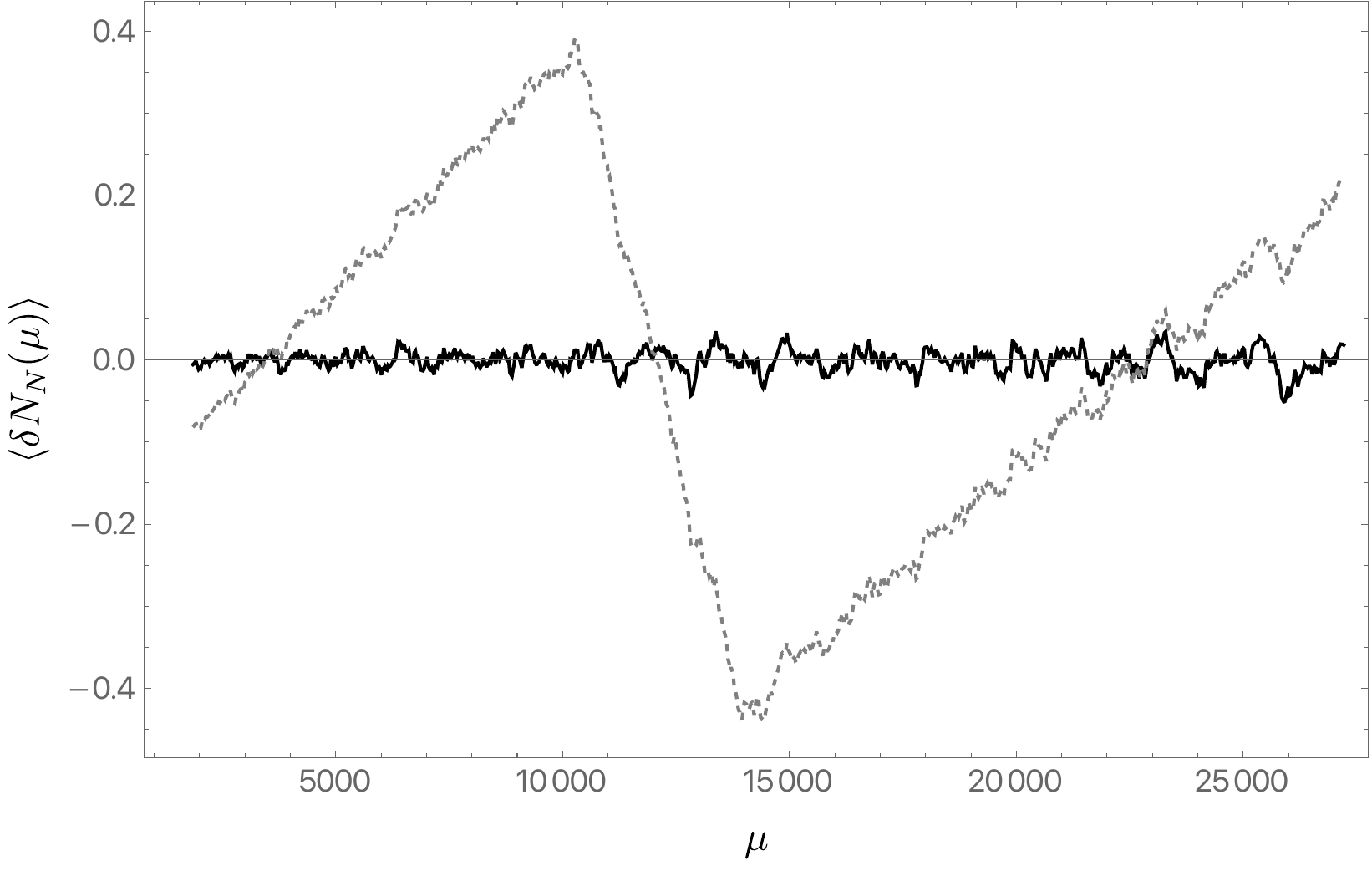}
    \caption{Black curve: $\langle \delta N_N(\mu)\rangle$ as a function of $\mu$. The grey dashed line is obtained from the same procedure, but with the five hundredth eigenvalue removed.}
    \label{fig:deltaNN}
\end{figure}

For Neumann waveforms the equivalent Weyl formula is not known to us. Fortunately, a fitting procedure can be used to test the data \cite{farmer2005maass}. Motivated by the analytic formula (\ref{eq:weylF}), we propose that the Weyl formula for the even case takes the following schematic form
\begin{equation}
\bar{N}_N(\mu)=\frac{1}{4 \pi }\left[\frac{\pi \mu }{6}+A_0 \sqrt{\mu }\,\log \mu+B_0 \sqrt{\mu }+C_0+\mathcal{O}(\mu^{-1/2})\right]\,,
\label{eq:weylE}
\end{equation}
where the subscript ${}_N$ is to `Neumann' and $\{A_0,B_0,C_0\}$ are to be fit. The first term in (\ref{eq:weylE}) is the same as for the Dirichlet sector. A simple $\chi^2$ fit yields
\begin{equation}
A_0\approx 0.0010(3)\,,\quad B_0 \approx -0.7043(6)\,,\quad\text{and}\quad C_0\approx 13.1661(2)\,.
\end{equation}
Using (\ref{eq:weylE}) and our 1200 numerical eigenvalues we can generate a moving average $\langle \delta N_N(\mu)\rangle$, defined analogously to (\ref{eq:dND}), using the 150 nearest neighbors, and plot the result. If \emph{no} eigenvalue has been omitted, $\langle \delta N_N(\mu)\rangle$ should oscillate around zero with a very small amplitude. 
The solid black like in Fig.~\ref{fig:deltaNN} shows that this is indeed the case.
In contrast,
the grey dashed line is obtained from the same procedure
(including a new fit with slightly different parameters) and dataset but with a \emph{single} eigenvalue, the five hundredth, removed. Inspecting the plot, it becomes evident that the data is missing an eigenvalue. Indeed, the curve's shape provides a hint as to the value of the absent eigenvalue. The fact that the solid black line consistently oscillates around zero with a small amplitude is very reassuring.

\section{Properties of the spectra and arithmetic quantum chaos}
\label{sec:properties}

We now turn to the statistical analysis of the spectrum of the cosmological billiard Laplacian. To this end, we compute the level spacing statistics in \S\ref{subsec:Nearest-neighbour level spacing} and the spectral form factor in \S\ref{subsec:Spectral form factor}. Since we are interested in the properties of the Hamiltonian conjugate to the time $\tau$, it is most natural to perform this analysis for the energy levels $\vep_n$ in \eqref{eq: wave eq H2}.
The level spacing statistics associated to the late time chaotic mixmaster dynamics has been considered in the past. These works, which we cite below, have exclusively focused on the Dirichlet sector. We will extend those investigations to the Neumann sector, numerically compute the spectral form factor for both sectors, and explicitly construct the Hecke operators relevant for the interpretation of these results. 
All plots in this section have been generated using the 2250 Dirichlet and 1200 Neumann sector eigenvalues that we have found using the methods described in the previous section. 

\subsection{Density of states}

The Weyl formulae (\ref{eq:weylF}) and (\ref{eq:weylE}) are expressed in terms of $\mu$. The corresponding density of states in terms of $\vep$ is then
\begin{equation}\label{eq:chain}
\bar \rho(\vep)= \left(\frac{{\rm d} \mu}{{\rm d} \ep}\right)\left.\frac{{\rm d}\bar{N}(\mu)}{{\rm d} \mu}\right|_{\mu=\vep^2+\frac{1}{4}}\,.
\end{equation}
In particular, in both cases the leading order behaviour at large energies is
\be\label{eq:density}
\bar \rho(\vep) = \frac{A}{2\pi} \vep +\mathcal{O}(\log \vep) \,,
\ee
where $A=\pi/6$ is the area of the small triangle defined in (\ref{eq:walls small triangle}). We recalled the origin of this leading term in footnote \ref{foot:count} above. Although we shall work at fixed energy throughout, we can make a brief comment about the canonical ensemble. The density of states (\ref{eq:density}) corresponds to the entropy
\be\label{eq:entropy}
S(\vep) = \log \frac{A \vep}{2 \pi} \,.
\ee
In \S\ref{sec:inhomo} below we note that the entropy will be much larger beyond a minisuperspace description.
From (\ref{eq:entropy}) the temperature obeys the equipartition-like relation
\be
T = \frac{{\rm d}\vep}{{\rm d}S} = \vep \,.
\ee

\begin{figure}[h]
    \centering
   \includegraphics[width=\textwidth]{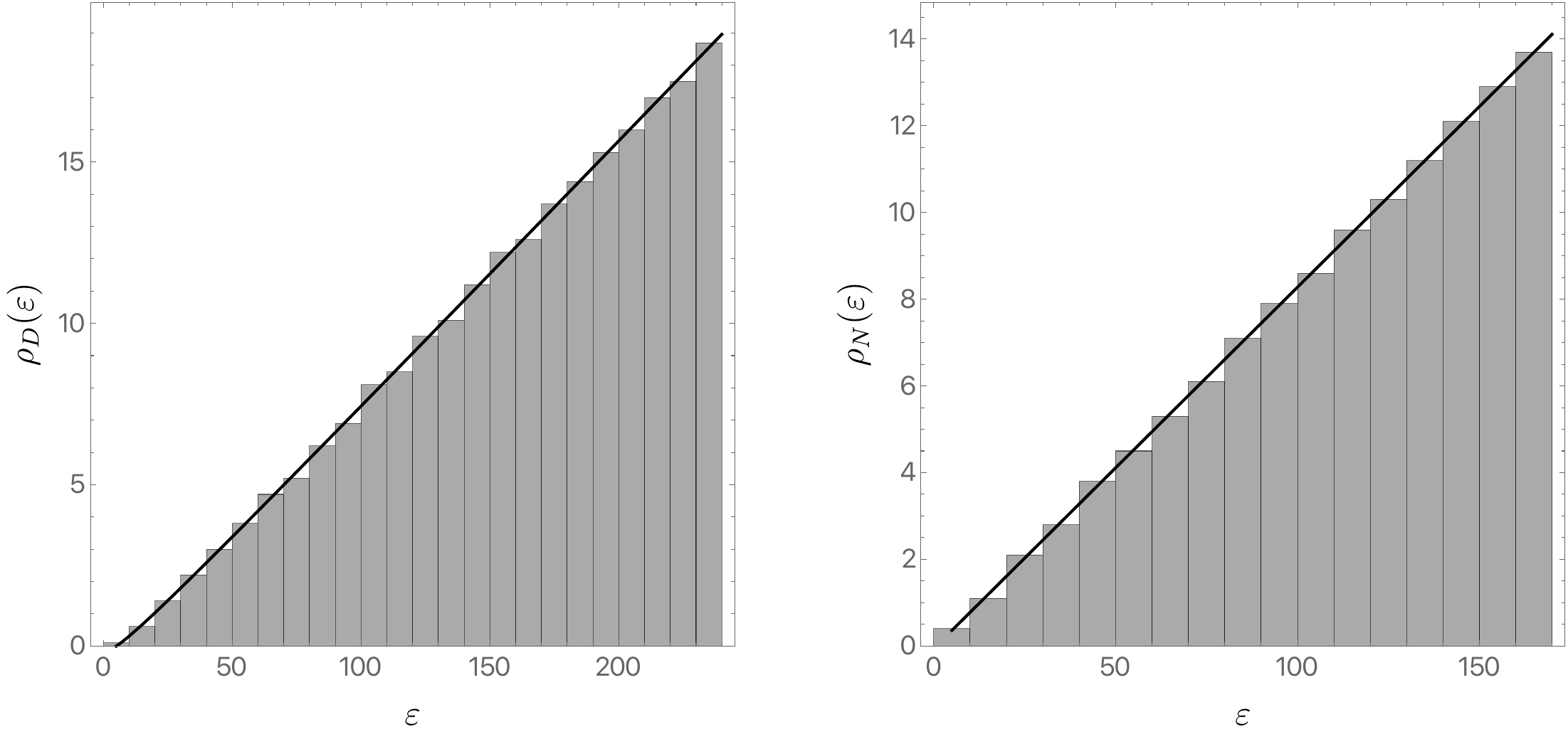}
    \caption{The density of states as a function of the energy $\vep$ for Dirichlet (left) and Neumann (right) waveforms. Solid black lines are the Weyl formulae (\ref{eq:weylF}) and (\ref{eq:weylE}), respectively.}
    \label{fig:energy_densities}
\end{figure}

We can compute the density of states from the numerical data, by constructing an intensity-type histogram, where the counting is divided by the bin width. The height of the histogram should follow the corresponding Weyl formulae for $\bar \rho(\vep)$. In Fig.~\ref{fig:energy_densities}, we plot the histogram of the eigenvalues for the Dirichlet (left panel) and Neumann (right panel) waveforms. The solid black lines are the Weyl formulae (\ref{eq:weylF}) and (\ref{eq:weylE}), respectively.

\subsection{Nearest-neighbour level spacing}
\label{subsec:Nearest-neighbour level spacing}

The spacing between neighbouring energy levels controls very late time dynamics. This dynamics is due to trajectories that remain nearby for a long time, and contains especially universal correlations. To isolate the universal features one must factor out the mean density of states, which is model-specific. This is done by locally rescaling the spectrum in a process called \emph{unfolding} \cite{dehesa1984mathematical,bohigas1991random}, that we now describe. 

The number of levels with energy less than or equal to $\vep$ is given by the staircase function $\mathcal{N}(\vep) \equiv N(\vep^2+1/4)$, which can be decomposed into a smooth and a fluctuating component:
\begin{equation}
\mathcal{N}(\vep)=\bar{\mathcal{N}}(\vep)+\delta \mathcal{N}(\vep)\,.
\end{equation}
Here the smooth component is given by the corresponding Weyl formula. Unfolding means that instead of the differences $\vep_{n+1} - \vep_n$, we compute the distribution of differences
\be\label{eq:sn}
s_n \equiv \bar{\mathcal{N}}(\vep_{n+1}) - \bar{\mathcal{N}}(\vep_n) \,.
\ee
In (\ref{eq:sn}) we see that $s_n$ is the average number of eigenvalues between $\vep_n$ and $\vep_{n+1}$. As $\vep_{n+1}$ is the eigenvalue following $\vep_n$ we have, by construction, that the average spacing $\bar s = 1$. In particular,
for closely spaced $\vep_n$ it follows that $s_n = \rho(\vep_n) (\vep_{n+1} - \vep_{n})$, and hence we can see explicitly that $s_n$ describes the energy differences locally rescaled so that the average $\bar s = 1$.

\begin{figure}[h]
    \centering
   \includegraphics[width=\textwidth]{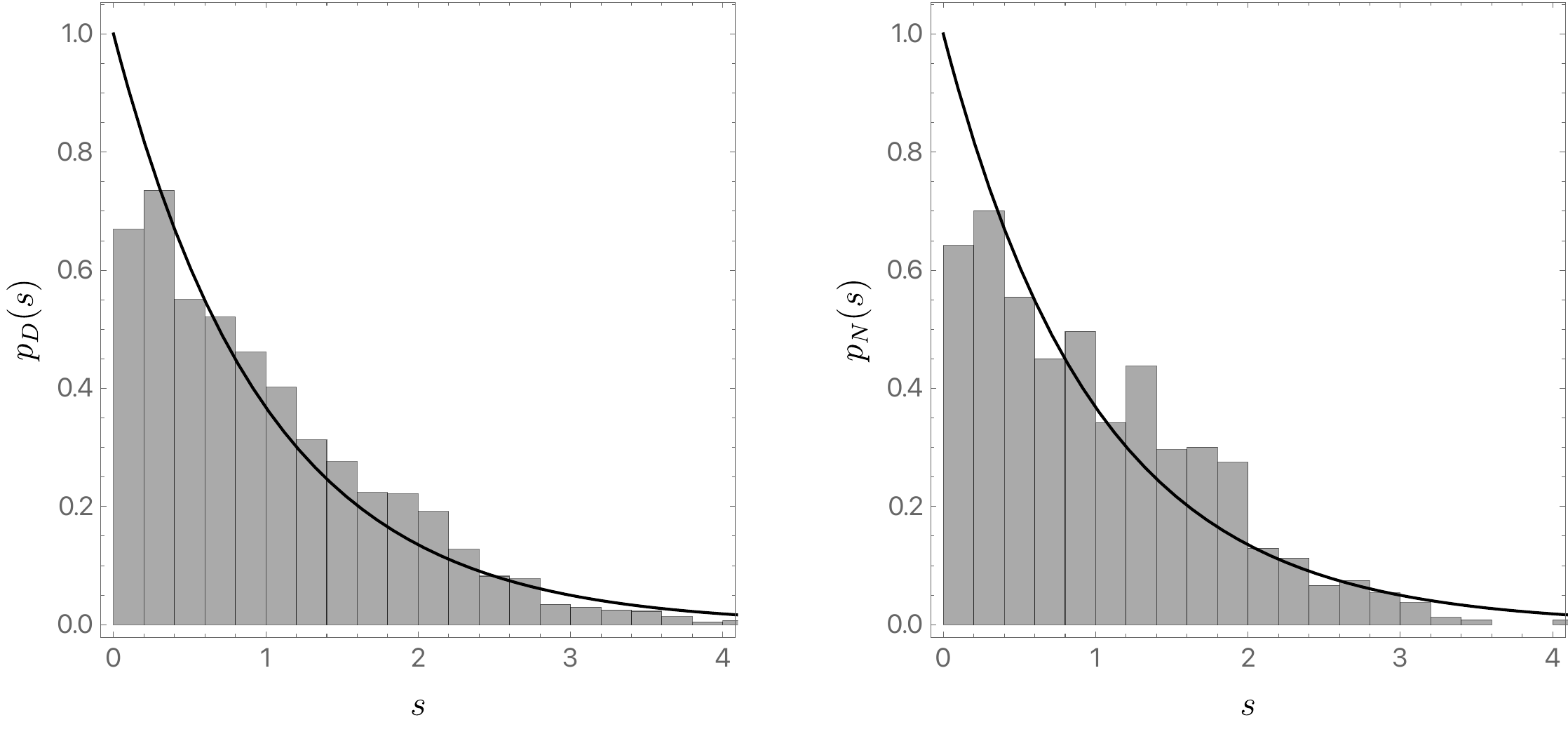}
    \caption{The nearest-neighbor spacing distribution $p(s)$ computed for the Dirichlet (left) and Neumann (right) waveforms. The solid black line is the Poisson distribution $p(s)=e^{-s}$.}
    \label{fig:poisson}
\end{figure}

The nearest-neighbor spacing distribution $p(s)\,\mathrm{d}s$ is defined to be the probability of finding a given $s_n$ in the interval $[s, s+\mathrm{d}s]$. We can extract these probabilities by constructing a PDF-type histogram from our unfolded numerical data.\footnote{An alternative method to unfold the data that does not rely on the Weyl formula is to obtain the density of states from the data itself \cite{Evnin:2018jbh}. Thus
\begin{equation*}
s_n = \frac{x_n}{{\rm mean}\,{x_n}} \quad \text{where} \quad
x_n\equiv \frac{\vep_{n+1}-\vep_{n}}{\vep_{n+\lfloor\sqrt{\Lambda}\rfloor}-\vep_{n-\lfloor\sqrt{\Lambda}\rfloor}}\,,\nonumber
\end{equation*}
where the mean is computed from the $x_n$ with $n$ in the range $n=\lfloor\sqrt{\Lambda}\rfloor+1,\ldots,\Lambda-\lfloor\sqrt{\Lambda}\rfloor$ and
$\Lambda$ is the total sample size. Here $x_n$ is the difference in energies rescaled by the density of nearby states, and the division by mean $x_n$ in the definition of $s_n$ ensures that the average $\bar s = 1$. We have checked that this more agnostic approach yields very similar results to the method described in the main text.} The results are displayed in
Fig.~\ref{fig:poisson},
which shows that the distribution is consistent with a Poisson distribution $p(s)=e^{-s}$. This is 
potentially confusing because the classical system is ergodic; quantum systems exhibiting time-reversal symmetry
and whose classical limit is ergodic typically have level spacing distributions well-described by the Wigner surmise $p(s)=\frac{\pi s}{2} e^{-\pi s^2/4}$ \cite{bohigas1984characterization}. A key feature of that distribution is level repulsion, which is clearly absent in Fig.~\ref{fig:poisson}. A Poisson distribution is instead generally associated to integrable systems \cite{berry1977level}. However, a Poisson level spacing distribution for the kind of hyperbolic billiards we are considering has been long understood to be a consequence of the arithmetic properties of the hyperbolic domain \cite{PhysRevLett.69.2188,PhysRevLett.69.1477,bolte1993some}. At the heart of this {\it arithmetic} quantum chaos is the existence of Hecke operators that commute with the Hamiltonian. We will discuss the Hecke operators shortly, but first consider a further probe of the energy spectrum.

\subsection{Spectral form factor}
\label{subsec:Spectral form factor}

The spectral form factor $K(\tau)$ encodes spectral correlations beyond nearest neighbour energy levels.
For a set of eigenvalues $\{\vep_k\}$, which must be in a given fixed $S_3$ symmetry sector, truncated at some upper value $k=\Lambda$,
\begin{equation}\label{eq:SFF}
K(\tau) \equiv \frac{1}{\Lambda}\left|\sum_{k=1}^{\Lambda} e^{-i\,\vep_k\,\tau}\right|^2=\frac{1}{\Lambda}\sum_{k,l=1}^{\Lambda}e^{-i\,(\vep_k-\vep_l)\,\tau}\,.
\end{equation}
The spectral form factor encompasses a disconnected and a connected contribution.
The disconnected part is determined by the density of states and hence does not contain new information. We now explain how to subtract the disconnected part using the Weyl formula for the density of states.

Define the (truncated) Fourier transform of the Weyl density
\begin{equation}
\tilde{\rho}(\tau) \equiv \int_{\vep_{1}}^{\vep_{\Lambda}}\bar{\rho}(\vep)\,e^{-i \vep \tau}\,\mathrm{d}\vep\,,
\end{equation}
where $\vep_{1}$ and $\vep_{\Lambda}$ are, respectively, the minimum and maximum values of $\vep$ in $\{\vep_{k}\}$. The chain rule (\ref{eq:chain}), along with the asymptotic formulae (\ref{eq:weylF}) and (\ref{eq:weylE}), gives $\bar{\rho}(\vep)$ up to order $\mathcal{O}(\log \vep/\vep^2)$ in each symmetry sector. For instance, in the Dirichlet sector
\begin{equation}
\bar{\rho}_{D}(\vep)\approx\frac{\vep}{12}-\frac{1}{2\pi}\log \vep-\frac{3}{4\pi}\log 2+\mathcal{O}(\log \vep/\vep^2)\,.
\end{equation}
The connected component is now defined as
\begin{equation}
\delta K(\tau) \equiv \frac{1}{\Lambda}\left|\sum_{k=1}^{\Lambda} e^{-i\,\vep_k\,\tau}-\tilde{\rho}(\tau)\right|^2\,.
\end{equation}

\begin{figure}[h]
    \centering
   \includegraphics[width=\textwidth]{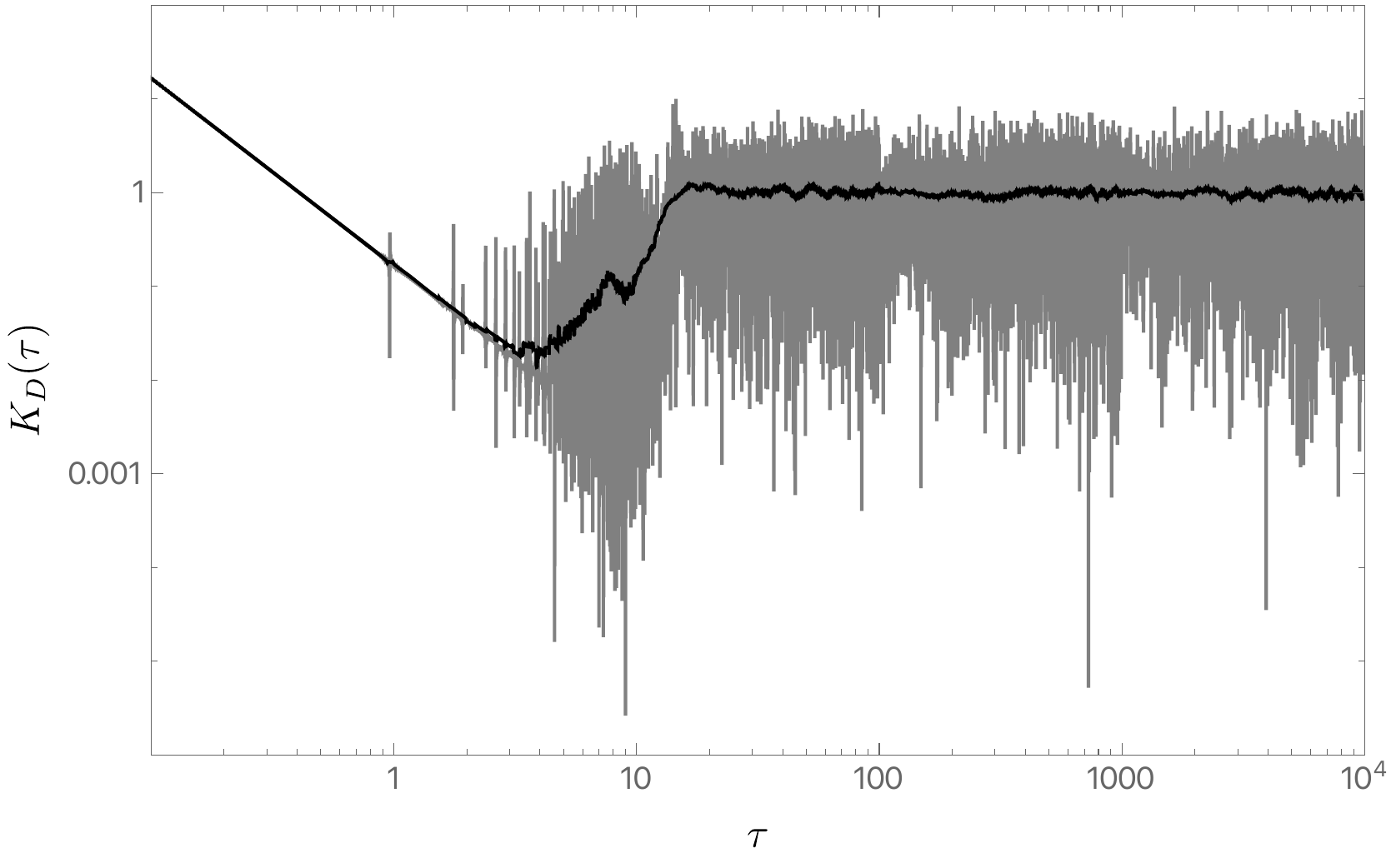}
    \caption{Log-log plot of the spectral form factor $K_D(\tau)$ as a function of $\tau$ for the Dirichlet sector. The grey points show the raw data on our sampling grid, and the black solid line marks the moving average for $\tau\geq0.8$ using the nearest 200 neighbouring elements.}
    \label{fig:spectralD}
\end{figure}

Fig.~\ref{fig:spectralD} shows a log-log plot of the spectral form factor $K_D(\tau)$ for the Dirichlet sector. The grey points are obtained from the raw data, with the spectral form factor sampled from a non-uniform grid in $\tau$, with four thousand points \emph{per decade}. The sampling starts at
at $\tau_{\min} = 10^{-4}$ and extends to $\tau_{\max} = 10^{4}$. The minimum and maximum values of nearest neighbors on the $\tau$ grid are then consistent, respectively, with the largest energy $\vep_{\Lambda}$ and the minimum energy difference between consecutive eigenvalues $\{\vep_{k}\}$ we used in the calculation. The solid black line shows, for $\tau\geq0.8$, a moving average of the same sampled data using the nearest 200 neighbouring elements. 

The small $\tau$ behaviour of the spectral form factor is entirely determined by the Weyl formula for the density of states $\bar{\rho}(\vep)$. That is to say, it is controlled by the disconnected part of the spectral form factor. In particular, since $\bar{\rho}(\vep)$ grows linearly in $\vep$ at large $\vep$, the leading small $\tau$ behavior of the spectral form factor is
\begin{align}
\Lambda\,K(\tau) & \approx|\tilde{\rho}(\tau)|^2  \label{eq:discon} \\
& = \frac{1-\cos (\Delta \vep\,\tau)}{72 \tau^4}+\frac{\vep_{\Lambda}^2+\vep_{1}^2}{144\tau^2}-\frac{\vep_{\Lambda}\vep_{1}}{72\tau^2}\cos (\Delta \vep\tau)-\frac{\Delta \vep\,\sin(\Delta \vep\,\tau)}{72 \tau^3}+\mathcal{O}(1)\,,\nonumber
\end{align}
where $\Delta \vep \equiv \vep_{\Lambda}-\vep_{1}$. The error term is estimated based on the subleading $\log \vep$ behaviour of the Weyl formula for the density of states \eqref{eq:density}. The inverse polynomial behavior in $\tau$ at small times in (\ref{eq:discon}), as well as the trigonometric modulation due to the finite range of energies, can be verified to be in excellent agreement with Fig.~\ref{fig:spectralD}, and also with the Neumann sector result Fig.~\ref{fig:spectralN} below, where the early time oscillations are more visible.

The Neumann spectral form factor, shown in Fig.~\ref{fig:spectralN}, is similar. The statistics is poorer as we do not accurately know as many eigenvalues. The sampling grid has been adjusted so that it is still consistent with $\vep_{\Lambda}$ and the minimum difference between consecutive eigenvalues $\{\vep_{k}\}$ in the Neumann sector. Due to the poorer statistics, we also have to adjust the number of nearest neighbouring elements used in the moving average. We now take the nearest 100 neighbouring elements, and again we only average for $\tau\geq0.8$.
\begin{figure}[h]
    \centering
   \includegraphics[width=\textwidth]{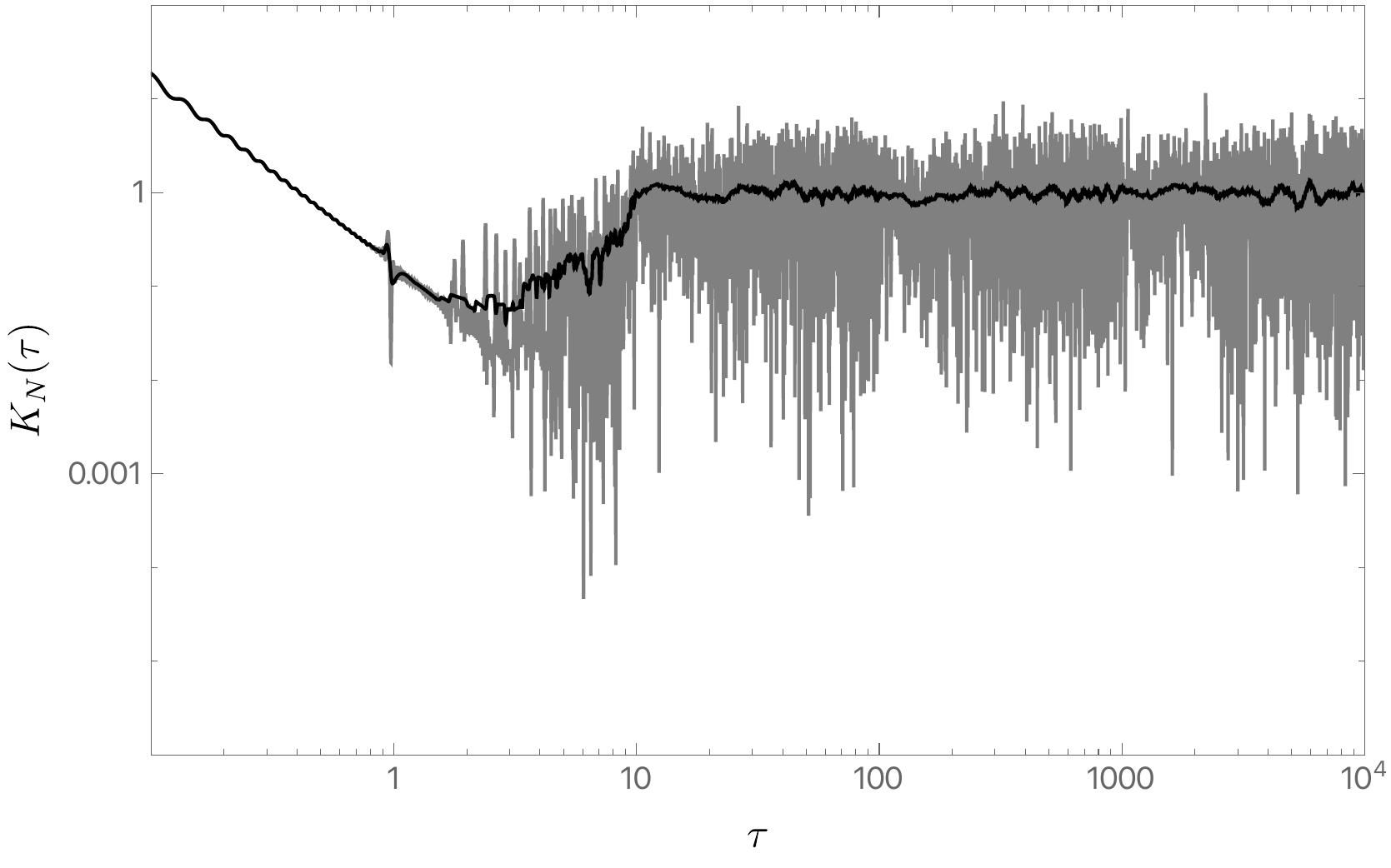}
    \caption{Log-log plot of the spectral form factor $K_N(\tau)$ as a function of $\tau$ for the Neumann sector. The grey points show the raw data on our sampling grid, and the black solid line marks the moving average using the nearest 100 neighbouring elements.}
    \label{fig:spectralN}
\end{figure}

The large $\tau$, saturated behavior of both Figs.~\ref{fig:spectralD} and \ref{fig:spectralN} is called the `plateau'. The saturation at $K(\tau)=1$ follows directly from the discreteness of the spectrum and the lack of degeneracies, so that only terms in (\ref{eq:SFF}) with $\vep_k = \vep_l$ contribute at late times.

\begin{figure}[h]
    \centering
   \includegraphics[width=\textwidth]{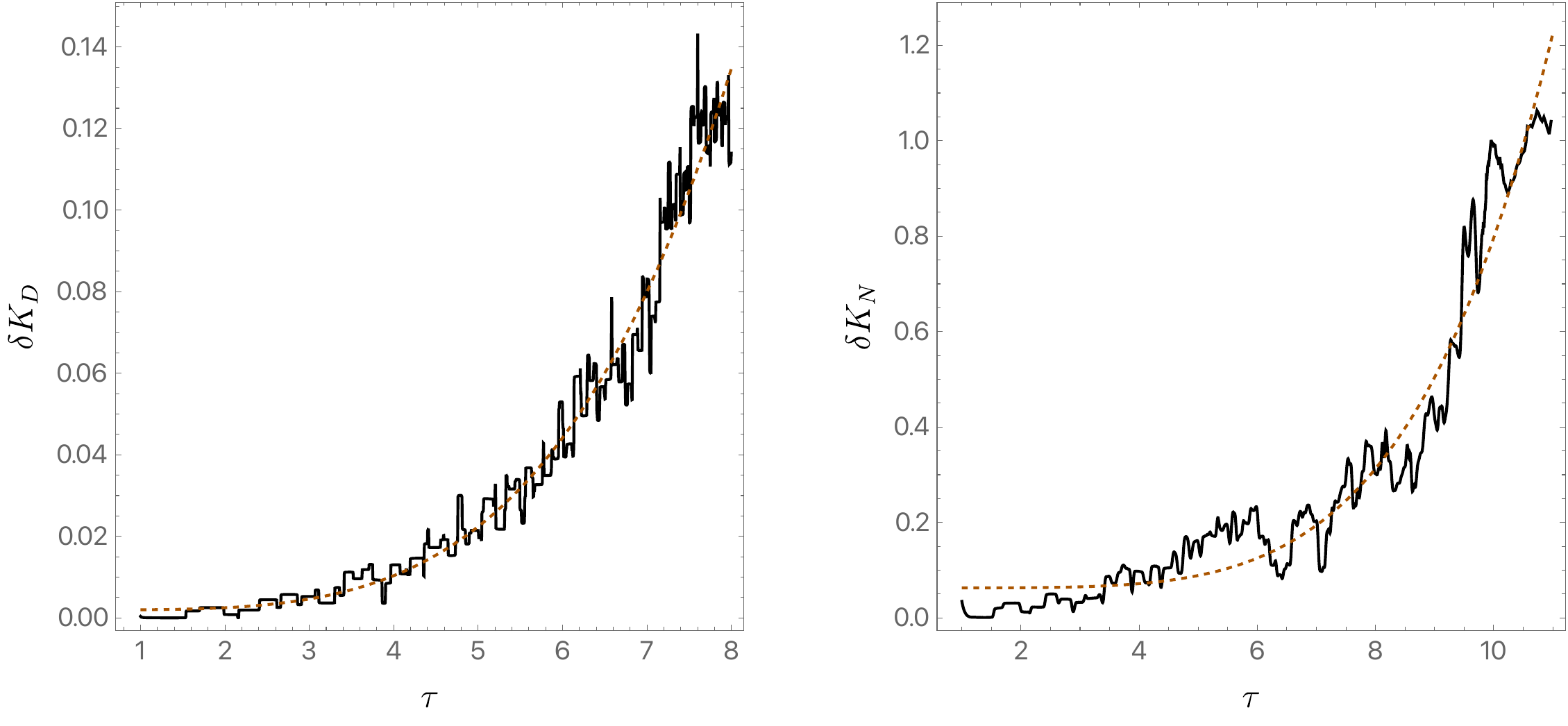}
    \caption{Connected spectral form factor $\delta K$ as a function of $\tau$ for the Dirichlet (left) and Neumann (right) sectors. The moving average is obtained as in Figs. \ref{fig:spectralD}  and \ref{fig:spectralN}, respectively. The dashed brown line shows the best fit to the exponential form (\ref{eq:fitS}).}
    \label{fig:deltaK}
\end{figure}

The region in between the early time decay and the late time plateau is called the `ramp' \cite{Cotler:2016fpe}. To examine the ramp more closely, we show the corresponding connected spectral form factor $\delta K(\tau)$ for each sector in Fig.~\ref{fig:deltaK}. These (linear axes) plots clearly show that the ramp does not display the linear growth characteristic of generic random matrix ensembles \cite{doi:10.1098/rspa.1985.0078}.
As with the Poisson level spacing distribution seen in the previous section, an anomalous ramp in the spectral form factor is known to be a characteristic of systems exhibiting arithmetic chaos  \cite{PhysRevLett.69.1477,bolte1993some,Raurich_1994,Aurich:1994eq}.
The behavior is again ultimately due to the presence of Hecke operators that express additional arithmetic symmetries of the system. We discuss Hecke operators in the following \S\ref{sec:heck}. The result of \cite{bolte1993some,Raurich_1994,Aurich:1994eq}, re-expressed in terms of our $\tau$ variable, is that there is an exponential ramp
\begin{equation}\label{eq:predict}
\delta K(\tau)\propto {\textstyle \frac{1}{\bar \rho(\vep_\Lambda)}} e^{\frac{1}{2} \tau}\,.
\end{equation}
This behaviour is predicted for $\tau$ after some initial time $\Delta \tau_e$ (roughly, the period of the shortest classical periodic orbit \cite{PhysRevLett.69.1477}), but smaller than the saturation time at which $K$ and $\delta K$ reach unity.

Motivated by (\ref{eq:predict}) we have performed a 
$\chi^2$-fit of the data to
\begin{equation}
f(\tau)=a_0+b_0 e^{c_0 \tau}\,.
\label{eq:fitS}
\end{equation}
For the Dirichlet sector, the fit is to data in the range $\tau\in(1,8)$, while for the Neumann sector the range is taken to be $\tau\in(1,11)$. The shorter range for the Dirichlet case is to avoid the wiggle visible in Fig.~\ref{fig:spectralD}. As far as we can tell, that wiggle is a feature of the full spectrum and not an artifact of the truncation. Indeed, it is only the early-time ramp that is expected to follow a simple behavior. The transition to the plateau is a non-universal regime and can exhibit features. On the other hand, the lower end of the fitting interval was estimated using the same approach as in section IIIB of \cite{Aurich:1994eq} and was found to be roughly around $\tau\sim1$ for the number of eigenvalues collected in both sectors. The fits yield $c_0\approx 0.55$ for the Dirichlet case and $c_0\approx 0.5$ for the Neumann case, in good agreement with the expectation \eqref{eq:predict}. These best fits are represented as the dashed brown curves in Fig.~\ref{fig:deltaK}.
The growth of $\delta K(\tau)$ indeed appears to be \emph{consistent} with an exponential trend. However, the available dynamical range is not parametrically large, and therefore the exponents quoted above do exhibit a certain sensitivity to the range of $\tau$ used in the fit and to the averaging procedure. As a test, we also attempted a fit to a power law $a_0 + b_0 \tau^{c_0}$ in the same range of $\tau$. The adjusted $\bar{R}^2$ was found to be somewhat better in both sectors for the exponential fits.

The crossovers between the three different regimes of the spectral form factor in Figs.~\ref{fig:spectralD} and \ref{fig:spectralN} can be determined from the early time behavior (\ref{eq:discon}), the ramp (\ref{eq:predict}) and the plateau $K(\tau) = 1$. The prefactor of the exponential ramp in
(\ref{eq:predict}) is $1/\bar \rho(\vep_\Lambda) \propto 1/\vep_\Lambda \propto 1/\sqrt{\Lambda}$, as the number of eigenvalues $\Lambda$ is obtained by integrating the density of states. We have verified that the prefactor of the exponential in our fit (\ref{eq:fitS}) is consistent with a $1/\sqrt{\Lambda}$ scaling. The crossover between disconnected behaviour and the exponential ramp therefore occurs at
\be
\tau_1 \sim \frac{1}{\Lambda^{1/8}} \,,
\ee
while the crossover from the exponential ramp and the plateau occurs at
\be
\tau_2 \sim \log \Lambda \,.
\ee
It is immediately noticed that the time window over which the ramp is observed is rather narrow, even for moderately large values of $\Lambda$.

We should emphasise that $\Lambda$ is a `UV' cutoff on superspace, not on space itself. That is to say, $\Lambda$ is related to potential finiteness of the quantum gravity Hilbert space. This is logically distinct from the Planck scale. 
We may recall from \S\ref{sec:erad} and \S\ref{sec:revisit} that $\tau$ is the double logarithm of the spatial volume. This means that the spatial volume can reach the Planck scale at fairly small values of $\tau$. If $\Lambda$ is large, therefore, higher derivative and stringy corrections to the gravitational dynamics become important before the finiteness of the Hilbert space becomes relevant.

An exponential ramp in the spectral form factor has also been found in integrable cousins of the SYK model \cite{Liao:2020lac, Winer:2020mdc}. This is quite consistent with the exponential ramp in our model being due to presence of many conserved quantities. It is interesting to note that the exponential ramp of the spectral form factors computed in \cite{Liao:2020lac, Winer:2020mdc} also exhibit features similar to the wiggle in Fig.~\ref{fig:spectralD}. In \cite{Winer:2020mdc}, the exponential ramp and wiggles thereon were argued to originate from a time-dependent symmetry breaking pattern that leads to an exponentially growing number of fluctuations around the saddle point as time evolves.

\subsection{Hecke operators}
\label{sec:heck}

Hecke operators are responsible for the integrable-like behaviour encountered above. Hecke operators for the modular group $SL(2,\mathbb{Z})$ are reviewed in \cite{Bogomolny:1992cj}. That discussion applies directly to our Dirichlet waveforms. To consider all of our symmetry sectors, however, we must describe the Hecke operators for the modular subgroup $\Gamma(2)$. Recall that this is the relevant arithmetic group for the large triangle with walls \eqref{eq: walls UHP}. Hecke operators for the principal congruence subgroups $\Gamma(N)$ are described in \cite{Harvey:2018rdc}. In this section we give the explicit action of the $\Gamma(2)$ Hecke operators on functions on the upper half plane. We show that they respect the boundary conditions \eqref{eq: walls UHP} and commute with the hyperbolic Laplacian.

There is an infinite set of $\Gamma(2)$ Hecke operators $\mathcal{T}_p$. It is sufficient to consider the case where $p$ is prime. The action of the $\mathcal{T}_p$ operators on a function on $H^2$ is
\begin{equation}
    \mathcal{T}_p \psi(z) \equiv \frac{1}{\sqrt{p}} \psi'(z) \equiv \frac{1}{\sqrt{p}} \sum_{\beta \in \Delta_{p}} \psi(\beta z)= \frac{1}{\sqrt{p}} \left[ \psi(pz) + \sum_{b=0}^{p-1}\psi\left(\frac{z+2b}{p}\right) \right] \,,
    \label{eq:heckeOperators}
\end{equation}
with
\begin{equation}
    \Delta_p = \Bigg\{
    \begin{pmatrix}
    p & 0 \\
    0 & 1
    \end{pmatrix}, 
    \begin{pmatrix}
    1 & 2 b \\
    0 & p
    \end{pmatrix}
    \vert  \, 0 \leq b \leq p-1 \Bigg\}\,.
    \label{eq:repsGamma2}
\end{equation}
The matrices in \eqref{eq:repsGamma2} do not belong to $\Gamma(2)$, and hence it is not immediate that the action \eqref{eq:heckeOperators} maps the space of $\Gamma(2)$ automorphic forms to itself. We will now recall how this works.

Consider the set of matrices
\begin{equation}\label{eq: Mp gamma(2)}
    M_p \equiv \left\{ \begin{pmatrix}
        a & b\\ c &d
    \end{pmatrix}  \mid (a,d  = 1 \land  b,c  = 0) \mod 2, \,  ad-bc=p\right\} \,.
\end{equation}
These are very similar to the elements of $\Gamma(2)$, see \eqref{eq: gamma(2)}. The difference is that the determinant is constrained to be $p$ instead of $1$. Such matrices can act on $\Gamma(2)$ by conjugation, preserving the unit determinant in $\Gamma(2)$. 
In Appendix \ref{sec:Hecke operators}, we show that any element $m_p \in M_p$ can uniquely be decomposed as
\be\label{eq:malpha}
m_p = \gamma \alpha_p \,,
\ee
with $\gamma \in \Gamma(2)$ and $\alpha_p \in \Delta_p$. 
The results in Appendix \ref{sec:Hecke operators}, in particular (\ref{eq:malpha}), lead to an equivalence relation on the elements of $M_p$: $m_p \sim m_p'$ if $m_p =\gamma m_p'$ for some $\gamma \in \Gamma(2)$. The corresponding equivalence classes $\{\left[\alpha_p\right] | \alpha_p \in \Delta_p\}$ partition $M_p$. The sum over elements of $\Delta_p$ in the Hecke operators \eqref{eq:heckeOperators} can also, therefore, be thought of as a redundant sum over elements of $M_p$.

From (\ref{eq:malpha}) we may show that the Hecke operators \eqref{eq:heckeOperators} preserve automorphicity. If $\psi$ is an automorphic waveform of $\Gamma(2)$ then
\begin{equation}
\psi'(\gamma z) = \psi'(z), \qquad \forall \gamma \in \Gamma(2) \,.
\end{equation}
This is proved as follows. From the definition (\ref{eq:heckeOperators}), $\psi'(\gamma z) = \sum_{\beta \in \Delta_p} \psi(\beta \gamma z)$. It is easily checked that $\beta \gamma \in M_p$ and hence, from (\ref{eq:malpha}), $\beta \gamma = \gamma' \beta'$. Here both $\gamma'$ and $\beta'$ depend on $\beta$. Thus $\psi'(\gamma z) = \sum_{\beta \in \Delta_p} \psi(\gamma' \beta' z) =  \sum_{\beta \in \Delta_p} \psi(\beta' z) = \sum_{\beta' \in \Delta_p} \psi(\beta' z) = \psi'(z)$, as required. The penultimate step uses the fact that $\beta$ and $\beta'$ are in a one-to-one relation, which follows from the uniqueness of the decomposition (\ref{eq:malpha}).
The Hecke operators \eqref{eq:heckeOperators} therefore map the set of automorphic waveforms on $\Gamma(2)$ to itself.

We must also check that the Hecke operators commute with the reflection $z \rightarrow -z^*$. This is necessary in order to restrict to odd Maa{\ss} waveforms that, we recall from \S\ref{sec:cong}, obey the vanishing boundary conditions on all three walls of the cosmological triangle. Given $\psi$ odd, we see that $\psi'$ is odd as follows
\begin{align}
\psi'(z)&= \psi(pz) + \sum_{b=0}^{p-1}\psi\left(\frac{z+2b}{p}\right) \nonumber \\
& = - \psi(-pz^*) - \psi\left(\frac{-z^*}{p}\right) - \sum_{b=1}^{p-1}\psi\left(\frac{-z^*-2b}{p}\right) \label{eq:refsym} \\
& =  - \psi(p(-z^*)) - \psi\left(\frac{-z^*}{p}\right) - \sum_{b=1}^{p-1}\psi\left(\frac{(-z^*)+2 (p-b)}{p}\right) = - \psi'(-z^*) \,. \nonumber
\end{align}
The first equality is the definition \eqref{eq:heckeOperators}, the second uses oddness $\psi(z) = - \psi(-z^*)$, the third uses the $\Gamma(2)$ invariance $\psi(z+2) = \psi(z)$.

Finally, we must check that the Hecke operators commute with the Laplacian. This step is straightforward since
the arguments of $\psi$ in \eqref{eq:heckeOperators} can be expressed as
$SL(2,\mathbb{R})$ transformations of $z$, which are isometries of the hyperbolic plane.
Furthermore, an argument similar to \eqref{eq:refsym} shows that the Hecke operators also commute with the symmetry group $S_3$. Putting all of the above together: the Hecke operators map odd $\Gamma(2)$ waveforms to themselves, in each $S_3$ symmetry sector. There must, therefore, exist a basis of eigenforms for our problem which simultaneously diagonalises the Hecke operators. In the remainder of this section we obtain an important consequence of this fact, called the Hecke relations.

Odd Maa{\ss} waveforms on $\Gamma(2)\backslash H^2$ can, in full generality, be expanded in terms of coefficients $d^n_m$ as
\be
\psi_n(x,y) = \sum_{m=1}^\infty d^n_m \sqrt{y} K_{i \vep_n}(m \pi y) \sin(m \pi x) \,,
\label{eq: maass waveform}
\ee
which satisfies Dirichlet boundary conditions at the two vertical walls $x=0$ and $x=1$. As described above, for the eigenvalues $\vep_n$ there exist coefficients $d_m^n$ so that the third Dirichlet boundary condition in \eqref{eq: walls UHP} is also satisfied. The earlier expressions \eqref{eq:oddcase} and \eqref{eq:evencase}, for the sectors obeying Dirichlet or Neumann boundary conditions at $x = \frac{1}{2}$, are recovered from \eqref{eq: maass waveform} by setting to zero the 
coefficients $d^n_m$ with odd or even $m$, respectively.
Recall from \S\ref{sec:cong} that individual states in the standard representation may also be taken to be either even or odd under $x \to 1 - x$. For the purposes of the discussion below we subsume these cases under our consideration of Dirichlet and Neumann boundary conditions, as all that is being used is parity under $x \to 1 - x$.

Let us choose the basis of Maa{\ss} waveforms to be simultaneously eigenfunctions of the Hecke operators. It turns out, then, that their Fourier coefficients are precisely the eigenvalues of the Hecke operators. Considering the Dirichlet and Neumann cases separately:
\begin{equation}
    \begin{cases}
 \mathcal{T}_p \psi_n(z) = d_{2p}^n \psi_n(z), & \qquad \qquad \text{(Dirichlet)} \\
 \mathcal{T}_p \psi_n(z) = d_p^n \psi_n(z).  & \qquad \qquad \text{(Neumann)}
\end{cases}
    \label{eq: Hecke operator eigenvalue}
\end{equation}
This can be seen from the action (\ref{eq:heckeOperators}) of the Hecke operator on the expansion \eqref{eq: maass waveform}:
\begin{align}
    \mathcal{T}_p \psi_n(z) = \sum_{m=1} d^n_m \sqrt{y} &K_{i \vep_n}(m \pi py)\sin(m \pi p x) + \nonumber \\
   & \sum_{m=1} \sum_{b=0}^{p-1} d^n_m \frac{\sqrt{y}}{p} K_{i \vep_n}\left(\frac{m \pi y}{p}\right)  \sin \left( \frac{m \pi(x+2b)}{p}\right) \,.
\end{align}
In the second term, one can expand the sine using trigonometric summation rules and note that the sum over $b$ vanishes except when $m$ is a multiple of $p$. Introducing $m = vp$:
\begin{equation}
    \mathcal{T}_p \psi_n(z) = \sum_{m=1} d^n_m \sqrt{y} K_{i \vep_n}(m \pi py)\sin(m \pi p x) + 
    \sum_{v=1}  d^n_{vp} \sqrt{y} K_{i \vep_n}\left(\pi v y \right)  \sin \left(v \pi x\right) .
\end{equation}
By assumption $\psi_n$ is an eigenfunction of the Hecke operator and hence
$\mathcal{T}_p \psi_n(z) = \lambda_p^n \psi_n(z)$. Equating the coefficient of the first Fourier mode on either side of this equation gives
\be
\lambda_p^n d_1^n =   d_p^n \quad (\text{Neumann}) \quad \text{or} \quad
\lambda_p^n d_2^n =   d_{2p}^n \quad (\text{Dirichlet}) \,.
\ee
It is customary to choose the normalisation so that the first nonzero coefficient is set to one. Thus $d_1^n = 1$ in the Neumann case and $d_2^n = 1$ in the Dirichlet case, and then \eqref{eq: Hecke operator eigenvalue} follows.

Identifying the remaining Fourier coefficients on either side of the eigenvalue equation yields the {\it Hecke relations} for $p$ prime and $v \in \mathbb{N}$:
\begin{equation}
\label{eq:heckerelations_cases}
\begin{cases}
d_{vp}^n = d_{v}^nd_{2p}^n - d_{v/p}^n \,, & \qquad \qquad \text{(Dirichlet)} \\
d_{vp}^n = d_{v}^nd_{p}^n - d_{v/p}^n \,. & \qquad \qquad \text{(Neumann)}
\end{cases}
\end{equation}
where the last term $d_{v/p}^n$ is only nonzero when $p$ divides $v$. In the Dirichlet sector, the first line in \eqref{eq:heckerelations_cases} is equivalent to a relation for the coefficients $c_m^n$ in \eqref{eq:oddcase}:
\begin{equation}
\label{eq:heckerelations_odd}
c_{vp}^n = c_{v}^n c_{p}^n - c_{v/p}^n.
\end{equation}
This is a special case of the general Hecke relations in which neither number need be prime: $c_s^n c_t^n =\sum_{d|(s,t)} c^n_{st/d^2}$, see e.g.~\cite{terras2013harmonic}. Using these relations, all coefficients in the Fourier expansion of the waveforms can be expressed in terms of the prime Fourier coefficients.\footnote{Expressing the general Fourier coefficient in terms of prime coefficients implies
that the $L$-function built using the $c^n_m$ admits the Euler product formula:
\be
L_n(s) \equiv \sum_{m=1}^\infty \frac{c^n_m}{m^s} = \prod_{p \in \mathbb{P}} \frac{1}{1 - c^n_p p^{-s} + p^{-2s}} \,.
\ee}

The existence of infinitely many conserved Hecke operators means that the integrable-like behaviour of the spectral statistics is not entirely suprising. One way to establish a more precise connection between the Hecke operators and the anomalous spectral statistics is via periodic-orbit theory \cite{Bogomolny:1992cj,bolte1993some,Aurich:1994eq}. In this description, the oscillation of the density of states about the Weyl formula is expressed in terms of a sum over periodic orbits. The key consequence of the arithmetic nature of the hyperbolic billiard domain is that there is an exponential degeneracy of periodic orbits of a given length. This is related to the Hecke operators as follows \cite{Bogomolny:1992cj}. Firstly, one shows that there is a one-to-one correspondence between conjugacy classes of $\Gamma(2)$ and classical periodic orbits. Secondly, the conjugacy classes that are connected to one another via conjugation by a matrix in \eqref{eq: Mp gamma(2)} --- recall that these matrices are not themselves in $\Gamma(2)$ --- correspond to different periodic orbits of the same length. We saw above that the matrices in \eqref{eq: Mp gamma(2)} are closely connected to the definition of the Hecke operators. In particular, the existence of the Hecke symmetry operators is a quantum mechanical manifestation of the large multiplicity of classical periodic orbits of equal length. The exponential degeneracy of periodic orbits of equal length is the cause of the exponential ramp in the spectral form factor \cite{PhysRevLett.69.1477,bolte1993some,Raurich_1994,Aurich:1994eq}.

\subsection{Waveform coefficients}

We have seen, around (\ref{eq:heckerelations_odd}) above, that the Hecke relations imply that only the prime Fourier coefficients are independent. These prime coefficients themselves are subject to various conjectures and theorems \cite{Sarnak1987, hejhal1993fourier, Steil:1994ue}. Perhaps the simplest to state is the Ramanujan-Petersson conjecture that
\begin{equation}
|c^n_p|\leq 2\quad\text{for }\text p \in\mathbb{P}.
\end{equation}
All the data that we report in the paper satisfies this bound.

The Sato-Tate conjecture is a more refined statement about the values attained by the prime Fourier coefficients within a given energy level. This conjecture asserts that, at fixed $n$, the sequence $\{c^n_p\}_{p\in\mathbb{P}}$ is equidistributed\footnote{A sequence $\{y_j\}_{j\in\mathbb{N}}$ is equidistributed with respect to ${\rm d} \mu(x)$ if
\begin{equation}
\lim_{N\to+\infty}\frac{1}{N}\sum_{n=1}^N f(y_n)=\int_{\mathbb{R}}f(x)\,{\rm d}\mu(x)\,,
\end{equation}
for all continuous functions $f$ with compact support on the real line.
} with respect to a Wigner semicircle:
\begin{equation}
\mathrm{d}\mu(x)=\left\{\begin{array}{cc}
\displaystyle \frac{1}{\pi}\sqrt{1-\frac{x^2}{4}}\,\mathrm{d}x&\quad\text{for}\quad|x|<2
\\
0 & \text{otherwise}
\end{array}\right.\,.\label{eq:wigner}
\end{equation}
The distribution ${\rm d} \mu(x)$ is independent of the energy level $n$. It is therefore a very universal property of the late interior dynamics, in a way that is reminiscent of the classical equilibrium distribution that we discussed previously around (\ref{eq:mu}). The Wigner semicircle (\ref{eq:wigner}) naturally suggests that the waveforms should admit a formulation in terms of random matrices.

\begin{figure}[h]
    \centering
   \includegraphics[width=0.48\textwidth]{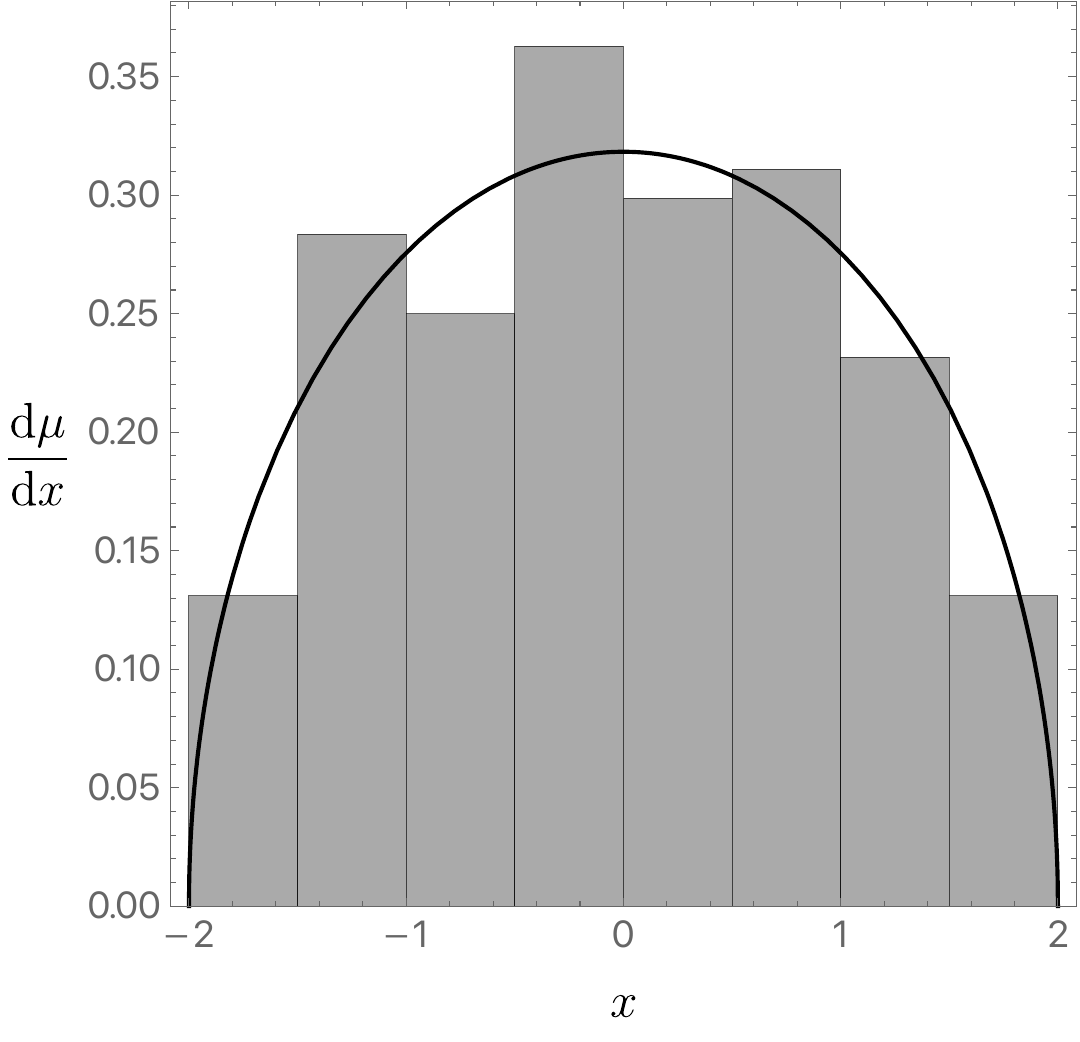}
    \caption{PDF-type histogram of the first 656 prime $c^1_p$. The solid black line is the conjectured Wigner semicircle distribution of the Sato-Tate conjecture.}
    \label{fig:c1p}
\end{figure}

We have attempted to validate (\ref{eq:wigner}) for the lowest,
$n=1$, eigenstate in the Dirichlet sector. Using the method outlined in \S\ref{sec:fewthousandcns} we accurately computed 4914 of the $c^1_m$. These include 656 prime coefficients, which we can use to test the Sato-Tate conjecture. Fig.~\ref{fig:c1p} shows ${\rm d}\mu/{\rm d}x$ obtained from a PDF-type histogram of these prime coefficients. The solid black line is the conjectured Wigner semicircle distribution. The histogram is seen to be consistent with the Wigner semicircle, within the statistical limitations of the data.

\section{Discussion}
\label{sec:inhomo}

\subsubsection*{Inhomogeneities and the smaller triangle}

In order to obtain an explicit realisation of mixmaster chaos in an asymptotically AdS setting we have considered a cohomogeneity-one Ansatz (\emph{i.e.}~the fields only depend on one coordinate) with a diagonal metric and no longitudinal component to the vector fields. This Ansatz consistently leads to solutions to the full equations of motion. However, clearly, it does not give a fully generic approach to the interior singularity. A generic solution will be inhomogeneous and will develop off-diagonal components in the metric.

The original BKL paper \cite{BKL} argued that a remarkable simplification was possible close to the singularity, in which different points of space decouple. Mathematically this is loosely the statement that spatial derivatives in the equations of motion become subleading compared to time derivatives. See e.g.~\cite{Damour:2002et, Garfinkle:2020lhb}. The metric at each point in space then independently undergoes billiard motion, similar to that described in this paper. 

Considering now the metric at a given point in space, the important difference between the billiards we have studied and the generic case is that off-diagonal metric components introduce additional so-called symmetry walls. These new walls dominate over some of the mixmaster walls \cite{Damour:2002et,belinski_henneaux_2017}.
Furthermore, the extra off-diagonal metric components do not lead to additional dimensions of motion in superspace.
The billiard motion is therefore bounded by a smaller triangle. This smaller triangle turns out to be precisely the smaller triangle that we considered in the analysis of the spectral properties of the late interior Hamiltonian, shown in green in figure \ref{fig:domain}. We may recall that this smaller triangle is half of the fundamental domain of the modular group $SL(2,\Z)$. Our discussion of the Dirichlet sector in \S\ref{sec:properties} therefore fully characterises the pointwise arithmetic chaos of the inhomogeneous case too.

We may also explain how the dynamics of Kasner epochs described in \S\ref{sec:bkl1} and \S\ref{sec:bkl2} is adapted to the smaller triangle. The transformation rule for the Kasner exponents due to the symmetry walls (which come together in the right angle of the triangle) is to permute two of the Kasner exponents \cite{belinski_henneaux_2017}. This type of bounce does not modify the Kasner parameter $u$. It follows that all mixing is induced by the bounces against the third wall, which is shared by the mixmaster billiard. These collisions again lead to the transformations (\ref{eq:unmap}) and (\ref{eq:ppp}) and hence to an ergodic dynamical system of Kasner eras governed by the Gauss map.

The dynamics of the large and small triangles have been further studied in \cite{Damour:2010sz, Lecian:2013cxa, Lecian:2013apa}.

\subsubsection*{Entropy and microscopics}

At a semiclassical level, the first consequence of developing pointwise billiard dynamics is that the entropy (\ref{eq:entropy}) will become
\be\label{eq:ent2}
S(\vep) = \frac{V}{(\ell_\text{Pl})^3} \log \frac{\vep}{12} \,.
\ee
Here $V$ is spatial volume of the interior slice, in units of a `Planck scale' cutoff $\ell_\text{pl}$. Even with the entropy (\ref{eq:ent2}) regulated at short distances, it still suffers from a long distance divergence because the volume is infinite. This large number of states is in tension with the finite black hole entropy, and
is a version of the `bags of gold' paradox for eternal AdS black holes \cite{Raju:2020smc}.

Presumably quantum gravitational or stringy effects must drastically reduce the number of independent degrees of freedom. Recent interesting discussions of this reduction include \cite{Chakravarty:2020wdm, Balasubramanian:2022gmo}. One might like, however, to understand the microscopic bulk mechanism at work. The emergent late interior Hamiltonian (indeed, at the classical level there is now a conserved Hamiltonian at each point in space) may make the near-singularity regime well-suited to ground a string or matrix theoretic description of the dynamics. An absolutely basic question is whether the arithmetic properties of the dynamics survive beyond the semiclassical gravity description. If the arithmetic structure does survive in the full theory, it may be possible to observe it from the boundary by looking back in time at a black hole that has fully evaporated. One simple setup in which to explore some of these issues may be three dimensional gravity, in which Maa{\ss} forms already make an appearance \cite{carlip_1998}.

\subsubsection*{Semiclassical physics near the singularity}

In our discussion of the WDW equation in \S\ref{sec:WDW} we focused on the spectrum of the Hamiltonian. It is also interesting to consider the time evolution of a semiclassical quantum state towards the singurality.
States that are strongly supported on classical solutions to the equations of motion are built by superimposing many eigenmodes. Even within a semiclassical regime, quantum mechanical wavepackets can spread and develop increasing variance. For the case of the Schwarzschild singularity, it was found in \cite{Hartnoll:2022snh} that wavepacket spreading does not lead to new effects: the variance of macroscopic metric functions remains small compared to their mean as the singularity is approached. This behaviour occurs because the configuration space is unbounded, which allows the mean to grow faster than the variance. However, in mixmaster cosmology we have seen that the configuration space is fully bounded by the cosmological billiard. Furthermore, the classical chaotic motion suggests that the corresponding quantum wavefunction will spread out over the allowed configuration space. This spreading will cause the variance to become comparable to the mean, leading to an inherently quantum mechanical interior state as the singularity is approached.

The ergodic spreading of the wavefunction over the hyperbolic billiard is an interesting quantum mechanical effect that can be studied without reference to microscopic degrees of freedom. For example, the quantum state should develop spatial isotropy --- this possibility was explored in early numerical work solving the WDW equation \cite{10.1143/PTP.75.59, 10.1143/PTP.76.67, PhysRevD.39.2426}. A possibly natural endpoint may be the supersymmetric solution found in \cite{PhysRevLett.67.1381}. We hope to return to this phenomenon in the future.

\section*{Acknowledgements}

We thank Nathan Benjamin, Frederik Denef, Diego Hofman, Lewis Sword and David Vegh  for helpful discussions and Holger Then for sharing unpublished data. The work of
all three authors is partially supported by STFC consolidated grants ST/T000694/1 and ST/X000664/1. S.A.H.~and M.D.C.~are partially supported by Simons Investigator award $\#$620869. S.A.H.~acknowledges the hospitality of the KITP while part of this work was underway. This research was supported in part by the National Science Foundation under Grants No. NSF PHY-1748958 and PHY-2309135.

\appendix

\section{Equations of motion}
\label{app:eqs}

The equations of motion for the metric and vector field components in (\ref{eq:met}) and (\ref{eq:vec}) are:
\begin{align}
4 z^2 e^{H} \left(\frac{e^{-H}F}{z^3}\right)' + \frac{12}{z^2}  = & \,
e^{2G} \mu_x^2 \phi_x^2 + e^{-2G} \mu_y^2 \phi_y^2 + z^2 e^{2H} (\phi_t')^2 \,, \label{eq:done} \\
\frac{4 F H'}{z} - 4 F (G')^2 = & \,\frac{e^{2H}}{F} \mu_t^2 \phi_t^2 + z^2 e^{2G} F (\phi_x')^2+ z^2 e^{-2G} F (\phi_y')^2  \,, \\
\frac{4F}{z} (z G')' - \frac{12 G'}{z} = & \,
- \frac{\mu_x^2 \phi_x^2}{2} z^3\left(\frac{e^{2G}}{z^2} \right)' + 
\frac{\mu_y^2 \phi_y^2}{2} z^3\left(\frac{e^{-2G}}{z^2} \right)' \nonumber \\
& \,- z^3 G' e^{2H} (\phi_t')^2 + z^2 e^{2G} F (\phi_x')^2 - z^2 e^{-2G} F (\phi_y')^2 \,, \\
z^2 F e^{-H} \left(e^{H} \phi_t' \right)' = & \,\mu_t^2 \phi_t \,, \label{eq:v1} \\
z^2 e^{H-2G}\left(F e^{-H+2G} \phi_x' \right)' = & \, \mu_x^2 \phi_x \,, \label{eq:v2} \\
z^2 e^{H+2G}\left(F e^{-H-2G} \phi_y' \right)' = & \, \mu_y^2 \phi_y \,. \label{eq:v3}
\end{align}
It is well-known that at late interior times the mass terms for the vector fields and the cosmological constant term, which is the second term in (\ref{eq:done}), become negligible. This can be verified numerically and is illustrated in the following Fig.~\ref{fig:test}.
\begin{figure}[h]
    \centering
   \includegraphics[width=0.9\textwidth]{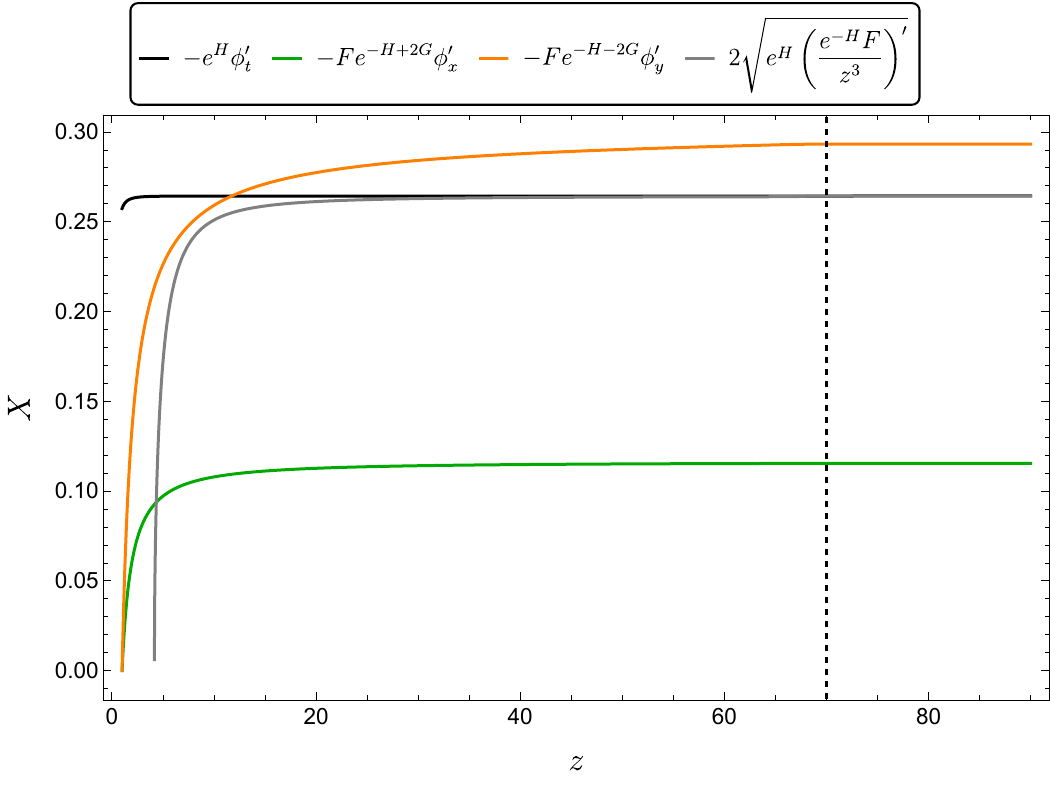}
    \caption{Irrelevance of the mass and cosmological constant terms towards the interior. The first three curves show that fluxes in the vector field equations (\ref{eq:v1}), (\ref{eq:v2}) and (\ref{eq:v3}) become conserved towards the interior. The final curve shows that the first term in (\ref{eq:done}) equals the final term at late interior times (this curve was made by using (\ref{eq:done}) to obtain an expression for $e^H(e^{-H}F/z^3)'$ and plotting that). The collapse of the Einstein-Rosen bridge is shown as a vertical dashed line for reference. Numerical parameters are as in Fig.~\ref{fig:continuity}.}
    \label{fig:test}
\end{figure}

It is instructive to consider the evolution of the spatial volume in the interior
\begin{align}
\frac{8 z^2 e^{H}}{\sqrt{-F}} \left(\frac{\sqrt{-F} e^{-H}}{z^3}\right)' = & - \frac{12}{z^2} - 4 (G')^2 - z^2 e^{-2G} (\phi_y')^2 - z^2 e^{2G} (\phi_x')^2 + \frac{z^2 e^{2 H}}{F} (\phi_t')^2 \nonumber \\
& - \frac{12}{z^2 F} - \mu_t^2 \phi_t^2 \frac{e^{2H}}{F^2}
+ \mu_x^2 \phi_x^2 \frac{e^{2G}}{F} + \mu_y^2 \phi_y^2\frac{e^{-2G}}{F} \,.
\end{align}
The terms on the right hand side in the top line are all manifestly negative in the interior where $F<0$. The terms in the bottom line are either positive or have a sign depending on the sign of the masses squared $\mu_i^2$. We have just argued above that the mass terms and the cosmological constant term (the first term on the bottom line) become negligible at late interior times. Therefore only the terms in the top line survive and the volume decreases monotonically towards the interior singularity.

Although we do not make use of this fact, we note that there is a conserved quantity that follows from the equations of motion above. This is the Noether charge associated with the rescaling symmetry of the action $z \to \lambda z$, $H \to H - 3 \log(\lambda)$, $\phi_t \to \lambda^2 \phi_t$, $\phi_x \to \phi_x/\lambda$, $\phi_y \to \phi_y/\lambda$ and with $F$ and $G$ invariant (cf.~\cite{Gubser:2009cg}). Namely,
\be
Q = 2 e^{H} \phi_t \phi_t' + e^{2G-H} F \phi_x \phi_x' + e^{-2G - H} F \phi_y \phi_y' - \frac{2 e^{H}}{z^2}\left(F e^{-2H} \right)'\,,
\ee
obeys
\be
\frac{{\rm d}Q}{{\rm d}z} = 0 \,.
\ee
We have not found this conserved quantity useful because it explicitly depends on the vector potentials $\phi_i$. In the far interior where the flux is conserved the potentials are given by integrals of the flux.

\section{Probing the interior with geodesics}
\label{sec:geodesic}

Solutions to wave equations in the exterior of a black hole obey analyticity properties that mean that they contain information about the interior \cite{Motl:2003cd, Fidkowski:2003nf, Festuccia:2005pi}.
However, the delicate nature of analytic continuation can make it difficult to access the near-singularity behaviour in this way. In the current setup, and as was the case previously in \cite{Hartnoll:2020rwq, Hartnoll:2020fhc}, the interior maximum in $g_{tt}$ that occurs prior to the chaotic mixmaster regime, see Fig.~{\ref{fig:continuity}}, prevents easy access to the far interior. We briefly review this effect.

If a large mass field is considered as a probe of the exterior, the wave equation reduces to a study of geodesics. The geodesics move in an effective potential given by the metric component $-g_{tt}$, with turning points where the frequency $\omega^2 = - g_{tt} M^2$, with $M$ the mass of the field. If the turning point coincides with a maximum of the potential, call it $-g_{tt}^o$, there is a divergence in the density of modes at the maximum, resulting in a pole in the Green's function for the field. For maxima outside of the horizon, with $g_{tt} < 0$, such a pole describes a long-lived oscillating quasinormal mode at real $\omega$ \cite{QNM}. For a maximum inside the horizon, the pole corresponds to an overdamped quasinormal mode with $\omega$ pure imaginary. For an elegant discussion of this case see \cite{Hartman:2013qma}. Precisely, an interior maximum is expected to lead to a non-oscillatory quasinormal mode that decays at the rate
\be
\Gamma = M \sqrt{g_{tt}^o} \,.
\ee
In \cite{Hartnoll:2020rwq} this predicted quasinormal mode was found explicitly in a black hole that came close to developing an inner horizon. In \cite{Hartnoll:2020fhc} an interior with additional bounces in $g_{tt}$ was considered, but only the first maximum, closest to the horizon, was found to have a corresponding quasinormal mode. We expect this to be the case here. The remaining poles, if they are present at all, are likely hidden on other sheets of the boundary Green's function.

\section{Additional detail on dynamics}

\subsection{Perturbations about Kasner}
\label{sec:kaspert}

Perturbations about the Kasner solution in (\ref{sec:kasner}) can be written as
\be
\dot g = v_n + \delta \dot g \,, \qquad \dot h = w_n + \delta \dot h \,.
\ee
By linearising the equations of motion (\ref{eq:deqs}) one finds the solutions
\be
\left(\delta \dot g, \delta \dot h\right) = \sum_{i=\{t,x,y\}} \left(a_{ni}, b_{ni} \right) e^{-3 p_n^i (\rho - \rho_{n}) + 2 V_{n}^i} \,. \label{eq:cor2}
\ee
Here $V_{n}^i$ are constants defined in (\ref{eq:pot2}).
The constants $a_{ni}$ and $b_{ni}$ can be obtained explicitly, up to an overall constant of integration.
The result (\ref{eq:cor2}) shows that the corrections $\delta \dot g$ and $\delta \dot h$ grow exponentially due to the fact that one of the Kasner exponents is negative.

\subsection{Bounce solution}
\label{sec:bouncesol}

The evolution of the metric components across the bounce can be solved for analytically. This is done by solving (\ref{eq:deqs}) with $\pa_g V$ and $\pa_h V$ taken constant. The solution shows explicitly the crossover between two Kasner regimes. Suppose that the bounce is off a wall of type $s_n =\{t,x,y\}$. One obtains
\be\label{eq:bsol}
|\dot h - 2 \partial_h V^{s_n}|^{2/3} |\dot h - w_{n}|^{1/(3 p_{n}^{s_n})}|\dot h - w_{n-1}|^{1/(3 p_{n-1}^{s_n})} = e^{\rho_n-\rho} \,.
\ee
Here $\rho_n$ is a constant of integration. There is an analogous expression for $\dot g$, following from (\ref{eq:linear}). From (\ref{eq:Wi}) we have $2 \partial_h V^t = -2$ and $2 \partial_h V^x = 2 \partial_h V^y = 1$.
The expression (\ref{eq:bsol}) was obtained previously for an Einstein-Maxwell bounce in \cite{Hartnoll:2020fhc}. In \cite{BKL} the bounce is instead solved directly for the functions $g$ and $h$ rather than their derivatives. Recall from (\ref{eq:cor2}) that in the approach to an $s_n$ bounce the exponent $p_{n-1}^{s_n} < 0$. Therefore from (\ref{eq:bsol}) one has $\dot h \to w_{n-1}$ as $\rho \ll \rho_n$ and $\dot h \to w_{n}$ as $\rho \gg \rho_n$. The constant $\rho_n$ is thus the location of the $n$th bounce, between Kasner epochs with exponents $w_{n-1}$ and $w_n$. There is some freedom in precisely where $\rho_n$ is placed, but so long as it is within the bounce regime, which is very short compared to the Kasner epochs, the precise location is unimportant. What is important is that the bounce solution extends to the Kasner epochs on both sides and hence can be matched up with the Kasner solutions on each side.

\subsection{Evolution of metric components within an era}
\label{sec:fiddly}

\begin{figure}[h]
    \centering
   \includegraphics[width=\textwidth]{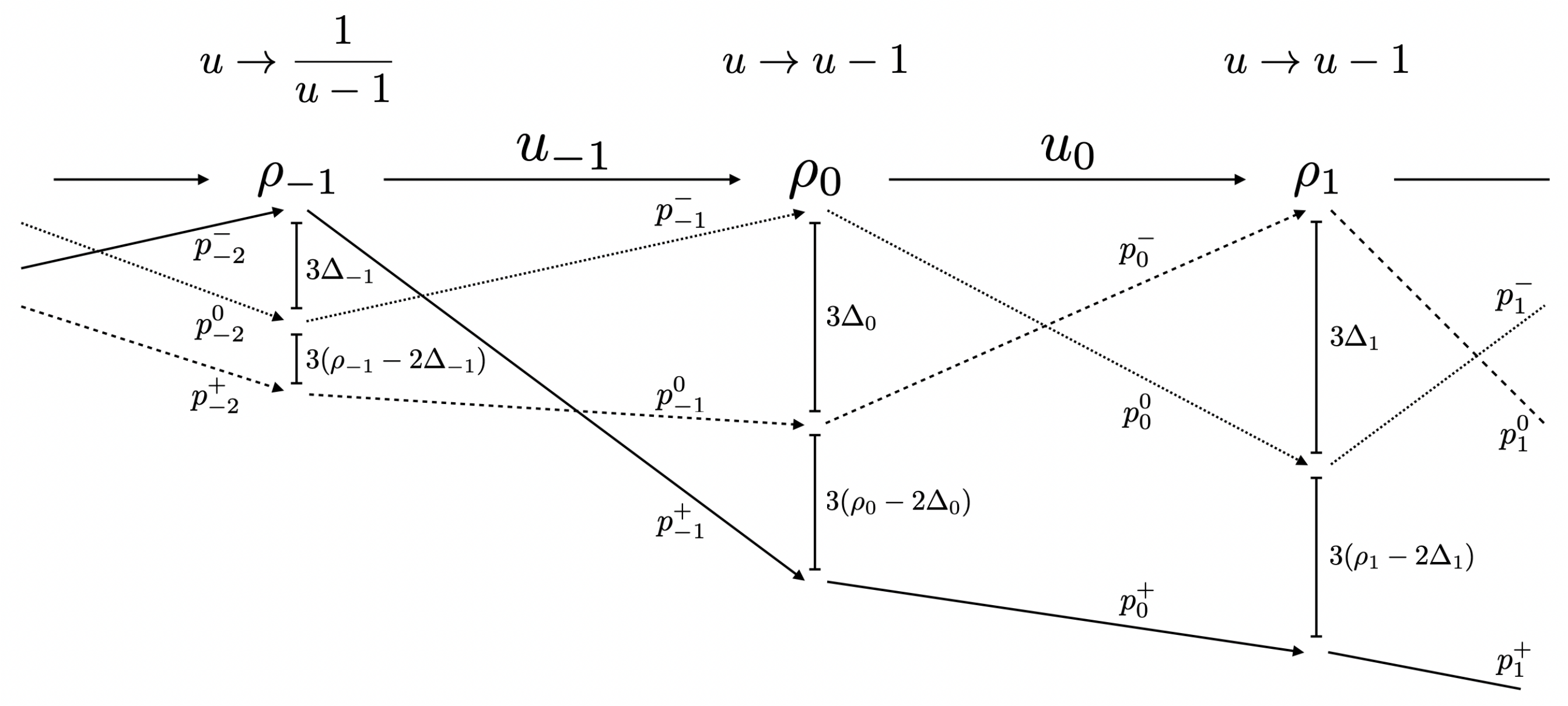}
    \caption{Kasner exponents and logarithm of the metric components at the start of a new era. The first bounce of the new era is at $\rho_0$ while the preceding `big bounce' is at $\rho_{-1}$. The dotted, dashed and solid lines follow the three metric components across epochs. The logarithm of the metric components attain values $0,-3 \Delta_n$ and $-3(\rho_n - \Delta_n)$ at the bounces. The Kasner exponents change behaviour between epochs according to (\ref{eq:ppp}).}
    \label{fig:era}
\end{figure}

In this Appendix we obtain a recursion relation for (one third of) the change in the logarithm of the growing metric component over the $n$th epoch
\be\label{eq:dn1}
\Delta_n \equiv - p_n^- (\rho_{n+1} - \rho_n) \,.
\ee
From the main text we know that the logarithm of the metric grows linearly in $\rho$ until it bounces at $\rho_{n+1}$. Furthermore, the logarithm of the metric reaches zero (to leading exponential accuracy) at $\rho_{n+1}$: the metric components reach $e^{-\rho_{n+1} + 2 V_{n+1}^i}$ at the $(n+1)$th bounce and for the dominant growing direction $2V_{n+1}^i = 2 V_{n+1} = \rho_{n+1}$, from (\ref{eq:rV}). In order to reach zero at $\rho_{n+1}$, the logarithm of this metric component must have been equal to $-3 \Delta_{n}$ at $\rho_n$, where it started growing. This is illustrated in Fig.~\ref{fig:era}.

Over the same epoch, the logarithm of the metric component with Kasner exponent $p^0_{n}$ is collapsing from $0$ to $-3 \Delta_{n+1}$. We know these limiting values because, according to the intra era transition rule in (\ref{eq:ppp}), the collapsing $p^0$ direction alternates with the growing $p^-$ direction throughout an era. We saw in the previous paragraph that the $p^-$ direction expands from $-3 \Delta_n$ to $0$, which must now connect onto the limiting values of the $p^0$ direction on either side. In order for the collapse from $0$ to $-3 \Delta_{n+1}$ to occur between $\rho_n$ and $\rho_{n+1}$ we must have
\be\label{eq:dn2}
\Delta_{n+1} = p^0_{n}(\rho_{n+1} - \rho_{n}) \,.
\ee
From (\ref{eq:dn1}) and (\ref{eq:dn2}) and repeatedly using the intra era transition $u_{n+1} = u_n - 1$ we obtain
\be\label{eq:Dk}
\Delta_{k-1} = - \frac{p^0_{k-2}}{p^-_{k-2}} \Delta_{k-2}  = \frac{1+u_{k-2}}{u_{k-2}} \Delta_{k-2} = \frac{1 + u_0}{1+x} \Delta_0 \,.
\ee
This gives the value attained by the direction with $p_{k-2}^0$ Kasner exponent at the subsequent big bounce at $\rho_{k-1}$. There is one more step left to obtain $\Delta_k$, but it must be treated separately because there is now a big bounce. Fig.~\ref{fig:era} may be helpful at this point.

From the inter era bounce rules, the second case in (\ref{eq:ppp}), we know that the $p_{k-2}^0$ direction becomes the growing direction $p^-_{k-1}$, while the direction $p^{-}_{k-2}$, whose logarithm of the metric reached zero at $\rho_{k-1}$, becomes the collapsing $p^+_{k-1}$ direction. This means that after the first epoch of the new era we may write
\be\label{eq:tilde}
\widetilde \Delta_k = -\frac{p_{k-1}^+}{p_{k-1}^-} \Delta_{k-1} = (u_{k-1}+1) \Delta_{k-1} = \frac{1+u_0}{x} \Delta_0 \,,
\ee
where in the final step we used the inter era transition $u_{k-1} = 1/(u_{k-2}-1)$. We put a tilde over $\widetilde \Delta_k$ in (\ref{eq:tilde}) because this is the value attained by the $p_{k-1}^+$ direction, while what we are aiming for, in order to reconnect to the next era, is the value attained by the $p_{k-1}^0$ direction (as this is the direction that becomes the growing $p_k^-$). However, from the metric (\ref{eq:metric}) we know that the sum of the logarithm of the three metric components must be $-3\rho$. At $\rho_k$ we know that the $p_{k-1}^-$ direction reached zero and the $p_{k-1}^+$ direction reached $-3\widetilde \Delta_{k}$. Therefore we have, using (\ref{eq:length}) and (\ref{eq:tilde}),
\be
\Delta_k = \rho_k - \widetilde \Delta_k = \rho_0 + \Delta_0 (k^2 +kx - 1) \,.
\ee
This is equation (\ref{eq:ddk}) in the main text.

\subsection{Behaviour at large volume}
\label{sec:large}

In the large volume limit, $\rho \to - \infty$, the variables $g$ and $h$ tend towards the minimum of the potential (\ref{eq:pot}). That is,
\be
g = \frac{1}{\sqrt{3}} \log \frac{f_x}{f_y} + \delta g \,, \qquad
h = \frac{1}{3} \log \frac{f_t^2}{f_x f_y} + \delta h \,.
\ee
In this limit the potential $V = \frac{1}{2} \log [3 (f_t f_x f_y)^{2/3}] + \frac{1}{2} \left(\delta g^2 + \delta h^2 \right) + \cdots$. Expanding the equations of motion (\ref{eq:deqs}) in $\delta g$ and $\delta h$, one obtains linear equations that are easily solved to give the leading asymptotic behaviour
\be\label{eq:deltag}
\delta g = A_g e^{\frac{1}{4} \rho} \sin \left(\frac{\sqrt{15}}{4} \rho + \Phi_g \right) \,.
\ee
Here $A_g$ and $\Phi_g$ are constants. The perturbation $\delta h$ has the same form as (\ref{eq:deltag}) but with independent constants of integration $A_h$ and $\Phi_h$. We see in (\ref{eq:deltag}) that $\delta g$ and $\delta h$ become exponentially small as $\rho \to - \infty$, showing that the linearisation of the equations is selfconsistent.

\section{Hecke operators and automorphic waveforms}
\label{sec:Hecke operators}

In this Appendix we prove the uniqueness of the decomposition \eqref{eq:malpha} of elements in $M_p$ in terms of matrices in $\Gamma(2)$ and $\Delta_p$, defined in  \eqref{eq:repsGamma2}. The exposition will closely follow \cite{Bogomolny:1992cj}, that discusses the modular group rather than $\Gamma(2)$.

We may first show that for every $m_p \in M_p$ there exists a $\gamma \in \Gamma(2)$ such that the product $\beta = \gamma m_p \in \Delta_p$. Let us write out the matrices explicitly,
\begin{equation}
   \gamma= \begin{pmatrix}
        a_1 &  b_1\\  c_1 &d_1
    \end{pmatrix} \qquad \text{and} \qquad m_p= \begin{pmatrix}
        a & b\\ c &d
    \end{pmatrix},
\end{equation}
so that
\begin{equation}
    \beta= \begin{pmatrix}
        a' & b'\\ c' &d'
    \end{pmatrix} = \begin{pmatrix}
        a_1a+b_1c & a_1b+b_1d\\ c_1a+d_1c &c_1b+d_1d
    \end{pmatrix} \,.
\end{equation}
From the definitions of $\Gamma(2)$ and $M_p$, we have that $\{a_1,d_1,a,d\}$ must be odd, while $\{b_1, c_1,b,c\}$ are even. We aim to choose the entries of $\gamma$ such that $\beta \in \Delta_p$. We can enforce $c'=0$ by choosing $c_1=\pm c/(c,a)$ and $d_1=\mp a/(c,a)$ with $(c,a)$ the biggest common divisor of $a$ and $c$. The sign should be chosen so as to ensure $d'>0$. Note that this is in line with the assumption that $c_1$ is even and $d_1$ odd. Moreover, with this choice $d'$ is found to be odd. The constraint on the determinant of $\gamma$ tells us that $a_1$ and $b_1$ need to be chosen so as to satisfy $a_1d_1-b_1c_1=1$.
Since $c_1$ and $d_1$ have been chosen to be relatively prime, B\'ezout's identity implies the existence of integers $a_1$ and $b_1$ which fulfill this requirement. In particular, $a_1$ is indeed found to be odd and this results in $a'$ being odd as well. Because $a'd' = p$ (as $\det \beta = \det \gamma \det m_p = p$), a prime, it must be that one of $a'$ or $d'$ equals $p$ and the other equals one. Now consider a pair $(\tilde a_1, \tilde b_1)$ that solves B\'ezout's identity. Any combination $(a_1, b_1) \equiv (\tilde a_1 + k c_1, \tilde b_1+kd_1)$ with $k$ integer is also a solution. We now show that we may choose $k$ such that $0\leq b'<2d'$.
From the definitions
\be
b' = \tilde a_1 b + \tilde b_1 d + k d' \,.
\ee
Here $\tilde a_1 b + \tilde b_1 d$ is a fixed integer. By subtracting off or adding enough copies of $d'$ (that is, by choosing $k$ appropriately) one can obtain $0 \leq b' < d'$. However, the $b'$ obtained in this way may be odd. In that case we add $d'$ to $b'$ by shifting $k$ by one. Because $d'$ is odd, this now leads to an even $b'$. In this way we obtain an even $b'$ with $0\leq b'<2d'$. This ensures that $\beta \in \Delta_p$.

Having demonstrated that any element in $M_p$ can be connected to an element in $\Delta_p$ by a matrix in $\Gamma(2)$, we now show that the decomposition \eqref{eq:malpha},
\be
m_p = \gamma \alpha_p \,,
\ee
is unique. Clearly, since $\alpha_p$ is invertible, $\gamma \alpha_p = \gamma' \alpha_p$ implies $\gamma = \gamma'$. Nonuniqueness would therefore require two matrices $\alpha_p \neq \alpha_p'$, related by an element of $\Gamma(2)$.
We now show that this is not possible, \emph{i.e.}~that $\gamma \alpha_p = \alpha_p'$ implies that $\gamma$ is the identity. Let 
\begin{equation}
   \alpha_p= \begin{pmatrix}
        a &  2b \\  0 & d
    \end{pmatrix}, \qquad  \alpha_p'= \begin{pmatrix}
        a' & 2b'\\ 0  & d'
    \end{pmatrix}, \qquad \text{and} \qquad \gamma =  \begin{pmatrix}
        a_1 & 2b_1\\ 2c_1  & d_1
    \end{pmatrix} \,.
\end{equation}
We have
\begin{equation}
    \gamma \alpha_p  = \begin{pmatrix}
        a_1 a &  2a_1b+2b_1d \\  2c_1a & 4c_1 b+d_1d
    \end{pmatrix} = \alpha_p'=  \begin{pmatrix}
        a' & 2b'\\ 0  & d'
    \end{pmatrix} .
\end{equation}
We aim to show that $\gamma$ is necessarily the identity matrix. Since the determinant of $\alpha_p$ is nonzero, $c_1$ must be zero. Hence, $a_1 d_1 = 1$ to ensure the determinant constraint for $\gamma$ and the two numbers are either both $-1$ or both $1$. However, both $d$ and $d_1 d$ are positive so that $a_1 = d_1 = 1$. Finally, $ 0\leq b'= b+b_1d < d$ while $ 0\leq b < d$ so that $b_1 = 0$. This proves the uniqueness of the decomposition \eqref{eq:malpha}.

\providecommand{\href}[2]{#2}\begingroup\raggedright\endgroup

\end{document}